\documentstyle[epsfig]{elsartrv}
\begin{document}
\newcommand{\be}{\begin{eqnarray}}
\newcommand{\ee}{\end{eqnarray}}

\begin{frontmatter}
\title{Physics of the Nucleon Sea Quark Distributions}
\author{R. Vogt\thanksref{doe}}
\address{ Nuclear Science Division, Lawrence Berkeley National Laboratory,  
Berkeley, CA 94720 \\ and \\
Physics Department, University of California at Davis, Davis, CA 95616}
\thanks[doe]{This work was supported in part by
the Director, Office of Energy Research, Division of Nuclear Physics
of the Office of High Energy and Nuclear Physics of the U. S.
Department of Energy under Contract Number DE-AC03-76SF00098.}

\begin{abstract}
Sea quark distributions in the nucleon have naively been
expected to be generated
perturbatively by gluon splitting.  In this case, 
there is no reason for the light 
quark and anti-quark sea distributions to be different.  
No asymmetries in the strange or heavy
quark sea distributions are predicted in the improved parton model.  However,
recent experiments have called these naive expectations into question.  A
violation of the Gottfried sum rule has been measured in several experiments,
suggesting that $\overline u < \overline d$ in the proton.  
Additionally, other measurements,
while not definitive, show that there may be an asymmetry in the strange and
anti-strange quark sea distributions.  These effects may require
nonperturbative explanations.  In this review we first discuss the
perturbative aspects of the sea quark distributions.  We then describe the
experiments that could point to nonperturbative contributions to the nucleon
sea.  Current phenomenological models that could explain some of these effects
are reviewed.
\end{abstract}
\end{frontmatter}

\section{Introduction}

Elastic scattering of the proton with electrons revealed that, unlike the
electron, the proton had a finite size \cite{epelas}.  Further experiments
with large momentum transfers
showed that, in addition to excitation of the proton into resonance states such
as the $\Delta^+$, the proton could break up into a large amount of debris.
Beyond the resonance region, it appeared as though the virtual photon probe was
no longer interacting with the proton as a whole but with point-like
constituents of the proton.  Early measurements of the structure functions
suggested that they were independent of the momentum transfer, $Q$ 
\cite{scaling}.
Out of these data emerged the parton model of the nucleon \cite{partons}.  
In this picture,
the proton has three valence quarks, two up, or $u$, quarks and one down, or
$d$ quark.  If these valence quarks, bound in the proton,
were its only constituents, there would be nothing further to tell.  However,
there is much more to the story.  When later experiments probed lower values of
the fraction of the proton momentum carried by the partons, $x$,
$Q^2$ scaling was found to be violated \cite{disdata}.  
Additionally, if the proton only contained valence quarks at
all values of $Q$, each valence quark should carry a third of the total proton
momentum.  This also proved to be untrue.  The average $x$ of the partons was
less than 1/3 and the parton density was correspondingly larger in this 
$x < 1/3$ region.
The low charged $x$ partons contributing to the proton structure functions are 
quark-anti-quark pairs, known as sea quarks.

The sea quarks are perturbatively
generated from gluon splitting, the gluons being the
mediators of the strong force.  Since the gluons 
are charge-neutral, they do not 
directly contribute to the proton structure functions.  When no interaction
occurs, the sea quarks and gluons serve to dress the bare valence quarks.
However, an interaction with a high energy probe such as a virtual photon
breaks the coherence of the proton wavefunction and hadronizes the sea quarks
and gluons.  The higher the energy of the probe, the smaller the value of $x$
that can be studied.  Currently the smallest values of $x$ have been reached in
$ep$ collisions at HERA \cite{zeus,h1} with a center of mass energy of 310 GeV.
These data 
show a strong increase of the sea quark and gluon densities at low $x$.

Inherently, the generation of sea quarks at low $x$ and $Q$ is a soft process
and cannot be calculated in perturbative Quantum Chromodynamics, QCD.
Typically a minimum perturbative momentum transfer is chosen at which the
valence quarks, sea quarks, and gluons are assigned distributions for further
evolution to higher $Q^2$.  Since these distributions are not calculable,
certain assumptions are made regarding the sea quarks.  Until recently, it was
always assumed that the sea quarks were all symmetric--there was no difference
between up, down and strange sea quarks and no difference between {\it e.g.}
the up and anti-up quarks in the sea.  In the last few years these assumptions
have been seriously challenged.  These challenges are the subject of this
review. 

\section{Generation of the perturbative sea}

\subsection{The improved parton model}

The parton model can be applied to the calculation of any cross section with
large momentum transfer, referred to as a `hard interaction'.  
At high energies, the hadron and parton masses are
neglected compared to the hard scattering scale $Q$.  Partons from the initial
state hadrons participating in the hard
scattering have momentum $xp^\mu$ where $x$ is the
fraction of the parent momentum carried by the parton, $0 \leq x \leq 1$, and
$p^\mu$ is the four-momentum of the initial hadron.  In the hard scattering,
$p^2 = 0$ since the hadron mass has been neglected.  
A hard cross section calculated in the parton model from collisions
of hadrons $A$ and $B$ to make final states $C$ and $D$ is schematically
\begin{equation}
\sigma_{AB \rightarrow CD}(p,p') \sim \int dx_1 dx_2 \frac{dz_1}{z_1}
\frac{dz_2}{z_2}
\widehat{\sigma}_{ij}(x_1p, x_2p') \phi_{i/A}(x_1) \phi_{j/B}(x_2) 
D_k^C(z_1) D_l^D(z_2) \, \, . 
\label{absigma}
\end{equation}
Here $\widehat{\sigma}_{ij}(x_1p, x_2p')$ is the lowest order partonic cross
section for partons $i$ and $j$ to make the final state $kl$ {\it i.e.} $gg
\rightarrow q \overline q$ for quark jet production.  The probability
density for finding parton $i$ in hadron $h$ with momentum fraction $x$ is 
$\phi_{i/h}(x)$.
To produce a final state hadron, $C$, from parton $k$ or $l$, the
partonic cross section is convoluted with a fragmentation function, $D_k^C(z)$,
that describes the hadronization of the parton.  The functions $\phi_{i/h}(x)$
which underlie all perturbative QCD calculations are the primary focus of this
review. 

The most transparent picture of the parton model is deep inelastic scattering
of an electron with a hadron.  In the hadron, partons exist in a virtual state
with definite momenta.  When the electron approaches the hadron, due to Lorentz
contraction and time dilation, the virtual state is frozen during the time it
takes for the electron to cross the hadron.  When the lepton transfers the
squared four-momentum $Q^2$ to the hadron via the virtual photon, the distance
of closest approach between the lepton and hadron is $\sim 1/Q$.  The ratio of
the ``area'' of the momentum exchange to the geometric area of the hadron,
$(1/Q^2)/\pi R_0^2$, measures the locality of the lepton probe.  When $Q^2$ is
large, this ratio is small, implying that the parton density in the hadron is
so dilute that only one parton interacts with the lepton.
The hadronization time scale of the partons produced in the hard scattering 
is taken to be long 
compared to that of the hard scattering.  Therefore, both the
initial and final states are decoupled from the hard interaction
\cite{cteqrmp}. 

Most of the existing information on the nucleon sea has been obtained from 
deep-inelastic scattering, DIS, and
lepton pair production by the Drell-Yan process.  The DIS and Drell-Yan
kinematics are briefly introduced,
following the review of Ref.~\cite{cteqrmp}.  While the
discussion remains at leading order for clarity, higher order calculations
are available, see {\it e.g.} \cite{hodis,hody}.
A typical deep-inelastic scattering reaction is
\begin{eqnarray}
l(p_l) + h(p_h) \rightarrow l'(p_{l'}) + X \label{dis}
\end{eqnarray}
where $l$ is a lepton such as 
an electron, muon or neutrino, $h$ is typically a
proton or a nucleon, and $X$ is the hadronic
final state. A virtual
vector boson such as a photon or a $W^\pm$ is exchanged in the
intermediate state.  DIS is an inclusive process so that the kinematics can be
specified by the measurement of the final lepton four momentum, 
$p_{l'}^\mu$.  The
momentum transfer is space-like, $q^\mu = p_l^\mu - p_{l'}^\mu$ and $-q^2 =
Q^2$.  The momentum fraction $x$, the Bjorken scaling variable, is the ratio of
the momentum transfer squared to the energy transferred from the lepton to the
hadronic system in the scattering,
\begin{eqnarray}
x = \frac{Q^2}{2p_h \cdot q} = \frac{Q^2}{2m_h \nu} \, \, . \label{xbj}
\end{eqnarray}
The second equality holds
in the target rest frame where $\nu = E_l - E_{l'}$ is the energy transferred
from the initial lepton.  The ratio of the energy
transfer to the total lepton energy in the target rest frame, $y$, 
is also useful,
\begin{eqnarray}
y = \frac{p_h \cdot q}{p_h \cdot p_l} = \frac{\nu}{E_l} \,\, . \label{ydef}
\end{eqnarray}
The mass of the hadronic final state, $W$, can be large when $x$ is fixed and
$Q^2$ is large since
\begin{eqnarray}
W^2 = m_h^2 + \frac{Q^2}{x} (1-x) \, \, . \label{wdef}
\end{eqnarray}

Depending on the initial lepton and the squared momentum transfer, the
exchanged boson can be a photon, $\gamma$, or one of 
the electroweak vector bosons, $W^+$, $W^-$, and $Z^0$.  When a $\gamma$ or 
$Z^0$ is exchanged, there is no charge transferred by the virtual vector 
boson and the reaction is said to be a neutral current process.  The exchange
of a $W^+$ or a $W^-$, on the other hand, is through a charged current.  The 
following discussion applies to both neutral and charged current processes.

The DIS cross section can be separated into leptonic and hadronic components so
that 
\begin{eqnarray}
d\sigma = \frac{d^3p_{l'}}{2s|\vec{p}_{l'}|} \frac{c_V^4}{4 \pi^2 (q^2 -
m_V^2)^2} L^{\mu \nu}(p_l,q) W_{\mu \nu}(p_h, q)  \label{dissig}
\end{eqnarray}
where $V = \gamma$, $W^\pm$, or $Z^0$ with mass $m_V$.  The vector 
coupling constants are
$c_\gamma = e$, $c_{W^\pm} = g/2 \sqrt{2}$, and $c_{Z^0} = g/2\cos \theta_W$
where $g = e/\sin \theta_W$ and $\theta_W$ is the Weinberg angle.  The boson
propagator is assumed to be $\propto (q^2 - m_V^2)^{-1}$.  

The lepton tensor is
\begin{eqnarray} L^{\mu \nu}(p_l,q) = \frac{1}{2}
{\rm Tr} \, [p_l\!\!\!\!/ \, \Gamma_{Vl}^\mu
(p_l\!\!\!\!/ - q\!\!\!/) \Gamma_{Vl}^\nu ] \label{leptens}
\end{eqnarray}
where the couplings of the vector bosons to the leptons are $\Gamma_{\gamma
l^\pm}^\mu = \gamma^\mu$, $\Gamma_{W^+ \nu}^\mu = \gamma^\mu (1 - \gamma^5)$,
$\Gamma_{W^- \overline \nu}^\mu = \gamma^\mu (1 + \gamma^5)$, and
$\Gamma_{Z^0}^\mu 
= \gamma^\mu (c_V^f - c_A^f \gamma^5)$ where $c_V^f$ and $c_A^f$ are the vector
and axial vector couplings for leptons and quarks of flavor $f$.  Since this
review focuses on values of $Q^2$ where the virtual photon contribution is 
much larger than that of the $Z^0$, the $Z^0$ will no longer be discussed.  
The lepton
spin average factor of 1/2 in Eq.~(\ref{leptens}) is absent in neutrino 
scattering.  

The hadron tensor is 
\begin{eqnarray}
W_{\mu \nu}(p_h,q) = 2 \pi^3
\sum_{\sigma, \, X} \langle h^\sigma(p_h)%, \sigma)
| j_\mu^{V \dagger}(0) | X \rangle \langle X| j_\nu^V(0)|h^\sigma(p_h)%,\sigma)
\rangle \delta^4(p_h + q - p_X) \,\,  \label{hadtens}
\end{eqnarray}
where the nucleon spin, $\sigma$, is averaged over and the final state, $X$, 
has
been summed over.  Symmetries allow the  hadron tensor to be written in terms 
of structure functions describing lepton scattering with point particles 
so that
\begin{eqnarray}
W_{\mu \nu}(p_h,q) & = & -\left(g_{\mu \nu} - \frac{q_\mu q_\nu}{q^2}
\right)W_1(x,Q^2) \label{wmunu} \\
&   & \mbox{} + \left(p_{h \, \mu} - q_\mu \frac{p_h \cdot q}{q^2} \right)
 \left(p_{h \, \nu} - q_\nu \frac{p_h \cdot q}{q^2} \right)
\frac{W_2(x,Q^2)}{m_h^2} \nonumber \\ &   & \mbox{}
- i \epsilon_{\mu \nu \lambda \sigma}
\frac{W_3(x,Q^2)}{m_h^2} \,\,  \nonumber
\end{eqnarray}
where parity conservation in strong interactions implies $W_3 = 0$ 
for virtual photons.  In the Bjorken limit, $\nu, \, Q^2 \rightarrow \infty$
for fixed $x$, the structure functions $W_i$ are functions of $x$ alone to
within logarithmic corrections in $Q^2$,
\begin{eqnarray}
F_1^{Vh}(x,Q^2) & = & W_1(x,Q^2)\, , \nonumber \\
F_2^{Vh}(x,Q^2) & = & \frac{\nu}{m_h} W_2(x,Q^2) \, , \label{fdef} \\
F_3^{Vh}(x,Q^2) & = & \frac{\nu}{m_h} W_3(x,Q^2) \,\, . \nonumber
\end{eqnarray}
The DIS cross section can then be expressed in terms of $x$, $y$, and
$F_i^{Vh}$ as 
\begin{eqnarray}
\frac{d\sigma^{lh}}{dx dy} & = & N^{lV} \left[ \frac{y^2}{2} 
2xF_1^{Vh}(x,Q^2) + 
\left( 1 - y - \frac{m_h x y}{2E} \right) F_2^{Vh}(x,Q^2) \right. \nonumber \\
&   & \mbox{} + \left. \delta_V \left(y -
\frac{y^2}{2} \right) xF_3^{Vh}(x,Q^2) \right] \, \, , \label{dissigpart}
\end{eqnarray}
where $E$ is the lepton energy, $m_h$ is the hadron mass, and $\delta_V = 0$
for $\gamma$, $-1$ for $W^-$, and +1 for $W^+$.  The normalizations are
\begin{eqnarray}
N^{l^\pm \gamma} & = & 8 \pi \alpha^2 \frac{m_h E}{Q^4} \label{normdefgam} \\
N^{\nu W^+} & = & N^{\overline \nu W^-} = \frac{\pi \alpha^2 m_h E}{2 \sin^4
\theta_W (Q^2 + M_W^2)^2} = \frac{G_F^2}{\pi} 
\frac{m_h E M_W^4}{(Q^2 + M_W^2)^2} 
\, \,  \label{normdefwpm}
\end{eqnarray} 
where $\alpha$ is the electromagnetic coupling constant.
The lepton-hadron cross section in Eq.~(\ref{dissigpart})
is equivalent to lepton scattering from
point-like objects.  These objects are the partons.

In the parton model, only charged partonic constituents of the hadron, 
the quarks and anti-quarks, 
couple to the
electroweak currents at leading order.  The DIS cross section is then
proportional to the elastic quark-lepton scattering cross section
multiplied by the probability of finding a quark or anti-quark
of flavor $f$ and momentum fraction $\xi$,
\begin{eqnarray}
\frac{d\sigma^{lh}(p_h,q)}{dE_{p_l'} d\Omega_{p_l'}} = \sum_f \int_0^1 d\xi
\frac{d\sigma^{lf}(\xi p_h,q)}{dE_{p_l'} d\Omega_{p_l'}} ( \phi_{f/h}(\xi)
+  \phi_{\overline f/h}(\xi) ) \, \, . \label{sigdiff}
\end{eqnarray}
The structure functions can then be directly related to the probability
densities by
\begin{eqnarray}
F_{1,3}^{Vh}(x) & = & \sum_f \int_0^1 \frac{d\xi}{\xi} 
F_{1,3}^{Vf} \left(\frac{x}{\xi}
\right) ( \phi_{f/h}(\xi)
+  \phi_{\overline f/h}(\xi) )  \\
F_2^{Vh}(x)     & = & \sum_f \int_0^1 d\xi F_2^{Vf} \left(\frac{x}{\xi}
\right) ( \phi_{f/h}(\xi)
+  \phi_{\overline f/h}(\xi) ) \,\,\, , \label{fancyfs}
\end{eqnarray} 
where $F_i^{Vf}$ are the parton-level structure functions 
calculated at leading order.  

The electromagnetic, neutral current scattering, case where $V = \gamma$ is 
discussed first.  Here, Eq.~(\ref{dis}) is $lh \rightarrow l X$ where
$l = e^\pm$ or $\mu^\pm$.
It can be shown \cite{cteqrmp} that $2F_1^{\gamma f}(x) = F_2^{\gamma 
f}(x) = Q_f^2
\delta(1-x)$ and that, subsequently, the hadronic structure functions  are
\begin{eqnarray}
2xF_1^{\gamma h}(x) = F_2^{\gamma h}(x) = \sum_f Q_f^2 x ( \phi_{f/h}(x)
+  \phi_{\overline f/h}(x) )
\label{plainfs} \, \, \, .
\end{eqnarray}
The simplified notation $\phi_{u/p}(x) = u_p(x)$, $\phi_{\overline
u/p}(x) = \overline u_p(x)$, {\it etc.} is now introduced
for the partons in the proton.
In this case, Eq.~(\ref{plainfs}) is typically written as
\begin{eqnarray}
2xF_1^{\gamma p}(x) = F_2^{\gamma p}(x) = 
\sum_f Q_f^2 x (f_p(x) + \overline f_p(x) ) \label{f2q} \, \,\, 
\end{eqnarray}
where $f = u, d, s, c,
\cdots$.  Note that $f(x)$ is the parton probability density while $x f(x)$ is
typically referred to as the parton momentum distribution.
The $u$ and $d$ quarks are already present in the proton since, in
the constituent quark model, the proton is a $uud$ state where the valence
quarks carry the bulk of the quark momentum at large $x$.  The sea quarks do 
not contribute to the baryon number of the hadron since they
are always produced in a virtual quark-anti-quark pair.  Therefore, for every
$\overline u_p$ there is a $u^s_p$ (denoted $u^s_p$ to avoid confusion with the
total $u_p$ quark density).  The $u$ and $d$ quark
densities include both valence and sea quark densities so that 
\begin{eqnarray}
u_p(x) & = & u^v_p(x) + u^s_p(x) \label{udist} \\
d_p(x) & = & d^v_p(x) + d^s_p(x) \label{ddist} \,\,\, .
\end{eqnarray}
The number of valence quarks in the proton
sums to three,
\begin{eqnarray} 
\int_0^1 dx \, u^v_p(x) & = & 2 \label{uvalsum} \\
\int_0^1 dx \, d^v_p(x) & = & 1 \label{dvalsum} \,\,\, ,
\end{eqnarray}
two up valence quarks and one down valence quark.

In addition to the proton parton distributions, DIS on targets such as
deuterium or other light nuclei measures the neutron distributions. Typically,
one assumes that the per nucleon structure function is {\it e.g.} for a 
deuterium target
\begin{eqnarray}
F_2^{\gamma {\rm D}}(x) = 
\frac{1}{2} \left( F_2^{\gamma p}(x) + F_2^{\gamma n}(x)
\right) \,\,\, , \label{f2d}
\end{eqnarray}
where the neutron parton densities are related to those in the proton by
charge symmetry so that
\begin{eqnarray}
u_p(x) & = & d_n(x) \,\,\,\,\, \,\,\,\,\, d_p(x) = u_n(x) \nonumber \\
\overline u_p(x) & = & \overline d_n(x) \,\,\,\,\, \,\,\,\,\, 
\overline d_p(x) = \overline u_n(x) \label{pvsn} \\
s_p(x) & = & s_n(x) \,\,\,\,\, \,\,\,\,\, \overline s_p(x) = \overline s_n(x)
\,\,\, \cdots \nonumber
\end{eqnarray}

Charge symmetry is also used to establish the valence quark sum rules for the
neutron, as in
Eqs.~(\ref{uvalsum}) and (\ref{dvalsum}).  The structure functions of other
baryons are unmeasured but could be inferred from relationships like those
of Eq.~(\ref{pvsn}).  Parton distribution functions have also been
determined with limited statistics for mesons.  In general, the valence quark
densities of the meson are assumed to be $\int_0^1 dx q^v_M(x) = 2$ where
$\overline u$ and $d$ are the valence quarks when 
$M = \pi^-$.  In this case, $u_{\pi^-}$ is a component of the pion sea.

An exchanged photon treats all charged partons identically.  Thus
the electromagnetic structure function of the proton, Eq.~(\ref{f2q}), 
is the sum over all parton flavors.  However, in neutrino-induced interactions,
to determine the momentum fraction of the proton carried by the interacting
parton, the reactions must proceed by charge 
current scattering for the final-state electron or muon to be detected.
Thus, the DIS reactions are {\it e.g.} $\nu_e h \rightarrow e^-
X$ with an exchanged $W^+$ and $\overline \nu_e h \rightarrow e^+ X$ with an 
exchanged $W^-$.  Now not all quarks and anti-quarks couple to the 
$W^+$ and $W^-$.  In the reaction $\nu_e h \rightarrow e^- X$, the incoming
$\nu_e$ ``decays'' to $W^+$ by $\nu_e \rightarrow W^+ e^-$.  Then
$W^+ d \rightarrow u$, $W^+ s \rightarrow c$, $W^+ \overline u \rightarrow 
\overline d$, and $W^+ \overline c \rightarrow 
\overline s$ so that the $W^+$ couples to the $d$, $s$, $\overline u$, and
$\overline c$ quarks in the proton.  Similarly, the $W^-$ emitted in the
$\overline \nu_e \rightarrow W^- e^+$ 
conversion can couple to the $\overline d$,
$\overline s$, $u$, and $c$ quarks in the proton.
Therefore, for a neutrino beam with exchanged $W^+$, 
the structure functions are
\begin{eqnarray}
F_2^{W^+ p}(x) & = & 2x \left[ d_p(x) + s_p(x) + b_p(x) + \overline u_p(x)
+ \overline c_p(x) \right] \label{f2wpp} \\
F_3^{W^+ p}(x) & = & 2 \left[ d_p(x) + s_p(x) + b_p(x) - \overline u_p(x)
- \overline c_p(x) \right] \label{f3wpp} \,\,\, . 
\end{eqnarray}
Note that the $b$ quark has been included in the proton sea while the still 
heavier top quark has not.
Likewise, the structure functions determined by an anti-neutrino beam with
exchanged $W^-$ are
\begin{eqnarray}
F_2^{W^- p}(x) & = & 2x \left[ \overline d_p(x) + \overline s_p(x) + \overline
b_p(x) + u_p(x) + c_p(x) \right] \label{f2wmp} \\
F_3^{W^- p}(x) & = & -2 \left[ \overline d_p(x) + \overline s_p(x) + \overline
b_p(x) - u_p(x) - c_p(x) \right] \label{f3wmp} \,\,\, . 
\end{eqnarray}

Some DIS data, particularly with neutrino beams, have been taken on nuclear 
targets heavier than deuterium.
Therefore it is useful to write 
the electromagnetic, neutrino and anti-neutrino structure functions in terms of
an isoscalar nuclear target.  (In an isoscalar target, $Z = N = A/2$, where
$A$ is the nuclear mass number, $Z$ is the number of protons and $N$ is the 
number of neutrons.)  Assuming no modifications of the parton densities in the
nucleus, the resulting structure functions per nucleon
are:
\begin{eqnarray}
  F_2^{\gamma N_0}(x) & =& \frac{5}{18} x 
   [ u_p(x) + \overline u_p(x) +d_p(x) +\overline d_p(x) 
+ \frac{8}{5}(c(x)+\overline c(x)) \nonumber \\  &   & \mbox{} + 
\frac{2}{5} (s_p(x) + \overline s_p(x) + b_p(x) + \overline b_p(x)) ] \, ,
\label{f2gamiso} \\
F_2^{W^+ N_0} (x) &=& x[u_p(x)+ \overline u_p(x) +d_p(x) +\overline d_p(x)
  + 2 s_p(x) + 2 b_p(x) + 2 \overline c_p(x)] \, , \label{f2wpiso} \\ 
xF_3^{W^+ N_0}(x) &=& x[u_p(x) + d_p(x) -\overline u_p(x) - \overline d_p(x) 
  +2 s_p(x) + 2 b_p(x) - 2 \overline c_p(x)] \, , \label{f3wpiso} \\ 
F_2^{W^- N_0} (x) &=& x[u_p(x)+ \overline u_p(x) +d_p(x) +
  \overline d_p(x) + 2 \overline s_p(x) + 2 \overline b_p(x) + 2 
c_p(x)] \, , \label{f2wmiso} \\
xF_3^{W^- N_0}(x) &=& x[u_p(x) + d_p(x) -\overline u_p(x) - \overline d_p(x) 
  -2 \overline s_p(x) - 2 \overline b_p(x) + 2 c_p(x)] \, . \label{f3wmiso}  
\end{eqnarray}

Since measurements of the lepton pairs produced in the
Drell-Yan process play an important role in the
determination of the nucleon sea, the Drell-Yan process \cite{drell}
is now introduced.
At leading order
a quark from the projectile annihilates with its corresponding anti-quark in the
target, producing a virtual photon which then decays to a lepton pair with
invariant mass $M$, $q \overline q \rightarrow
\gamma^\star \rightarrow l^+ l^-$.
The Drell-Yan cross section involves a convolution of the parton 
distributions of both the projectile and the target hadrons.  
The partonic cross section for Drell-Yan 
production is
\begin{eqnarray}
\frac{d\widehat{\sigma}}{dM} = \frac{8 \pi \alpha^2}{9M} Q_f^2 
\delta(\widehat{s} - M^2)
\end{eqnarray}
where $\widehat{s} = x_1x_2s$, $x_1$ is the fractional momentum carried by the
projectile parton, $x_2$ is the fractional momentum of the target parton, 
and $s$ is the square of the hadron-hadron center
of mass four-momentum. 
To obtain the hadroproduction cross section as a function of
pair mass the partonic cross
section is convoluted with the quark and anti-quark densities evaluated at 
scale $M$,
\begin{eqnarray}
\frac{d\sigma^{\rm DY}}{dM} & = & 
\frac{8 \pi \alpha^2}{9M} \int_0^1 dx_1 dx_2 \, \delta(\widehat{s} - M^2)
\sum_f Q_f^2 [f_p(x_1,M^2) \overline f_p(x_2,M^2) \nonumber \\
&   & \mbox{} + \overline f_p(x_1,M^2) f_p(x_2,M^2) ] \, \, .
\end{eqnarray}

Besides deep-inelastic scattering and Drell-Yan pair production, prompt photon,
heavy quark, quarkonium, and jet production are among the processes
calculable in perturbative QCD.  To obtain the hadronic cross sections, the
partonic cross section must convoluted with the projectile and target parton
densities.  For such a convolution to be reliable, the
hadronic cross sections must be factorizable into the process specific
short-distance cross section and the parton distributions which cannot be
obtained from a first principles calculation and must be assumed to be
universal \cite{css}, as in Eq.~(\ref{absigma}).  
Since the parton distributions are properties of the
hadron wavefunction and thus nonperturbative in nature, a universal set of
parton densities for a given hadron is typically obtained
from fits to DIS and other data.  How this is generally done is described in
the next sections.

\subsection{$Q^2$ Evolution}

Deep inelastic scattering data are available over a wide range of $x$ and $Q^2$
\cite{disdata,zeus,h1}.
In the naive parton model, scaling was assumed, {\it i.e.} the parton
distributions were thought to be independent of the $Q^2$ of the virtual
vector boson \cite{scaling}.  However, further measurements \cite{disdata}
soon showed that this was not the case.
To obtain a reliable set of parton densities at all $x$ and $Q^2$, a
description of their evolution in $x$ and $Q^2$ must be available.  
While this is relatively simple at fixed $x$ when $Q^2$ is large, the small $x$
behavior of the parton densities is more challenging, as will be discussed 
later.

Only tree-level (leading order) calculations have been described so far.
However, many processes have now been calculated at least to next-to-leading
order in QCD.  Next-to-leading order calculations involve both real and virtual
corrections to the leading order diagrams.  At any order in perturbation
theory, the hadronic cross section factorizes into a perturbatively calculable
partonic cross section and the parton distributions in the hadron, as in
Eq.~(\ref{absigma}).  Factorization provides a scheme for separating the short
distance partonic cross section from the nonperturbative parton distribution
functions, generally referred to as a factorization scheme.  The momentum 
scale at which the separation occurs is the factorization scale.  

There are two commonly used factorization schemes, the DIS and the
$\overline{\rm MS}$ schemes.  In the DIS scheme, $F_2^{Vh}$ is defined so that
all corrections are absorbed into the quark and anti-quark distributions
order-by-order in perturbation theory.  Thus $F_2^{Vh}$ is trivial.  However,
only $F_2^{Vh}$ enjoys this simplicity 
while all other structure functions must be
corrected order-by-order in $\alpha_s$.  Additionally, the gluon distribution
is not fixed in this scheme.  The more commonly used $\overline{\rm MS}$ or
minimal subtraction scheme  defines the parton distributions in terms of parton
creation and annihilation operators acting on the hadron wavefunction.  The
next-to-leading order corrections to the hard scattering are somewhat more
complex but the $\overline{\rm MS}$ scheme is more physical and includes the
gluon through evolution.  

In addition to the factorization scheme and scale dependence,
another scale, the renormalization scale, is required to remove the ultraviolet
divergences that occur in the calculation of the partonic cross section.  In
the $\overline{\rm MS}$ scheme, a uniform mass or energy scale is assigned to
all diagrams required to calculate the cross section, helping to regulate the
divergences.  Such a scale assignment automatically introduces a running
coupling $\alpha_s(\mu_R^2)$ 
which allows the cross section to remain independent
of the renormalization scale $\mu_R$.

There is no reason why the factorization and renormalization scales should be 
different and they are indeed usually chosen to be the same.  Choosing the two
scales to be equal in {\it e.g.}\ the $\overline{\rm MS}$ scheme ensures that
the parton distributions can be calculated with the local operators and
divergences in the cross sections can be handled simultaneously.
The cross section, a physical quantity, should not depend on the choice of
scheme or scale and would not if it could be calculated to all orders in
perturbation theory.  However, since the expansion of the cross section in
powers of the strong coupling constant is truncated at finite order,
a dependence on the scale and scheme remains.  The scale at which the hard
scattering occurs is then generally the one at which the parton distributions
are evaluated.

An important byproduct of factorization is that masuring an observable such as
$F_2^{\gamma p}$ at one value of $Q^2$ such as $Q_0^2$ enables one to predict
the behavior of $F_2^{\gamma p}$ at another value of $Q^2$ via evolution.
In the evolution from $Q_0^2$ to $Q^2$,
a parton of momentum fraction $x$, probed by a virtual photon with momentum
transfer $Q_0^2$, can split into two partons with lower
momentum fractions, $y$ and $z$, with $y+z = x$.  For a virtual photon to
resolve a charged parton of momentum $y$,  $y<x$, it must have a larger
momentum transfer than $Q_0^2$, $Q^2 > Q_0^2$.  When $x$ and $Q^2$ are not
too small, the Dokshitzer-Gribov-Lipatov-Altarelli-Parisi (DGLAP) equations
\cite{DGLAP} describe how the parton distributions are determined at 
$Q^2 > Q_0^2$ starting from an initial distribution at a low scale $Q_0^2$.
Evolution in $Q^2$ links the charged quark and anti-quark distributions with
the neutral partons, or gluons, since quarks emit gluons while gluons can
split into $q \overline q$ pairs or into two gluons.  The gluon is thus defined
through its evolution in the $\overline{\rm MS}$ scheme.  
At leading order, the DGLAP equations are
\begin{eqnarray}
\frac{dq_i(x,Q^2)}{d \log Q^2} & = & \frac{\alpha_s(Q^2)}{2 \pi} \int_x^1
\frac{dy}{y} \left[ q_i(y,Q^2)P_{qq} \left( \frac{x}{y} \right) + g(y,Q^2)
P_{qg} \left( \frac{x}{y} \right) \right] \label{qdglap} \\
\frac{dg(x,Q^2)}{d \log Q^2} & = & \frac{\alpha_s(Q^2)}{2 \pi} \int_x^1
\frac{dy}{y} \left[ \sum_i q_i(y,Q^2)P_{gq} \left( \frac{x}{y} \right) + 
g(y,Q^2) P_{gg} \left( \frac{x}{y} \right) \right] \,\,\, , \label{gdglap}
\end{eqnarray}
where the sum $i = 1 \ldots 2n_f$ runs over quarks and anti-quarks of all
flavors.  The  probability that parton $i$ has emitted parton $j$ is
proportional to the splitting functions $P_{ij}$.  The strong coupling 
constant, $\alpha_s$,
appears because of the quark-gluon and gluon-gluon vertices.  The parton
distributions in Eqs.~(\ref{sigdiff}) - (\ref{f3wmp}) all have had the $Q^2$
dependence suppressed.  The suppression of the $Q^2$ argument of the parton
distribution functions is common throughout this review.

While the evolution of the parton distributions can be described by
perturbative QCD, some initial set of parton distributions must be defined at
scale $Q_0^2$.  This initial scale is somewhat arbitrary but can be thought of
as the boundary between perturbative and nonperturbative physics.  The initial
scale is usually in the range $0.5 < Q_0^2 < 5$ GeV$^2$. 

\subsection{Experimental studies}

Although a great deal of precise information is generally obtainable from DIS,
particularly since the parton densities of only one hadron are involved, 
DIS does not
distinguish between quark flavors.  Other processes are also needed to
identify the effects of individual parton flavors.  A large collection of data
are used to fit the parton distributions at the initial scale $Q_0$.  The
DGLAP equations then evolve the parton densities
to higher scales.  A list of some
relevant 
measurements and which parton distributions they illuminate is given below:
\begin{itemize}
\item $F_2^{\gamma p}(x,Q^2)$ from $e^\pm p$ interactions at the HERA collider
\cite{zeus,h1}, charged partons at small $x$;
\item $F_2^{\gamma p}(x,Q^2)$ and $F_2^{\gamma {\rm D}}(x,Q^2)$ 
from $\mu$ beams by
NMC \cite{NMC} and BCDMS \cite{BCDMS}, charged partons;
\item $F_2^{W^\pm N}(x,Q^2)$ and $xF_3^{W^\pm N}(x,Q^2)$ from $\nu$ beams
on nuclear targets \cite{CCFRfig,CDHSW,CDHS}, charged partons;
\item prompt photons, $qg \rightarrow \gamma q$, gluon distribution;
\item Drell-Yan production, $q \overline q \rightarrow \gamma* \rightarrow l^+
l^-$, sea quark distributions;
\item $W^\pm$ and $Z^0$ production in $p \overline p$ colliders, $u$ and $d$
distributions --- {\it e.g.} $\sigma_{W^\pm}/\sigma_{Z^0} \propto d/u$;
\item $Z^0$ pole measurements in $e^+ e^-$ collisions \cite{LEP},
intrinsic scale, $\Lambda_{\rm QCD}$, of
$\alpha_s$.
\end{itemize}
Note that combinations of $F_2^{Vh}$ and $F_3^{Vh}$ can pick out certain parton
densities.  For example, using Eqs.~(\ref{f2wpp})-(\ref{f3wmp}),
$xF_3^{W^+ p} + xF_3^{W^- p} + F_2^{W^+ p} - F_2^{W^-
p} = xd_p^v(x,Q^2)$.  In fact, if both $p$ and $n$ targets are used with the
assumption of charge symmetry and an SU(3) flavor 
symmetric sea, $\overline u_p =
\overline d_p = \overline s_p$ with $\overline c_p = \overline b_p = 0$, the
parton distributions are overdetermined and consistency checks can be made.
It turns out, as discussed in this review, that there
is no sea quark SU(3) flavor symmetry.

\begin{figure}[htbp]
\setlength{\epsfxsize=\textwidth}
\setlength{\epsfysize=0.5\textheight}
\centerline{\epsffile{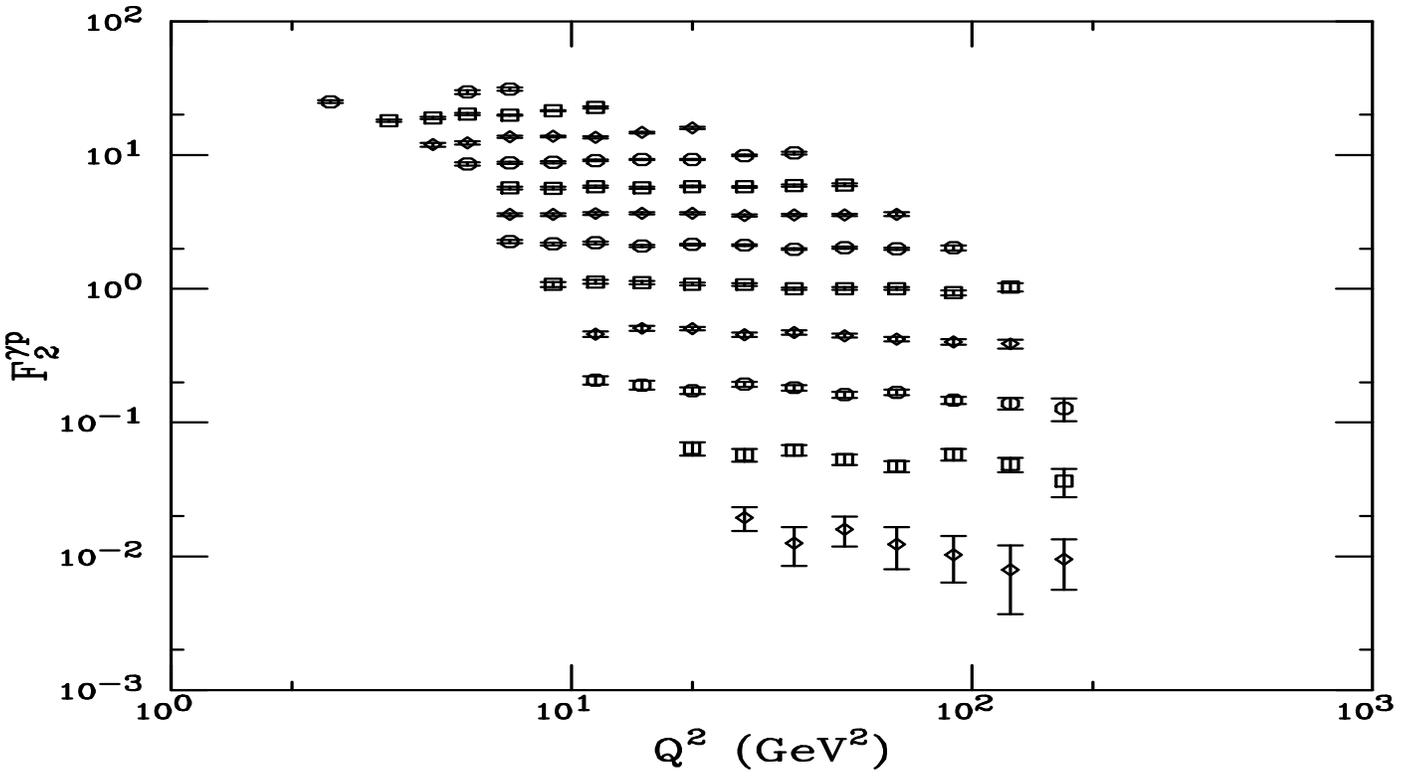}}
\caption[]{The proton structure function $F_2^{\gamma p}$ as a function of
$Q^2$ for muon beams with energies 120, 200, 240, and 280 GeV measured by 
the European Muon Collaboration \protect\cite{EMC}. 
From top to bottom the values of
the $x$ bins are:  0.0175, 0.03, 0.05, 0.08, 0.125, 0.175, 0.25, 0.35, 0.45,
0.55, 0.65, and 0.75. To separate the data, $F_2^{\gamma p}$ in each $x$ bin is
scaled by a factor of 1.5 from the next higher $x$ bin.  Therefore, only the
highest $x$ data (lowest points) have the correct scale.  Only the statistical
uncertainty is shown.} 
\label{emcfig}
\end{figure}

\begin{figure}[htbp]
\setlength{\epsfxsize=\textwidth}
\setlength{\epsfysize=0.5\textheight}
\centerline{\epsffile{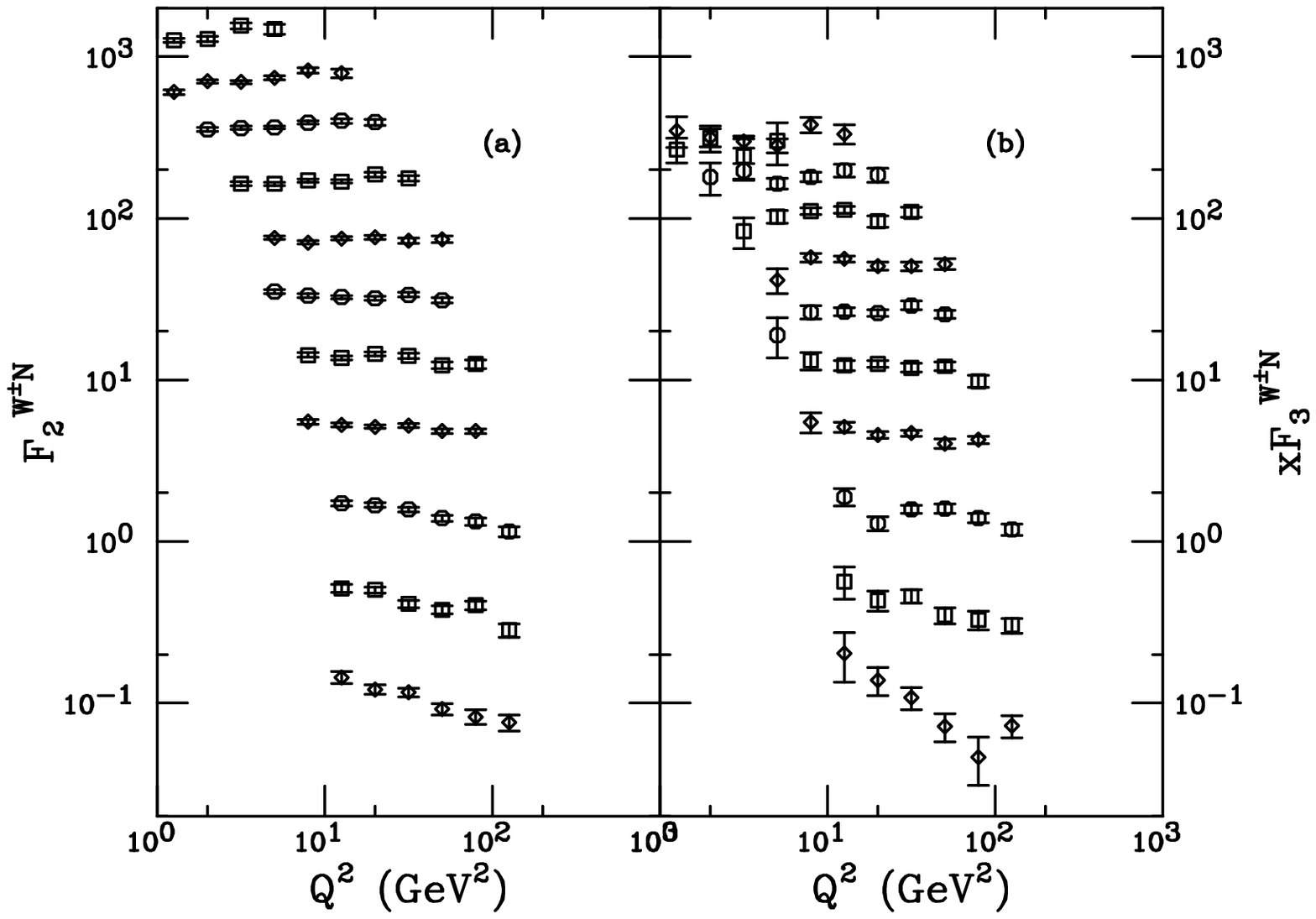}}
\caption[]{The nucleon structure functions $F_2^{W^\pm N}$ (a) and 
$x F_3^{W^\pm N}$ (b) as a function of
$Q^2$ \protect\cite{CCFRfig} from neutrino-iron deep-inelastic scattering.
The $x$ bins from top to bottom are: 0.015, 0.045, 0.08, 0.125,
0.175, 0.225, 0.275, 0.35, 0.45, 0.55, and 0.65. To separate the data, the 
structure functions in each $x$ bin are
scaled by a factor of 2 from the next higher $x$ bin.  Therefore, only the
highest $x$ data (lowest points) have the correct scale.  Only the statistical
uncertainty is shown.} 
\label{ccfrfig}
\end{figure}

\begin{figure}[htbp]
\setlength{\epsfxsize=\textwidth}
\setlength{\epsfysize=0.5\textheight}
\centerline{\epsffile{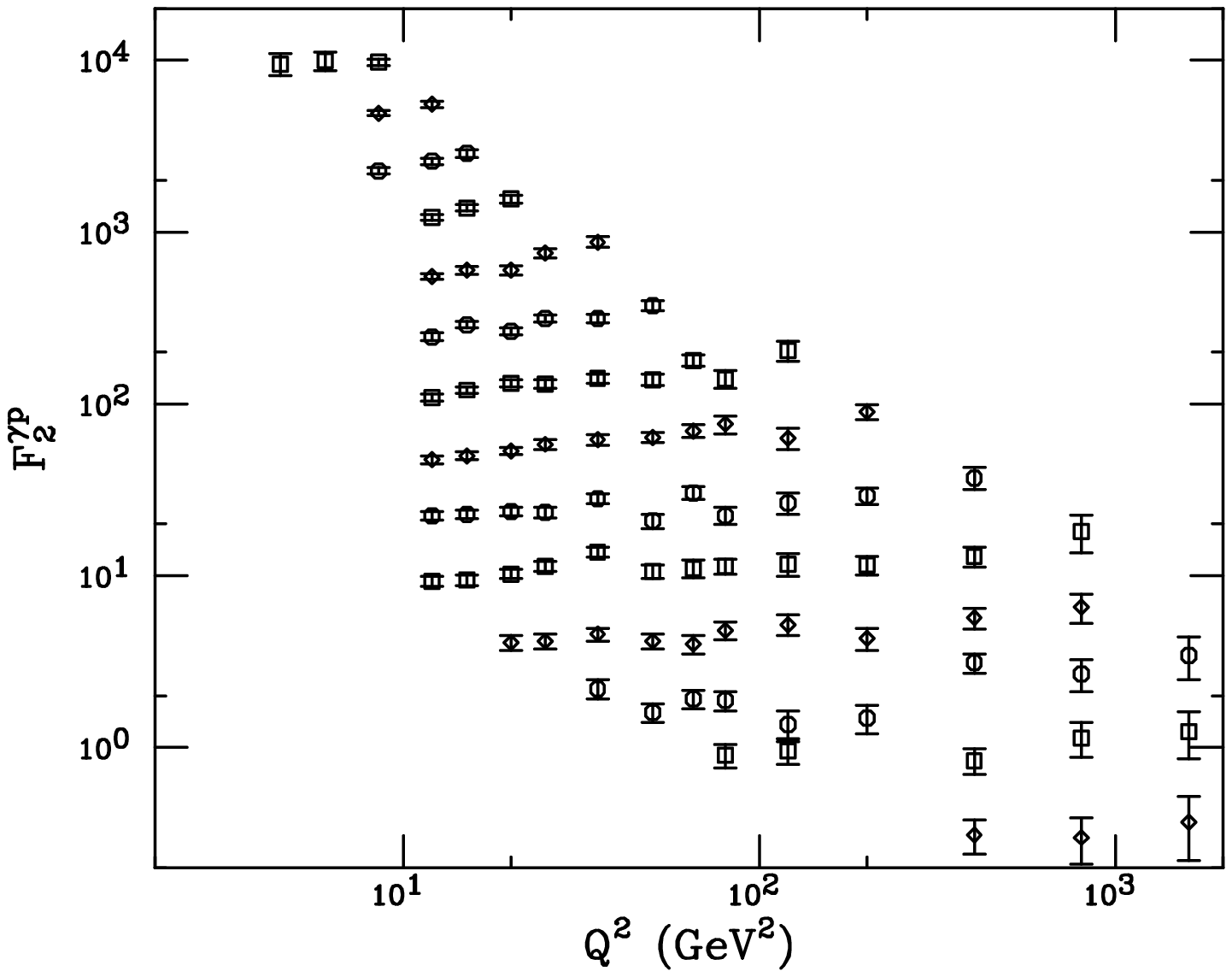}}
\caption[]{The proton structure function $F_2^{\gamma p}$ as a function of
$Q^2$ at the HERA collider (26.7 GeV electrons on 820 GeV protons)
\protect\cite{h1}.  From top to bottom the values of
the $x$ bins are:  0.000178, 0.000261, 0.000383, 0.000562, 0.00075, 0.000825,
0.00133, 0.00237, 0.00421, 0.0075, 0.0133, 0.0237, 0.0421, 0.075, and 0.133.
To separate the data, the structure function in each $x$ bin is
scaled by a factor of 2 from the next higher $x$ bin.  Therefore, only the
highest $x$ data (lowest points) have the correct scale.  Only the statistical
uncertainty is shown. } 
\label{herafig}
\end{figure}

Some characteristic data on $F_2^{\gamma p}$ from muon beams at 120, 200, 240,
and 280 GeV on hydrogen targets at CERN and the neutrino structure functions
$F_2^{W^\pm N}$ and $xF_3^{W^\pm
N}$ from $\nu$ and 
$\overline \nu$ beams on an iron target at Fermilab are shown
in Figs.~\ref{emcfig} and \ref{ccfrfig} respectively.  Several values of $x$
are selected and the $Q^2$ dependence of each $x$ bin is shown.  
Note that in the intermediate $x$ range, $0.125 < x < 0.35$, little $Q^2$
dependence is observed, prompting the scaling assumption from the early data
taken in this $x$ region \cite{scaling}.  More recently, the HERA $e p$
collider at $\sqrt{s} = 310$ GeV has provided data over a wider range of $x$
and $Q^2$ than previously available.  A sample of the HERA data
is shown in Fig.~\ref{herafig}.  The scaling violations only somewhat apparent
in the fixed target experiments are more obvious at HERA.

Since pure neutrino beams are unavailable, neutrino experiments with secondary
beams collect data 
from both neutrino and anti-neutrino-induced events from all possible neutrino
energies.  The resulting measured structure functions are a weighted average
of the $\nu$ and $\overline \nu$ induced events,
{\it e.g.}\ $F_2^{W^+ N_0}$ and $F_2^{W^- N_0}$.  If $\beta \equiv 
N_\nu/(N_\nu
+ N_{\overline \nu})$ is the fraction of DIS events induced by neutrinos, the
experimentally determined structure function is 
\begin{eqnarray}
  F_2^{W^\pm N_0}(x)& = &\beta F_2^{W^+ N_0} (x) + (1-\beta ) 
F_2^{W^- N_0}(x)\nonumber\\ 
  &=& \frac{1}{2} [F_2^{W^+ N_0}(x) + F_2^{W^- N_0}(x)] \nonumber \\
&  & \mbox{} +
\frac{1}{2} (2\beta -1) [F_2^{W^+ N_0}(x) - F_2^{W^- N_0}(x)] \, \, . 
\label{f2nuave}  
\end{eqnarray}
If $\beta = 1/2$ or $F_2^{W^+ N_0} = F_2^{W^- N_0}$, implying $s_p(x) = 
\overline s_p(x)$, the second term in Eq.~(\ref{f2nuave}) vanishes.  From 
the CCFR data \cite{CCFRLO},
$\beta = 0.83$ \cite{AWT} so that $F_2^{W^\pm N_0}$
is dominated by $F_2^{W^+ N_0}$ or neutrino-induced production.  In an analysis
of opposite sign dilepton production, they found that to next-to-leading order,
$s_p(x) = \overline s_p(x)$ within their experimental errors \cite{CCFRNLO}.  
(These results will be discussed in more detail in Sec. 3.2.1.)
However, since $\beta > 1/2$, the associated uncertainties are significant.
By comparing the neutrino and electromagnetic
structure functions, one can form the relation,
\begin{eqnarray}
\frac{5}{6} F_2^{W^\pm N_0}(x) - 3 F_2^{\gamma N_0}(x) & = &
\frac{1}{2} x [s_p(x) +
\overline s_p(x)] \nonumber \\ 
&   & \mbox{} + \frac{5x}{6} (2\beta -1)[s_p(x) - \overline s_p(x)] \, \, .
\label{f2diff}
\end{eqnarray}
If $s_p(x) = \overline s_p(x)$, then Eq.~(\ref{f2diff}) 
could be used to determine
the strange quark distribution.  However, the strange quark distribution
extracted in this fashion does not agree with independent experimental results
\cite{AWT}.

To obtain the parton densities, a global analysis of all available
data is performed.  This analysis involves taking data from many different
processes measured over a range of $x$ and $Q^2$ and making a multi-parameter 
fit to all the data, as described below.  First, it is necessary to solve the
coupled integro-differential evolution equations, Eqs.~(\ref{qdglap}) and 
(\ref{gdglap}) and at leading order or their next-to-leading order, NLO,
counterparts,  numerically to be able to
calculate the parton densities at the $Q^2$ appropriate to the measurement.
The data to be used in the fit are then chosen.  Typically only data sets that
can provide the best constraints on the parton densities are included except
when the data indicate significant new physics at work.  Since the evaluation
is done at fixed order in perturbation theory, a factorization scheme is chosen
and all the evaluations of the parton densities are done for the same order of
perturbation theory.  It is very important that this be done consistently
when calculating all processes considered.  A starting distribution at
scale $Q_0^2$, typically of the form 
\begin{eqnarray}
x f(x,Q_0^2) = A_0 x^{A_1} (1-x)^{A_2} P(x) \, \,  \label{pdfform}
\end{eqnarray}
is assumed
where $P(x)$ is a smooth polynomial function, the exponent $A_1$ determines the
small $x$ behavior, and $A_2$ governs the large $x$ behavior.  These initial
distributions are then evolved to the appropriate scale of the data and used to
calculate the cross section or structure function at the same $x$ and $Q^2$ as
the data.  The goodness of the fit, its $\chi^2$, is calculated from a
comparison to all the data in the fit and the
parameters in {\it e.g.}\
Eq.~(\ref{pdfform}) adjusted in each iteration to minimize the
$\chi^2$.  The final fits are generally made available either in tables with
some interpolation formula used to obtain the densities at any $x$ and $Q^2$
or as parameterized functions of $x$ and $Q^2$.  Parton distribution functions 
determined in this manner are only as
good as the data they fit and may fail to match new data outside the $x$ and
$Q^2$ range of the old measurements.  As new data are taken, the parton
distribution functions are thus continually updated.

An example of updating the parton distribution functions as new data appear
is the changing starting assumption regarding the degree of flavor 
symmetry in the sea.  In early
global analyses, an SU(3) flavor symmetric sea was assumed, see {\it e.g.}
Ref.~\cite{DO}.  However, it was shown that the strange sea is smaller than the
light quark sea so that $2 \overline s_p/(\overline u_p + \overline d_p) \sim
0.5$ \cite{CDHS,CCFRs}.  Thus, the sea could at most be SU(2) flavor
symmetric.  It has been further shown that the SU(2) flavor symmetry may not
hold, as discussed in the next section.  Newer fits of the parton distribution
functions such as those in Refs.~\cite{cteq5,mrst,grv94} 
include these results in the sea quark
analysis. 

Several different groups have been involved in the global analysis of the
parton distributions.  The CTEQ (Coordinated Theoretical-Experimental Project
on QCD) collaboration \cite{cteq5} and
Martin, Roberts, and Sterling (MRS) and their collaborators \cite{mrst}
have produced a variety of
parton distribution functions which improve with each round of data. Gl\"{u}ck,
Reya, and Vogt (GRV) \cite{grv94} 
have taken a rather different route, beginning at a 
low initial scale with valence-like sea quark and gluon distributions without 
trying to make a comprehensive fit.  A library of all available parton
distribution functions can be found in the program package PDFLIB available
from the CERN program libraries \cite{pdflib}.  
Care must be taken however to use only the
most recent sets, matching the appropriate order of the parton densities and
hard scattering cross section in scale and scheme.

\begin{figure}[htbp]
\setlength{\epsfxsize=\textwidth}
\setlength{\epsfysize=0.5\textheight}
\centerline{\epsffile{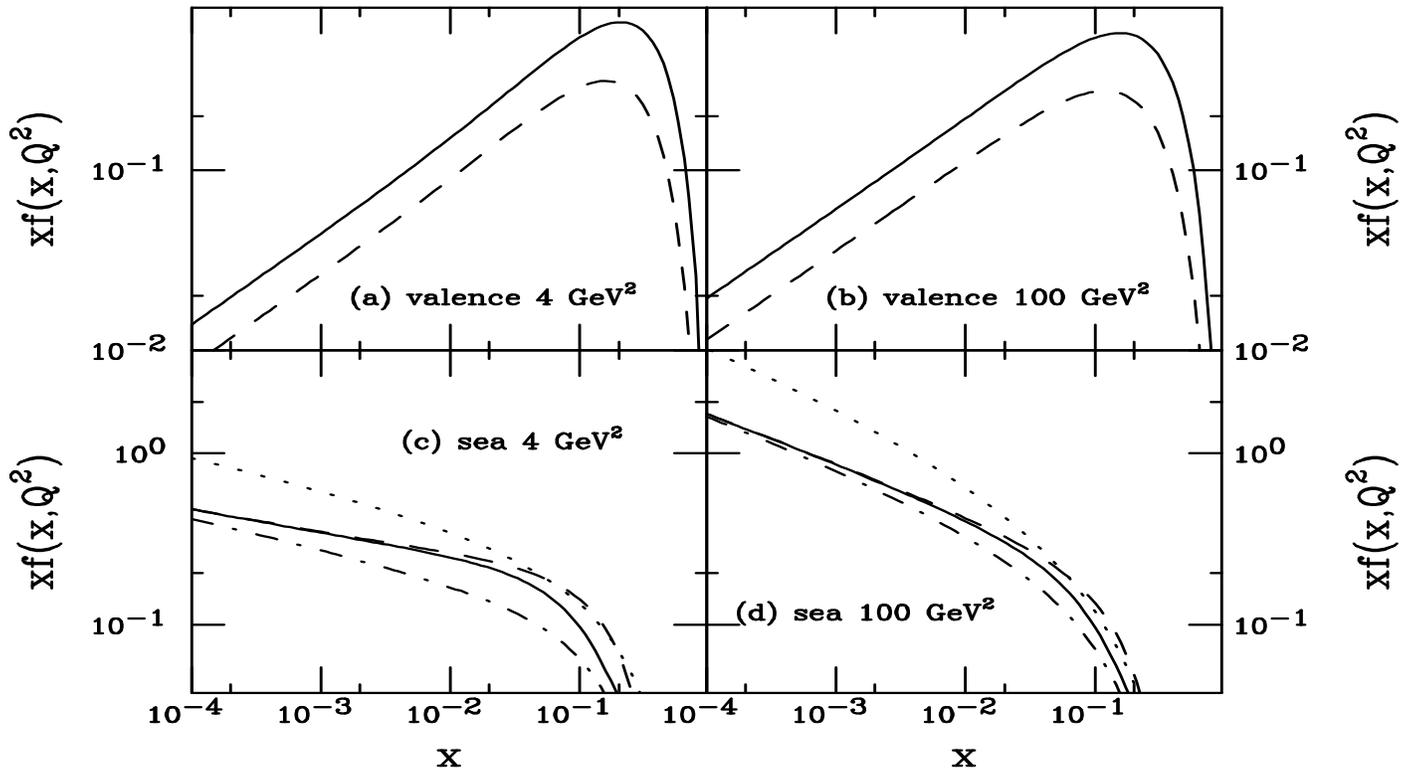}}
\caption[]{The CTEQ5 NLO $\overline{\rm MS}$ 
scheme proton parton distributions are
given at $Q^2 = 4$ and 100 GeV$^2$.  The up and down valence distributions are
shown in the solid and dashed lines respectively in (a) and (b).  The up, down
and strange sea distributions and the gluon distributions are given in the
solid, dashed, dot-dashed and dotted curves respectively in (c) and (d). } 
\label{cteqpdf}
\end{figure}

\begin{figure}[htbp]
\setlength{\epsfxsize=\textwidth}
\setlength{\epsfysize=0.5\textheight}
\centerline{\epsffile{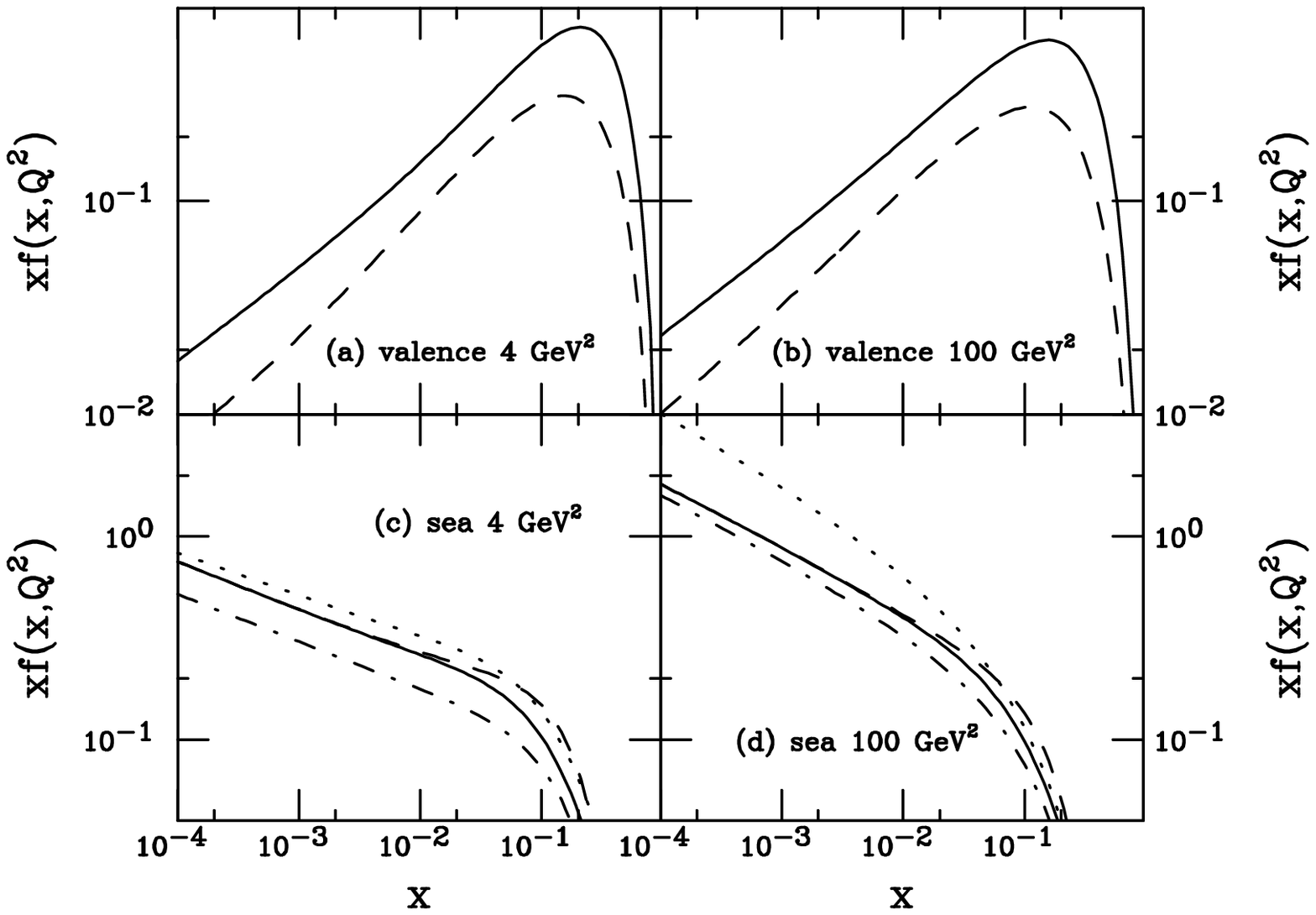}}
\caption[]{The MRST NLO $\overline{\rm MS}$ 
scheme proton parton distributions are
given at $Q^2 = 4$ and 100 GeV$^2$.  The up and down valence distributions are
shown in the solid and dashed lines respectively in (a) and (b).  The up, down
and strange sea distributions and the gluon distributions are given in the
solid, dashed, dot-dashed and dotted curves respectively in (c) and (d). } 
\label{mrspdf}
\end{figure}

\begin{figure}[htbp]
\setlength{\epsfxsize=\textwidth}
\setlength{\epsfysize=0.5\textheight}
\centerline{\epsffile{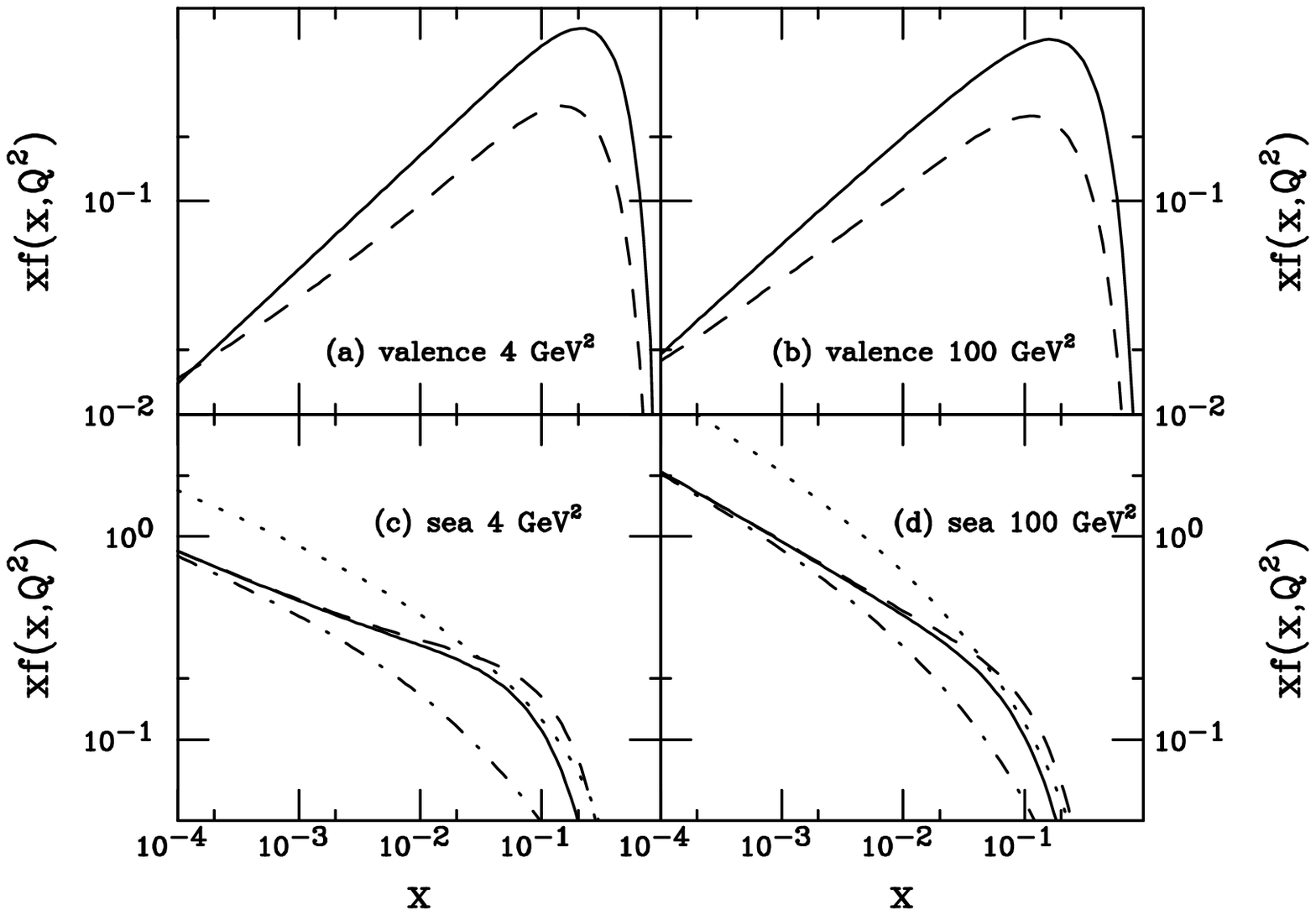}}
\caption[]{The GRV 94 HO
NLO $\overline{\rm MS}$ scheme proton parton distributions are
given at $Q^2 = 4$ and 100 GeV$^2$.  The up and down valence distributions are
shown in the solid and dashed lines respectively in (a) and (b).  The up, down
and strange sea distributions and the gluon distributions are given in the
solid, dashed, dot-dashed and dotted curves respectively in (c) and (d). } 
\label{grvpdf}
\end{figure}

In Figs.~\ref{cteqpdf}, \ref{mrspdf} and \ref{grvpdf}, the parton distributions
obtained in recent global fits to  NLO 
by CTEQ \cite{cteq5}, MRST \cite{mrst}, and
GRV \cite{grv94} are shown as a function of $x$ for two different values
of $Q^2$.  In each case, the sea quark and gluon evolution is stronger than
the valence evolution.  The evolution is faster at low $x$ than at intermediate
and large $x$.  While the $\overline u_p$ and $\overline d_p$ distributions
are different for $x > 0.01$, as discussed shortly,
they are nearly equal at low $x$.  The strange
quark distributions are generally smaller than the light quark distributions
but by $x \sim 10^{-4}$, $\overline s_p \approx \overline u_p$.  The gluon
distribution, least constrained by data, 
evolves more rapidly in $Q^2$ than do the charged parton distributions.

\subsection{Aspects of the small $x$ sea}

As already discussed, the parton distributions are determined in
practice from fits to DIS and related data by parameterizing the distributions
at a $Q_0$ large enough for perturbative QCD to be applicable.  Then the 
distributions are evolved to higher $Q^2$ using the DGLAP equations,
Eqs.~(\ref{qdglap})-(\ref{gdglap}).  Evolution to higher $Q^2$ when $x$ is not
too small should give reliable results.  However, extrapolation to small $x$ is
more difficult.  Early expectations based on Regge theory suggested that the 
sea quark and gluon distributions should become constant as $x \rightarrow 0$.
The HERA data have clearly shown that this is not true, as seen in
Fig.~\ref{herafig}.  In this section, the way this behavior can arise is
explained. 

\begin{figure}[htbp]
\setlength{\epsfxsize=0.5\textwidth}
\setlength{\epsfysize=0.5\textheight}
\centerline{\epsffile{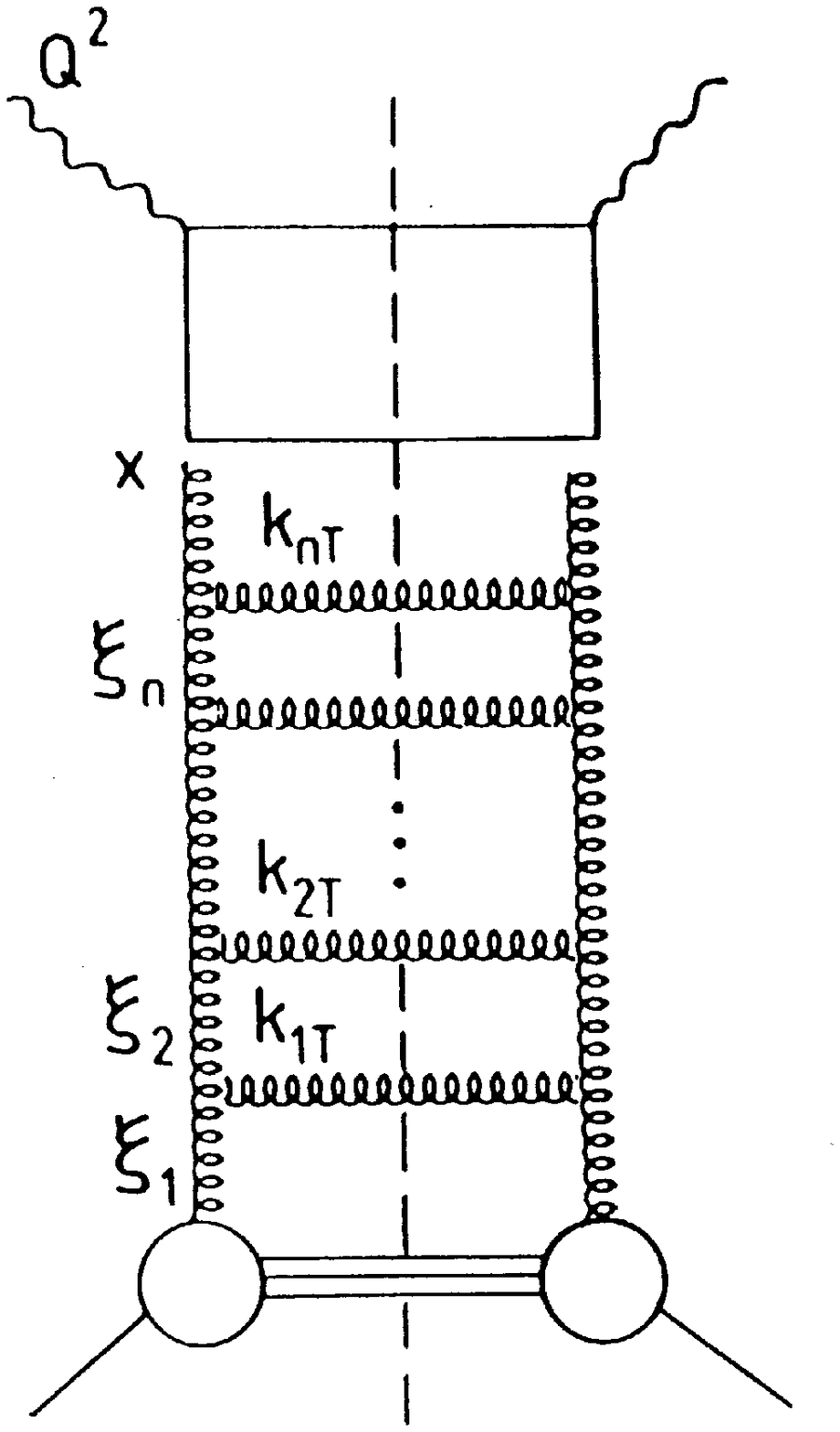}}
\caption[]{Ladder diagram corresponding to evolution of parton distribution
functions.  At large $Q^2$, DGLAP evolution sums diagrams with strong ordering
in $k_T$, $Q^2 \gg k_T^2 \gg k_{nT}^2 \gg \cdots \gg k_{1T}^2 \gg Q_0^2$.  At
small $x$, the leading contribution comes from ordering in $\xi$, 
$x \ll \xi_n \ll \cdots \ll \xi_1$ but with the gluon $k_T$ unordered.
Reproduced from Ref.~\cite{tobb} with permission of the publisher. } 
\label{gluladd}
\end{figure}

Gluon ladder diagrams such as the one in Fig.~\ref{gluladd}, taken from
Ref.~\cite{tobb}, represent 
the behavior of DIS scattering.  
In the regime of finite $x$ and large $Q^2$,
the DGLAP equations hold and the dominant contribution to the evolution of the
sea and gluon distributions come from the region of strongly ordered transverse
momentum, $Q^2 \gg k_T^2 \gg k_{nT}^2 \gg \cdots \gg k_{1T}^2 \gg Q_0^2$.  

When $x$ is small and $Q^2$ remains large, strong ordering in the longitudinal
momentum, $x \ll \xi_n \ll \cdots \ll \xi_1$, is as important as the
transverse momentum ordering.  The
probability of emitting the $i^{\rm th}$ gluon on the ladder is proportional
to $\alpha_s [dx_i/x_i]  [dk_{iT}^2/k_{iT}]^2$.  
The nested integrations over $x$
and $k_T$ generate the leading order behavior of the sea quark and gluon
distribution, assuming $\alpha_s$ is constant,
\begin{eqnarray}
xf(x,Q^2) \simeq \exp \left[ 2 \left[  \frac{3 \alpha_s}{\pi} \ln \left(
\frac{1}{x} \right) \ln Q^2 \right]^{1/2} \right] \, \, , \label{dlla}
\end{eqnarray}
known as the double leading logarithm approximation, DLLA.  The name DLLA 
implies
that only the terms proportional to $\ln (1/x) \ln Q^2$ 
multiplying each power of $\alpha_s$
are kept in the expansion of Eq.~(\ref{dlla}).  
In this case, $xf(x,Q^2)$ increases
as  $x \rightarrow 0$.  The increase is faster than any power of $\ln (1/x)$
but slower than any power of $x$.

In the case of small $x$ and finite $Q^2$, only the longitudinal momentum is
now strongly ordered and the full $k_T$ phase space must be integrated over
rather than retaining only the leading terms in $\ln Q^2$.  A counterpoint
to the DGLAP equations at small $x$
is the Lipatov or BFKL \cite{BFKL} equation that evolves
a starting distribution $f(x_0,k_T^2)$ at $x_0$ downwards in $x$ by 
integrating over the $k_T$ phase space,
\begin{eqnarray}
x \frac{\partial f(x,k_T^2)}{\partial x} = \int dk_T'^2 K(k_T, k_T')
f(x,k_T'^2) \, \, . \label{bfkl}
\end{eqnarray}
The kernel $K$ includes real and virtual gluon emission.  The Lipatov equation
is related to the sea quark or gluon distribution before the integration over
transverse momentum has been performed, $f(x,k_T^2) = \partial (xg(x,Q^2))/
\partial \ln Q^2 |_{Q^2 = k_T^2}$.  The hadronic cross section is 
typically the partonic cross section convoluted with an $x$-dependent parton
distribution, as in Eq.~(\ref{absigma}).  However,
in this case, the partonic cross section is convoluted with a
$k_T$-dependent parton density so that $\sigma \sim \int dx \, [dk_T^2/k_T^2] 
f(x,k_T^2) \widehat{\sigma}$.  
If $\alpha_s$ is fixed, an approximate analytical solution
may be found for $x \rightarrow 0$,
\begin{eqnarray}
xf(x,Q^2) \sim h(Q^2) x^{-\lambda} \,\, ; \,\,\,\,\,\, \lambda = \frac{12
\alpha_s}{\pi} \ln 2 \,\, . \label{bfklsol}
\end{eqnarray}
The predicted value of $\lambda$ is $\sim 0.5$ when $\alpha_s$ is fixed.
Allowing $\alpha_s$ to run with $Q^2$
introduces a dependence on the cutoff of the $k_T^2$
integration.

The increase in the sea quark and gluon distributions at small $x$ in
Eq.~(\ref{bfklsol}) cannot
continue indefinitely because the hadron will become so dense that the partons
can no longer be considered to be free.  The growth must eventually be
suppressed, most likely by gluon recombination \cite{glurecomb}.  It is unknown
at what value of $x$ this ``shadowing'' becomes important.  For example, an
anti-quark with distribution $\overline q (x,Q^2)$ has an apparent transverse
size of $1/Q$ but a smaller longitudinal size $1/xp$ with proton
momentum $p$ in a frame where $xp \gg Q$.  The scattering cross section of the
sea quarks and gluons is proportional to the square of the transverse size,
$\widehat{\sigma} \sim
\alpha_s(Q^2)/Q^2$.  When the area covered by the total number of sea
quarks and gluons, $n$, scattering with the cross section $\widehat{\sigma}$, 
approaches the transverse area of
the proton, $\pi r_p^2$, or $n \widehat{\sigma} \sim \pi r_p^2$,
shadowing is no longer negligible and introduces an additional
term in the evolution equations \cite{hot,muqui}.

In practice, the parton distributions are only fit with DGLAP evolution with
$A_1$ chosen in accord with the BFKL behavior.
The starting sea quark and gluon distributions in Eq.~(\ref{pdfform}) have been
found to have $A_1 \sim -0.3$.  While this is a faster growth than Regge theory
suggested, it is lower than the idealized BFKL result of $A_1 \sim -0.5$.  
It is difficult
to see from these results whether or not shadowing is important because a full
next-to-leading order evolution scheme incorporating DGLAP, DLLA, and BFKL is 
not available over the entire $x$ and $Q^2$ range.  

\subsection{Perturbative sea summary}

The naive parton model and its extensions have been extremely successful in
describing a large number of processes.  Both the data and the NLO perturbative
calculations used to model the data are now extremely precise over a wide range
of $x$, from $2 \times 10^{-4}$ to 0.75.  The parton distribution functions
extracted from global analyses by a number of groups
agree well with the data and with
each other over their range of validity and will continue to be refined as new
data are taken.

\section{Possible nonperturbative contributions to the sea}

So far, only the generation of the perturbative sea has been discussed.
However, since the parton distributions are fit rather than derived from first
principles and the perturbative evolution cannot go backwards and remain
stable, the true nature 
of the nucleon sea is, at its heart, unknown.  While the 
valence and sea quark distributions have been measured as a whole in DIS, and
the valence and sea quark distributions can be separated from one another, 
the individual sea quark distributions are more difficult to tease out from 
bulk measurements.  Therefore, simplifying assumptions have been made about the
proton sea. Recently, experiments have been
challenging the standard assumptions of the parton model such as SU(2) flavor 
and charge symmetry.  The rest of this review is devoted to some of these 
results and their interpretations.

\subsection{The Gottfried sum rule--$\overline u$ vs. $\overline d$}

Assuming flavor symmetry, proton and neutron parton distributions can be
related.  Valence quark counting suggests that the neutron $d$ distribution and
the proton $u$ distribution should be equal, $d_n = u_p$ at all $x$ and $Q^2$.
Similar assumptions can be made 
for the other parton distributions, see Eq.~(\ref{pvsn}).
Therefore, unless flavor symmetry breaking is discussed, $u$, $d$, $\cdots$
can be assumed to refer to the proton parton distributions.  The difference
between the proton and neutron structure functions can then be written as
\begin{equation}
\frac{1}{x} \left( F_2^{\gamma p}(x) - F_2^{\gamma n}(x) \right) =  
                \frac{1}{3} \, [ u^v(x) - d^v(x) ]
              + \frac{2}{3} \, [ \overline u(x) - \overline d(x) ]
\ .
\label{eqn: F2P-M}
\end{equation}
Using Eqs.~(\ref{uvalsum}) and (\ref{dvalsum}) for the proton valence quark
distributions, after integrating both sides of Eq.~(\ref{eqn: F2P-M}) over $x$,
one obtains 
\begin{equation}
\int_0^1 \frac{dx}{x} \, 
[ F_2^{\gamma p}(x) - F_2^{\gamma n}(x) ] 
                   = \frac{1}{3} + 
      \frac{2}{3} \int_0^1 dx \, [ \overline u(x) - \overline d(x) ]
\ \ \ .
\label{eqn:GINT}
\end{equation}
If the sea is flavor symmetric, $\overline u = \overline d$ and
the second term vanishes.  This is the Gottfried sum rule
\cite{GOTT}:
\begin{equation}
\int_0^1 \frac{dx}{x} \, 
 [ F_2^{\gamma p}(x) - F_2^{\gamma n}(x) ] = \frac{1}{3} 
\ \ \ .
\label{eqn: GOTTFRIED}
\end{equation}
The assumption of light anti-quark flavor symmetry is a serious one and if the 
sum rule is found to be violated, the nucleon sea could be SU(2) flavor
asymmetric. 

\subsubsection{Current experimental studies}

Because the integral of Eq.~(\ref{eqn: GOTTFRIED}) is over all $x$, the 
small $x$ region could have a significant contribution
to the sum rule.  Therefore the determination of any possible violation is only
partial and depends strongly on the extrapolation of the parton distributions
to low $x$.  This was especially true when the proton parton distribution
functions were not well constrained at low $x$ by other measurements.
The minimum $x$ is restricted by the lepton-beam energy, $E$, to be
$x_{\rm min} =Q^2/2 M E$, where
$Q^2$ should not be smaller than a few GeV$^2$
in order for the parton model to be applicable.  To quantify the discussion,
the Gottfried integral can be defined as 
\begin{equation}
I_G(x_{\rm min},x_{\rm max}) \equiv \int_{x_{\rm min}}^{x_{\rm max}} 
\frac{dx}{x} \, [ \, F_2^{\gamma p}(x) - F_2^{\gamma n}(x) \, ] 
\,\,\, . \label{gottint}
\end{equation}
Generally two values of $I_G$ are given for each measurement, one with $x_{\rm
min}$ and $x_{\rm max}$ limited by the experiment, the other extrapolated over
the entire $x$ range.  The integral over all $x$ depends on how the missing
small $x$ region is accounted for.  The results are summarized in
Table~\ref{gottdat}. 
When the Gottfried integral is given more than once for the same collaboration,
the second result is obtained from a reanalysis of the earlier data.

\begin{table}[tbp]
\begin{center}
\begin{tabular}{ccccc}
Measurement & $x_{\rm min}$ & $x_{\rm max}$ & $I_G(x_{\rm min},x_{\rm max})$ &
$I_G(0,1)$ \\ \hline
SLAC \cite{SLAC75} & 0.02 & 0.82 & $0.20 \pm 0.04$          & $-$  \\
EMC \cite{EMC83}   & 0.03 & 0.65 & $0.18 \pm 0.01 \pm 0.07$ & $0.24 \pm 0.02
\pm 0.13$ \\
EMC \cite{EMC87}   & 0.02 & 0.80 & $0.197 \pm 0.011 \pm 0.083$ & $0.235 
\, _{-0.099}^{+0.110}$ \\
BCDMS \cite{BCDMS} & 0.06 & 0.80 & $0.197 \pm 0.006 \pm 0.036$ & $-$  \\
NMC \cite{NMC91}   &0.004 & 0.80 & $0.227 \pm 0.007 \pm 0.014$ & $0.240 \pm
0.016$ \\
NMC \cite{NMC94}   &0.004 & 0.80 & $0.221 \pm 0.008 \pm 0.019$ & $0.235 \pm
0.026$ \\ \hline
\end{tabular}
\end{center}
\caption[]{Deep inelastic scattering measurements of the Gottfried sum rule.
The Gottfried integral, Eq.~(\protect\ref{gottint}), is given for each
measurement and for the full $x$ range when available.}
\label{gottdat}
\end{table}

\begin{figure}[htbp]
\setlength{\epsfxsize=\textwidth}
\setlength{\epsfysize=0.5\textheight}
\centerline{\epsffile{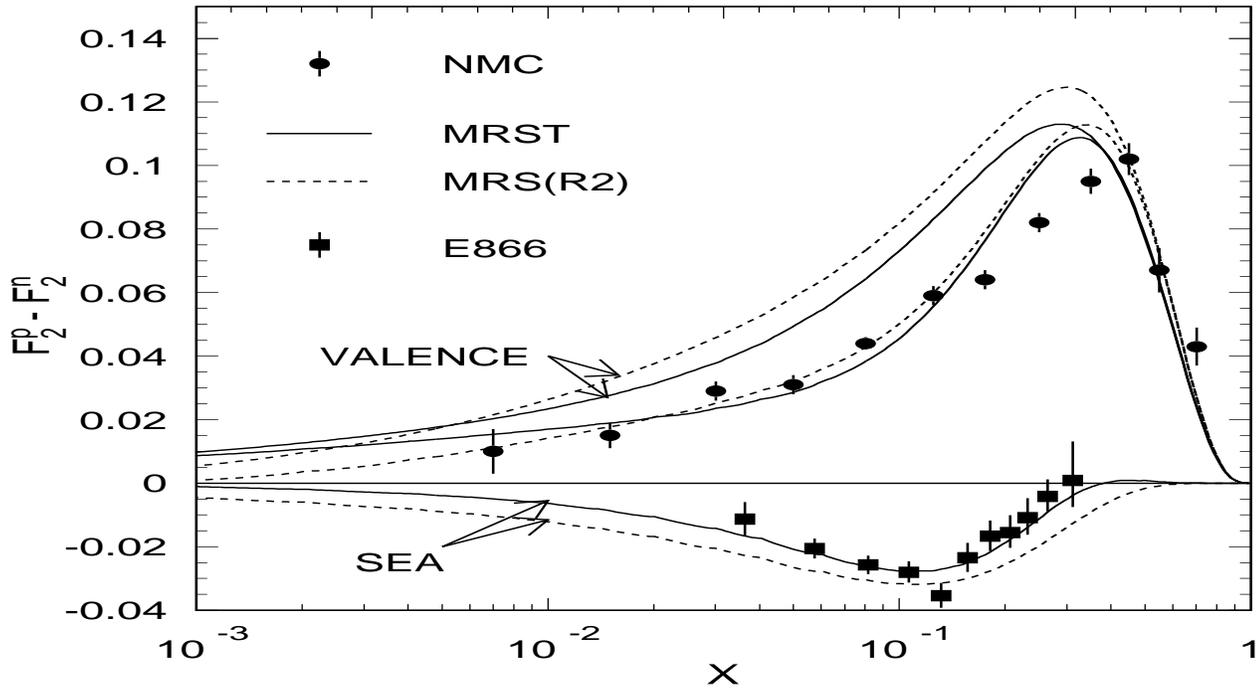}}
\caption[]{The difference $F_2^{\gamma p} - F_2^{\gamma n}$ measured by NMC
\cite{NMC94} at $Q^2 = 4$ GeV$^2$ compared with calculations using the MRS
(R2) \cite{mrsr2} (dashed) and the MRST \cite{mrst} (solid)
parton distributions.  The E866
\cite{E866dy} Drell-Yan results on the sea quark contribution to the difference
are shown at the same $Q^2$.  For each set of parton distributions, the upper
curve is the valence contribution to the difference, the lower curve is the sea
contribution to the difference and the center curve is the sum of the two.
Reproduced from Ref.~\cite{mcgmp} with permission from the {\em Annual 
Review of Nuclear and Particle Science}, Volume 49 $\copyright$ 1999 by 
Annual Reviews.} 
\label{nmcdiff}
\end{figure}

Typical DIS targets used to measure the structure functions are hydrogen and
deuterium.  The neutron distribution is then generally
extracted taking into account the
momentum smearing effects due to Fermi motion in the deuteron.  
However, the NMC analysis obtained the difference from the relation
\begin{equation}
F_2^{\gamma p} -F_2^{\gamma n}= 2 \, F_2^{\gamma {\rm D}} \, 
\frac{1-F_2^{\gamma n}/F_2^{\gamma p}}{1+ F_2^{\gamma n}/F_2^{\gamma p}}
\ \ \ ,
\label{eqn:NMCINTEG}
\end{equation}
where the ratio $F_2^{\gamma n}/F_2^{\gamma p} = 2F_2^{\gamma {\rm 
D}}/F_2^{\gamma p} -1$ is determined 
from the NMC data \cite{NMC91,NMC94}.  The absolute value of $F_2^{\gamma {\rm
D}}$ is fit from data. 
Nuclear corrections such as Fermi motion
were not taken into account.  The difference measured by NMC is shown in 
Fig.~\ref{nmcdiff}.

The data in Table~\ref{gottdat} indicate $I_G(0,1) < 1/3$, 
a clear violation of the Gottfried sum
rule.  When the first data appeared, the low $x$ parton distribution functions
were not known from other sources.  Therefore attempts were made to satisfy
SU(2) flavor symmetry by changing the shape of the valence $u$ and $d$
distributions in the unmeasured region, see {\it e.g.}\ Ref.~\cite{KMRS}.
However, the lower $x$ NMC data \cite{NMC91,NMC94} showed that this idea was
untenable.  Other attempts to understand the sum rule violation centered
around the extraction of the neutron distribution from deuterium or other light
nuclear data.  The nuclear parton distribution functions are different from
free proton distributions \cite{Arn}.  At small $x$, there is a depletion of
$F_2^{\gamma A}$ in nuclear targets, known as shadowing, which was typically
neglected in deuterium.  The effect has been studied in a variety of models
including vector meson dominance, Pomeron and meson exchange models and 
has been found to contribute only $\delta I_G(0,1) = -0.02$ \cite{BK,MST} 
to the Gottfried sum rule.  Note that since the shadowing contribution, $\delta
I_G(0,1)$ above, is negative, correcting for the effect in deuterium increases
the Gottfried sum rule violation.
Therefore, an SU(2) flavor asymmetry in the nucleon sea
is the most likely explanation of the observed effect with $\overline d >
\overline u$.

Unfortunately DIS data can only address the difference between the light
anti-quark distributions, usually written as
$\overline d - \overline u$ for a positive quantity.  The
individual anti-quark distributions cannot be studied separately.  An
independent measurement of $\overline d/\overline u$ is needed.  The Drell-Yan
process provides an elegant way to obtain the ratio $\overline d/\overline u$
by comparing Drell-Yan production on proton and deuterium targets \cite{ellis}.
Then
\begin{eqnarray}
\sigma_{p {\rm D}}^{\rm DY} &= & \sigma_{pp}^{\rm DY} + \sigma_{pn}^{\rm DY} \\
& \propto & \frac{1}{9} \left[ (4u(x_1) + d(x_1))(\overline u(x_2) + \overline
d(x_2)) \right. \nonumber \\ &   & \left. \mbox{} + (4 \overline u(x_1) 
+ \overline d(x_1))( u(x_2) + d(x_2)) \right]
\label{sigpddy}
\end{eqnarray}
assuming charge symmetry.  When $x_1$ is large, $\overline u(x_1),
\overline d(x_1) \approx 0$ so that
\begin{eqnarray}
\sigma_{p {\rm D}}^{\rm DY} & 
\approx & \frac{1}{9} (4u(x_1) + d(x_1))(\overline 
u(x_2) + \overline d(x_2))
\label{sigpddyx1l} \\
\sigma_{pp}^{\rm DY} & \approx & \frac{1}{9} (4u(x_1) \overline u(x_2) + 
d(x_1)\overline d(x_2))
\label{sigppdyx1l} \,\, .
\end{eqnarray}
The ratio of the $p {\rm D}$ and $pp$ 
Drell-Yan cross sections when $x_1 \gg x_2$ is
\begin{equation} 
\left. \frac{\sigma_{p {\rm D}}^{\rm DY}}
           {2\sigma_{pp}^{\rm DY}}
\right|_{x_1\gg x_2} \approx\frac{1}{2} \,
\frac{1 + \frac{1}{4} \left[ d(x_1)/u(x_1)\right] }
     {1 + \frac{1}{4} \left[ d(x_1)/u(x_1) \right] \left[ \overline d(x_2)/
\overline u(x_2)\right] }
      \left( 1 + \frac{\overline{d}(x_2)}{\overline{u}(x_2)} \right).
\label{eq:4}
\end{equation}
When
$\overline{d}=\overline{u}$, the ratio is 1.

The first test of SU(2) flavor symmetry with the Drell-Yan
process was made by the Fermi\-lab E772 collaboration.
They compared the production of Drell-Yan muon pairs from isoscalar
targets to that from a neutron-rich target and set
constraints on the difference between $\overline{u}$ and $\overline{d}$ in the
range $0.04\leq x\leq 0.27$~\cite{mcgaughey}.  Later, the CERN
NA51 collaboration \cite{baldit} carried out a comparison of the
Drell-Yan yield from hydrogen and deuterium at a single
value of $x$ with a 450~GeV proton beam and found
\begin{equation}
\left. \frac{\overline{u}_{p}}{\overline{d}_{p}} 
\right|_{\langle x \rangle=0.18} = 0.51\pm 0.04\pm 0.05.
\label{pdovpp}
\end{equation}

The most recent test has been made by the Fermilab E866 collaboration. 
They measured the Drell-Yan yield
from an 800~GeV proton beam on liquid deuterium and hydrogen
targets and extracted $\overline{d}/\overline{u}$ and 
$\overline{d}-\overline{u}$ in
the proton for $0.020 < x < 0.345$. 
The resulting ratio of the Drell-Yan cross section per nucleon,
Eq.~(\ref{pdovpp}), is given in 
Table~\ref{tab:1} as a function of $x_1$ and $x_{2}$~\cite{xf,cnb}.
The $J/\psi$ and $\Upsilon$ resonance regions, $M_{\mu^+\mu^-} < 4.5$
~GeV and $9.0 < M_{\mu^+\mu^-} < 10.7$~GeV respectively were 
excluded from the analysis.  The data show
that the Drell-Yan cross section per nucleon in $p {\rm D}$ interactions
exceeds the Drell-Yan $pp$
cross section over a range of $x_{2}$.  

\begin{table}[tb]
\begin{center}
\begin{tabular}{ccccccc}
$\langle x_2\rangle$ & $\langle x_F \rangle$ & $\langle x_1\rangle$ &
$\langle p_T\rangle$ & $\langle M_{\mu^+\mu^-}\rangle$  
& $\sigma_{p {\rm D}}^{\rm DY}/2\sigma_{pp}^{\rm DY}$
& $\overline d(x_2)/\overline u(x_2)$ \\ \hline
0.036 & 0.537 & 0.573 & 0.92 & 5.5 & 1.039 $\pm$ 0.017 & $1.091 \pm 0.037$ \\ 
0.057 & 0.441 & 0.498 & 1.03 & 6.5 & 1.079 $\pm$ 0.013 & $1.194 \pm 0.031$ \\ 
0.082 & 0.369 & 0.451 & 1.13 & 7.4 & 1.113 $\pm$ 0.015 & $1.298 \pm 0.039$ \\ 
0.106 & 0.294 & 0.400 & 1.18 & 7.9 & 1.133 $\pm$ 0.020 & $1.399 \pm 0.057$ \\ 
0.132 & 0.244 & 0.376 & 1.21 & 8.5 & 1.196 $\pm$ 0.029 & $1.664 \pm 0.096$ \\ 
0.156 & 0.220 & 0.376 & 1.21 & 9.3 & 1.124 $\pm$ 0.035 & $1.494 \pm 0.119$ \\ 
0.182 & 0.192 & 0.374 & 1.20 & 9.9 & 1.091 $\pm$ 0.043 & $1.411 \pm 0.142$ \\ 
0.207 & 0.166 & 0.373 & 1.19 & 10.6& 1.098 $\pm$ 0.055 & $1.476 \pm 0.195$ \\ 
0.231 & 0.134 & 0.365 & 1.18 & 11.1& 1.055 $\pm$ 0.067 & $1.397 \pm 0.250$ \\ 
0.264 & 0.095 & 0.359 & 1.18 & 11.8& 0.967 $\pm$ 0.067 & $1.178 \pm 0.239$ \\ 
0.312 & 0.044 & 0.356 & 1.12 & 12.8& 0.881 $\pm$ 0.141 & $0.937 \pm 0.539$ \\ 
\end{tabular}
\end{center}
\caption{Ratio of Drell-Yan cross sections in deuterium to hydrogen, 
$\sigma_{p {\rm D}}^{\rm DY}/2\sigma_{pp}^{\rm DY}$ and the ratio
$\overline d/\overline u$
as a function of $x_2$.  The average of the kinematic variables for each
bin is also tabulated.  The units of $\langle p_T\rangle$ and 
$\langle M_{\mu^+\mu^-}\rangle$  are GeV.
Note that $\langle x_1\rangle = \langle
x_F\rangle + \langle x_2\rangle$.  Only statistical errors are shown.  There
is a 1\% systematic uncertainty in all points.  Modified 
from Ref.~\cite{E866dy}. Copyright 1998 by the American Physical Society.}
\label{tab:1}
\end{table}

\begin{figure}[htbp]
\setlength{\epsfxsize=0.8\textwidth}
\setlength{\epsfysize=0.4\textheight}
\centerline{\epsffile{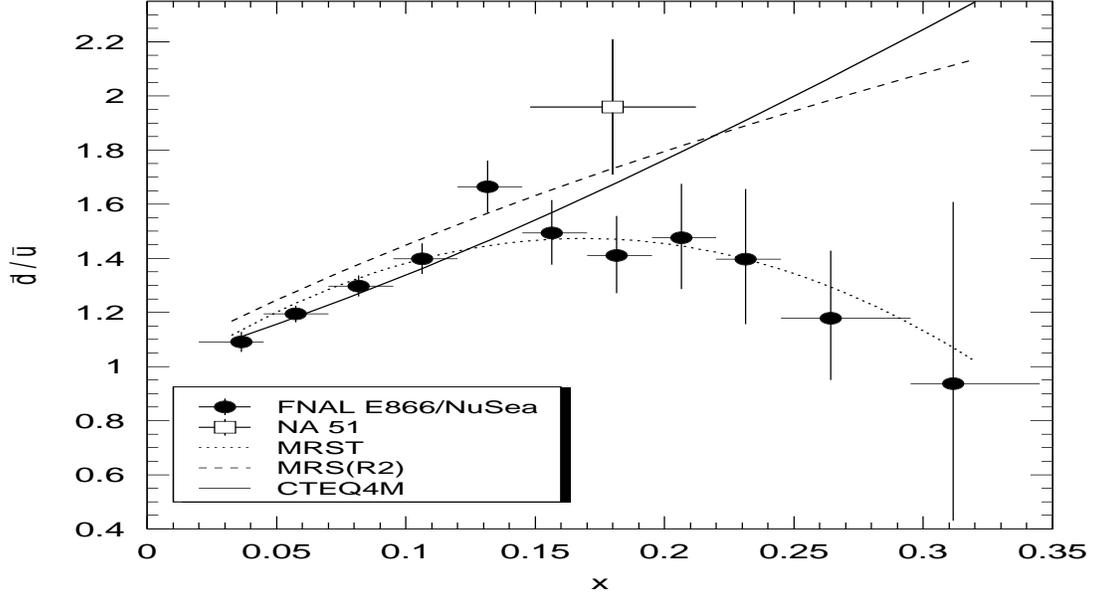}}
\caption[]{The ratio of $\overline{d}/\overline{u}$ in the proton as a function
of $x$ extracted from the Fermilab E866 cross section ratio
\protect\cite{E866dy} along with the NA51 result \protect\cite{baldit}. The
solid curve shows the CTEQ 4M ratio and the dashed curve is the MRS (R2) ratio.
The MRST ratio is shown in the dotted curve.  The E866 error bars are
statistical only with an additional systematic
uncertainty of $\pm0.032$ not indicated.  
Reproduced from Ref.~\protect\cite{mcgmp} with permission from the {\em Annual 
Review of Nuclear and Particle Science}, Volume 49 $\copyright$ 1999 by 
Annual Reviews.} 
\label{fig:dbub}
\end{figure}

Since all the data do not satisfy $x_1
\gg x_2$, as necessary for Eq.~(\ref{eq:4}) to hold, the ratio $\overline
d(x_2)/\overline u(x_2)$ 
was extracted iteratively by calculating the leading order Drell-Yan
cross sections using a set of parton distribution functions as
input and adjusting $\overline{d}/\overline{u}$ until the calculated
ratio agreed with the measured one.  In this analysis the light anti-quark sum,
$\overline{d}+\overline{u}$, the valence, and heavy quark
distributions obtained in global analyses of CTEQ 4M \cite{cteq4} and MRS (R2)
\cite{mrsr2} were assumed to be valid.   
The extracted $\overline{d}/\overline{u}$
ratio is shown in Fig.~\ref{fig:dbub} along with the corresponding CTEQ 4M
and MRS (R2) ratios. 
For consistency, when $x_1 \le 0.345$,
the projectile ratio $\overline{d}(x_1)/\overline{u}(x_1)$ was assumed to be
equal to that in the target proton while for $x_1 > 0.345$, $\overline
d(x_1)/\overline u(x_1) \equiv 1$ was assumed.  When this high $x_1$ assumption
was relaxed, there was a negligible difference in the low $x_2$ results and
only a 3\% change in the largest $x_2$ bin.  No significant dependence on the
parton distribution set was observed.

A qualitative feature of the data, which contradicts the parton distributions
studied, is the rapid decrease of $\overline d(x)/\overline u(x)$ to 
unity beyond $x=0.2$.  At the same value of $x$ as measured by NA51,
$x = 0.18$, the $\overline{d}/\overline{u}$ ratio is somewhat smaller than that
obtained by NA51 \cite{baldit}.  Although the average value of
$M_{\mu^+\mu^-}^2$ is different for the two data sets, the effects of evolution
on the ratio is expected to be small.

\begin{figure}[htbp]
\setlength{\epsfxsize=0.8\textwidth}
\setlength{\epsfysize=0.4\textheight}
\centerline{\epsffile{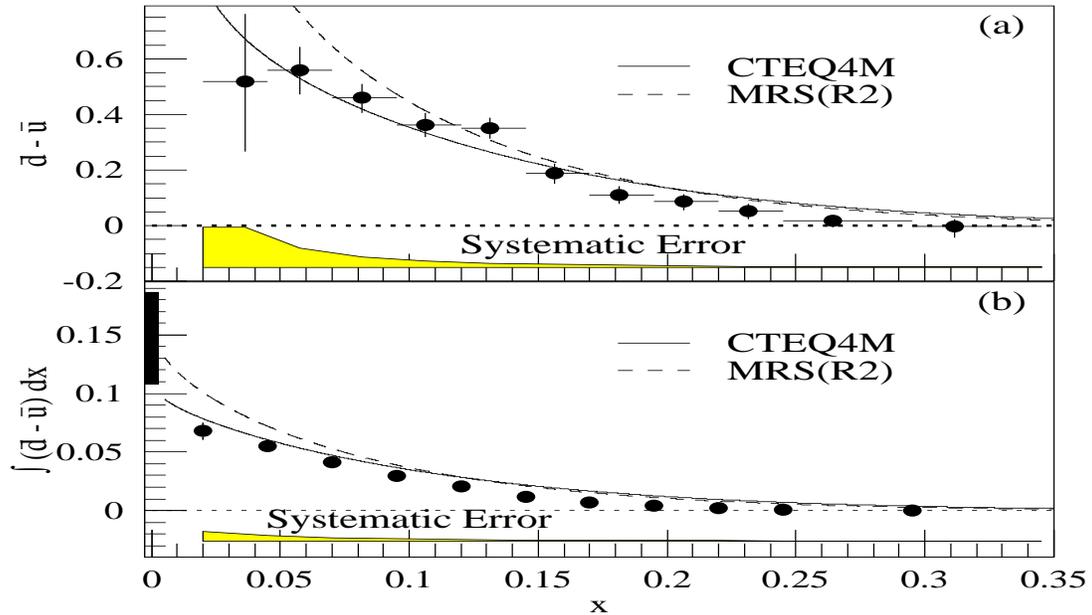}}
\caption[]{Fermilab E866 values for (a) $\overline{d}-\overline{u}$ and (b)
the integral
$\int^{0.345}_{x}\left(\overline{d}-\overline{u}\right) dx'$ in the proton
as a function of $x$.  The curves represent the corresponding CTEQ 4M (solid) 
and MRS (R2) (dashed)
results.  The bar at $0.147\pm0.039$ on the left axis in
(b) shows the value obtained by NMC for the integral from 0 to 1. 
Reproduced from Ref.~\protect\cite{E866dy}. Copyright 1998 by the American
Physical Society.}
\label{dmu}
\end{figure}

To address the Gottfried sum rule violation, the extracted
$\overline{d}/\overline{u}$ ratio is used with the CTEQ4M value of
$\overline{d} + \overline{u}$ to obtain $\overline{d}-\overline{u}$ at the
average $M_{\mu^+\mu^-}$ of the entire data set, $M_{\mu^+\mu^-} = 7.35$
GeV,
\begin{eqnarray}
\overline d(x) - \overline u(x) = \frac{\overline d(x)/\overline u(x) -
1}{\overline d(x)/\overline u(x) + 1} \left[ \overline u(x) + \overline
d(x) \right] \, \, . \label{e866diff}
\end{eqnarray}
The integral of $\overline{d}-\overline{u}$
between $x_{\rm min}$ and 0.345 is calculated.  Both the difference
$\overline{d}-\overline{u}$ and the integral 
are shown in Fig.~\ref{dmu}.  The
integral of the data is 
\begin{eqnarray}
\int_{0.02}^{0.345} dx [\overline d(x) - \overline u(x)]  = 0.068\pm 0.007 \pm
0.008 \, \, ,
\label{diffint}
\end{eqnarray}
compared to 0.076 and 0.1 for the CTEQ 4M and 
MRS (R2) sets integrated over the
same region.  When the range $10^{-4}<x<1$ is considered, CTEQ 4M gives a 
value of 0.108 for the integral while the MRS (R2) result is 0.160.  Above $x =
0.345$, both the CTEQ 4M and MRS (R2) distributions contribute less
than 0.002 to the total integral.  However,
significant contributions to the integral from the parton distribution
functions arise in the unmeasured
region below $x = 0.02$, 30-50\%, depending on the set.
Such a large $\overline{d} /\overline{u}$ asymmetry cannot arise from
perturbative effects~\cite{ross}, especially since global fits to the parton
distribution functions are tuned to accommodate the NMC, NA51, 
and, most recently,
the E866 results ({\it e.g.}\ the MRST distributions \cite{mrst} shown in the
dotted curve in Fig.~\ref{fig:dbub}).  

A more complete picture of the individual contributions to $F_2^{\gamma p} -
F_2^{\gamma n}$ emerges from the NMC and E866 data.  
Equation~(\ref{eqn: F2P-M})
gives the decomposition of the difference into valence and sea components.
Figure~\ref{nmcdiff} shows the NMC data on the total difference as a function
of $x$ along with the E866 contribution to the difference, $\propto \overline
u - \overline d$.  The curves show the total difference, the valence 
difference, $\propto u^v_p - d^v_p$, and the sea difference for two sets of
parton distributions:  MRS (R2), available before the E866 data, and MRST,
which included the E866 data in the fit.  Other relevant data in the MRST fit
include an improved $\alpha_s$ determination from the CCFR data, the final NMC
data on $F_2^{\gamma p}$, $F_2^{\gamma {\rm D}}$, and $F_2^{\gamma {\rm
D}}/F_2^{\gamma p}$, and more
precise information on the $u/d$ ratio at large $x$ from the Fermilab Tevatron
\cite{mrst}.  The effect of these improvements on the global analysis is clear
from the plot.  While both parton distributions fit the NMC data, only the
MRST set agrees with the E866 data.  Note that reducing the sea contribution to
Eq.~(\ref{eqn: F2P-M}) also significantly reduces the valence difference.

\begin{figure}[htbp]
\setlength{\epsfxsize=0.8\textwidth}
\setlength{\epsfysize=0.4\textheight}
\centerline{\epsffile{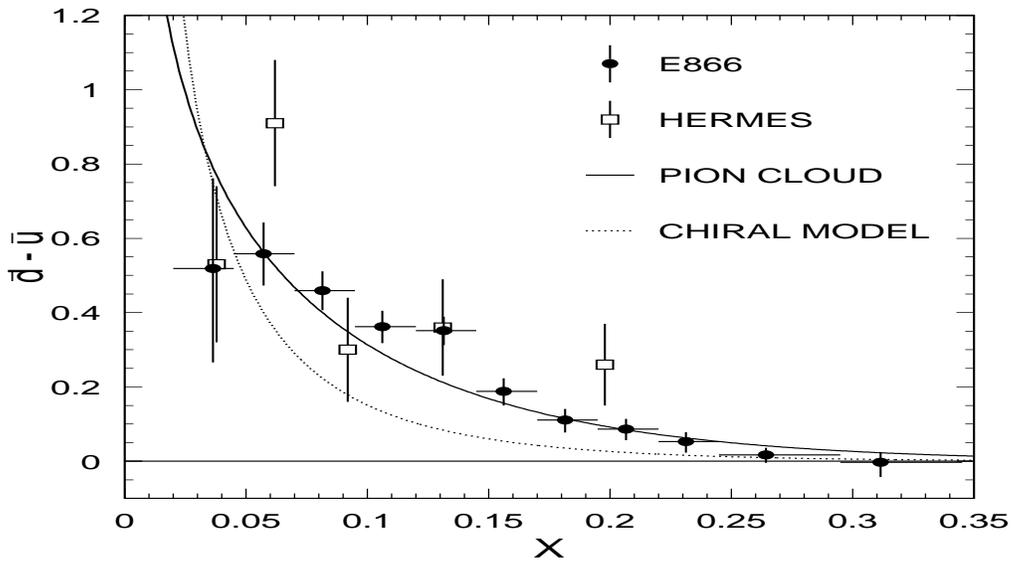}}
\caption[]{The E866 difference $\overline d - \overline u$ at $\langle Q^2
\rangle = 54$ 
GeV$^2$ \cite{E866dy} is compared with predictions of meson cloud and chiral
models.  The HERMES data \cite{HERMES} are also shown.
Reproduced from Ref.~\protect\cite{mcgmp} with permission from the {\em Annual 
Review of Nuclear and Particle Science}, Volume 49 $\copyright$ 1999 by 
Annual Reviews.} 
\label{dmumodel}
\end{figure}

The most recent experimental study of flavor asymmetries and the 
Gottfried sum rule is by the HERMES collaboration \cite{HERMES} at HERA. 
They measured the
charged pion yields in semi-inclusive DIS with a 27.5 GeV positron beam 
on hydrogen, deuterium, and $^3$He gas jet targets in the kinematic range 
$0.02 < x < 0.3$ and $1 < Q^2 < 10$ GeV$^2$.  Final-state hadron
production requires the fragmentation function, $D_i^h(z)$, for parton $i$ to
produce hadron $h$ with momentum fraction $z$ of the initial parton momentum.
The fragmentation functions $D_i^{\pi^\pm}(z)$ were extracted from the $^3$He
data and assumed to be independent of the target.  The HERMES collaboration 
determined the ratio
\begin{eqnarray}
\frac{\overline d(x) - \overline u(x)}{u(x) - d(x)} = \frac{J(z) \left[ 1 -
r(x,z) \right] - \left[ 1 + r(x,z) \right]}{J(z) \left[ 1 -
r(x,z) \right] + \left[ 1 + r(x,z) \right]} 
\label{dbmuboumd}
\end{eqnarray}
where
\begin{eqnarray}
r(x,z) & = & \frac{d\sigma_p^{\pi^-}(x,z)/dz - d\sigma_n^{\pi^-}(x,z)/dz}{d
\sigma_p^{\pi^+}(x,z)/dz - d\sigma_n^{\pi^+}(x,z)/dz} \label{rxz} \\
J(z) & = & \frac{3}{5}  \left( \frac{1 + D_u^{\pi^-}(z)/D_u^{\pi^+}(z)}{1 - 
D_u^{\pi^-}(z)/D_u^{\pi^+}(z)} \right) \, \, . \label{jz}
\end{eqnarray}
The ratio $r(x,z)$ was determined from the hydrogen and deuterium data.  No $z$
dependence was observed in the measurement of the ratio in
Eq.~(\ref{dbmuboumd}).  Using the GRV 94 LO \cite{grv94} parameterization of
$u(x) - d(x)$, they found
\begin{eqnarray}
\int_{0.02}^{0.3} dx \, (\overline d(x) - \overline u(x))  = 0.107\pm 0.021 \pm
0.017 \, \, ,
\label{diffintherm}
\end{eqnarray}
consistent with the E866 results, Eq.~(\ref{diffint}), even though the
E866 results are obtained at $\langle M_{\mu^+ \mu^-}^2 \rangle = 
\langle Q^2 \rangle = 54$ GeV$^2$ while
$\langle Q^2 \rangle = 2.3$ GeV$^2$ in HERMES, as shown in
Fig.~\ref{dmumodel}.  The model calculations will be described later.

An experiment following up the E866 results but with a lower energy proton 
beam, 120 GeV, has been proposed at Fermilab \cite{P120}.  This experiment
would be able to increase the statistics at $x>0.1$ and obtain results at 
larger $x$ than available with the higher energy beam.

Models which might account for SU(2) flavor asymmetry include meson clouds, 
Pauli blocking, chiral models, and charge symmetry breaking.  These
models will be discussed in the following sections.  For a more extensive
review of Gottfried sum rule violating mechanisms, see Ref.~\cite{Kumano}.

\subsubsection{Meson cloud models}

Clouds of virtual mesons were shown to play a role in the dynamics of chiral
models of nucleon structure such as the cloudy bag.  In these models,
the proton wavefunction can fluctuate into {\it e.g.}\ $\pi^+ n$ and $\pi^0 p$.
These virtual pion states have been used to explain the $\Delta$ decay width
and the charge square radius of the neutron \cite{Kumano}.  
While the proton wavefunction is
in the virtual $\pi N$ state, the valence quarks and anti-quarks of the virtual
state can contribute to the proton parton distributions.  The original model
\cite{SULL} has been developed further recently to include a number of
meson-baryon fluctuations with different form factors and meson-baryon-nucleon
couplings at each vertex.  Applications not only include the $\overline
d/\overline u$ asymmetry but also quark and anti-quark asymmetries between
strange and heavier quarks.  We introduce the model here and will discuss other
applications in later sections.

The amplitude of the Sullivan process is obtained from the DIS amplitude by
replacing the $\gamma^\star p$ vertex in the DIS formalism by {\it e.g.}\
a $\pi NN$ vertex
\begin{equation}
{\cal M}_\mu = \, <X|e J_\mu (0)|\pi> \, 
                   \frac{1}{p_\pi^2 - m_\pi^2} \, 
                   F_{\pi NN} (t) \, 
         \overline u(p') \, i \, g_{_{\pi NN}} \, \widetilde \phi_\pi^{\, *} 
                             \cdot \widetilde\tau \, 
                             \gamma_5 \, u(p)   
\ \ \ ,
\end{equation}
including the form factor, $F_{\pi NN}(t)$, coupling constant, $g_{_{\pi NN}}$,
and $\pi -N$ isospin coupling $\widetilde \phi_\pi^{\, *} \cdot 
\widetilde\tau$.  Replacing $p$
by $p_\pi$ and $M$ with $m_\pi$ in Eq.~(\ref{wmunu}), one obtains 
\begin{eqnarray}
F_2^{\gamma \pi}(x) & = & | \widetilde \phi_\pi^{\, *} 
                      \cdot \widetilde\tau  |^2
         \, \frac{g_{_{\pi NN}}^2}{16 \pi^2} \,
                  \int _x^1 dy \, y  \, F_2^\pi  \left( \frac{x}{y} \right)
                 \int_{-\infty}^{t^N_m} \frac{-t \, dt}{(t-m_\pi^2)^2}\, 
         \left | F_{\pi NN} (t) \right |^2 \nonumber \\
 & = & | \widetilde \phi_\pi^{\, *} 
                      \cdot \widetilde\tau  |^2
                  \int _x^1 dy \, F_2^\pi  \left( \frac{x}{y} \right)
        f_{MB}(y) \, \, ,
\label{eqn:f2pi}
\end{eqnarray}
where $t^N_m=-m_N^2y^2/(1-y)$.  Thus $F_2^{\gamma \pi}$ is the convolution of
the pion structure function with the
light-cone momentum distribution of the pion, $f_{MB}(y)$
\cite{ADELAIDE,JULICH,HM,INDIANA,KFS}, normalized by the isospin coupling.
The same formalism is used for quark and anti-quark distributions.

The direct interaction of a 
photon with the meson cloud surrounding a
nucleon does not contribute to the
Gottfried sum. However, it does not necessarily follow that the mesons do not
contribute to the Gottfried sum and the $\overline u-\overline d$
distribution, as discussed below.  
Two general approaches have been used depending
on whether \cite{ADELAIDE,JULICH} or not \cite{HM,INDIANA,KFS} the interaction
of the recoil baryon with the virtual photon is included.  Both are in
principle the same.  The main differences are in the number of meson and baryon
states included.

The largest number of contributions has been considered in Ref.~\cite{JULICH}.
They include $\pi N$, $\rho N$, $\omega N$, $\sigma N$, $\eta N$, $\pi\Delta$,
$\rho\Delta$, $K\Lambda$, $K^*\Lambda$, $K\Sigma$, $K^*\Sigma$,
$KY^*$, and $K^* Y^*$ states coupling to the nucleon.  In most models 
however, only the $\pi N$ \cite{HM} and $\pi \Delta$ \cite{INDIANA}
nucleon couplings are included.  Since the pions alone do not contribute to the
Gottfried sum rule directly, the $\pi NN$ and $\pi N \Delta$ contributions to
the sum rule can be written as \cite{ADELAIDE}
\begin{eqnarray}
\lefteqn{F_2^{\gamma p}(x) - F_2^{\gamma n}(x) = 
Z \left \{ \frac{x}{3} [u^v(x)-d^v(x)]
\right.} \\
 &  & \mbox{} \left.
-\frac{1}{3} \int_0^{1-x} \frac{dy}{1-y} \left( \frac{x}{3} f_{\pi N}(y)
\left [ u^v(y') - d^v(y') \right ] 
-\frac{1}{2} \frac{10 x}{3} f_{\pi \Delta} (y) d^v (y') \right) 
\right \}
\ \ \ . \label{f2pmf2n}
\end{eqnarray}
where $y' = x/(1-y)$ and $f_{\pi N(\Delta)}(y)$ are meson-baryon 
light-cone momentum
distributions.  The valence normalization $Z$ is defined as
$Z=1/(1+N_\pi+\Delta_\pi)$ where the 
probability of finding a pion in {\it e.g.}\
the $\pi NN$ state is
$N_\pi=\int_0^1 dy f_{\pi N}(y)$.

The light-cone momentum distribution of the virtual meson in the nucleon
is the sum over the meson-baryon momentum distributions
\begin{equation}
f_M(y) \, = \, \sum_B f_{MB}(y) \label{fmtot}
\ \ \ ,
\end{equation}
with 
\begin{equation}
f_{MB} (y) = \frac{g_{MNB}^2}{16\pi^2} \, y
                      \int_{-\infty}^{t_m^B} dt \,
                      \frac{{\cal I}(t,m_N,m_B)}{(t - m_M^2)^2}
                      \, [F_{MNB} (t)]^2
\ \ \ 
\label{eqn:fmb}
\end{equation}
as defined from Eq.~(\ref{eqn:f2pi}).
The meson-baryon momentum distributions $f_{MB}(y)$ are the probabilities of
finding meson $M$ in an $MB$ configuration in the nucleon with fraction $y$ 
of the nucleon momentum in
the infinite momentum frame.  These distributions must be related to the baryon
distributions in the nucleon in the same configuration, $f_{BM}(y)$, by
\begin{equation}
f_{MB}(y) = f_{BM}(1-y)
\label{probsym}
\end{equation}
to conserve probability \cite{MST99}.
The $t$-dependent function ${\cal I}(t,m_N,m_B)$ has two forms,
\begin{eqnarray}
{\cal I}(t,m_N,m_B) = \left\{ \begin{array}{ll} - \, t+(m_B-m_N)^2 
            & \mbox{ for $B \in$ {\bf 8}} 
\label{fmbint} \\  \frac{\big[ \, (m_B+m_N)^2-t \, \big]^2 \,
         \big[ \, (m_B-m_N)^2-t \, \big]}
         {12 \, m_N^2 \, m_B^2 }
            & \mbox{ for $B \in$ {\bf 10}} \ , \end{array} \right.
\end{eqnarray}
depending whether the baryon $B$ is part of the octet
or the decuplet.
The upper limit of the integral when $B \neq N$ is given by
\begin{equation}
t_m^B = m_N^2 \, y - \frac{m_B^2 \, y}{1-y} \label{mcmuplim}
\ .
\end{equation}
Several functional forms, including monopole, dipole, and exponential, 
have been assumed for the meson-baryon vertex 
form factor,
\begin{eqnarray}
F_{MNB}(t) = \left\{ \begin{array}{ll} \frac{\Lambda_m^2-m_M^2}{\Lambda_m^2-t}
    & \mbox{monopole} \label{eqn:monopolef} \\
    \left(\frac{\Lambda_d^2-m_M^2}{\Lambda_d^2-t}\right)^{\!\!2}
    & \mbox{dipole} \label{eqn:dipolef} \\
    e^{(t-m_M^2)/\Lambda_e^2}
    & \mbox{exponential} \label{eqn:exponf} \, \, \, . \end{array} \right.
\end{eqnarray} 
It has recently been pointed out that in the covariant formulation of
Eq.~(\ref{eqn:fmb}) with $t$-dependent form factors, 
the probability conservation of Eq.~(\ref{probsym}) could
not always be achieved \cite{MST99}.  However, if the calculation is done with
light-cone wavefunctions, {\it e.g.}\ the dipole form of $F_{MNB}(t)$ above is
replaced by $[ (\Lambda_d^2 + m_M^2)/(\Lambda_d^2 + s_{\pi N}) ]^2$ where
$s_{\pi N}$ is the square of the pion-nucleon center of mass energy,
probability conservation is satisfied \cite{MST99}.

Studies have been made of the hardness of the form factor.
A hard form factor with a monopole cutoff
$1.0 < \Lambda_m < 1.4$ GeV is needed to 
explain the deuteron $D$-state admixture and nucleon-nucleon 
scattering \cite{Kumano} although more recent $NN$ potential models allow for
softer cutoffs, $\Lambda_m \sim 0.8$ GeV \cite{Holtom}. 
Softer cutoffs are also generally obtained in quark models,
$\Lambda_m\approx 0.6$ GeV in the cloudy-bag model
\cite{CLOUDY} and $0.7 < \Lambda_m < 1.0$ GeV 
in flux-tube type models \cite{FLUX}.
An even softer cutoff, $\Lambda_m < 0.5$ GeV \cite{FMS}, arises from
analyses of the flavor asymmetric distribution 
$(\overline u+\overline d)/2-\overline s$.
However, later analyses of this flavor asymmetric distribution 
with more recent data resulted in a slightly
larger cutoff, $\Lambda_m\approx 0.6$ GeV \cite{INDIANA}, 
which could be consistent with the quark model estimates. 
As more mesons and baryons are added to the calculation, the cutoff becomes 
larger, $\Lambda_m\approx 0.74$ GeV, because
the probability to find the bare nucleon is reduced
due to the presence of the meson clouds.
The dipole and exponential cutoffs are related to the monopole cutoff by 
$\Lambda_m=0.62\Lambda_d = 0.78 \Lambda_e$ \cite{INDIANA}.

Mesonic contributions to an anti-quark distribution in the nucleon
are given by the convolution of the corresponding  valence anti-quark
distribution in a meson with the light-cone momentum distribution
of the meson in the $MB$ state.
The contributions are given by the equation 
\begin{equation}
x \, \overline q_N(x,Q^2) = \sum_{MB} \alpha_{MB}^q 
                               \int_x^1 dy \, f_{MB}(y) \, \frac{x}{y} \, 
                               \overline q_M \left( \frac{x}{y},Q^2 \right)
\ \ \ ,
\label{eqn:CONV1}
\end{equation}
where the summations are over combinations of
meson $M=(\pi,\, K)$ and baryon
$B=(N,\, \Delta,\, \Lambda,\, \Sigma,\, \Sigma^*)$ states
and $\alpha_{MB}^q$ are spin-flavor SU(6) Clebsch-Gordon
coefficients. 
Equation~(\ref{eqn:CONV1}) 
corresponds to Eq. (\ref{eqn:f2pi}) for $F_2^{\gamma \pi}$.

From Eq.~(\ref{f2pmf2n}), the Gottfried sum becomes
$I_G=(Z/3)(1-N_\pi/3+5\Delta_\pi /3)$.
According to this equation, the failure of the sum rule is due not to 
the photon interaction with the virtual pion but to
the interaction with the recoil baryons.

When contributions from light meson and baryon states beyond $\pi N$ and $\pi
\Delta$ are included in
Eq.~(\ref{f2pmf2n}), the Gottfried integral becomes
\begin{equation}
I_G= \frac{1}{3} \, (Z+\sum_i A_i) \ , \ \ \ 
{\rm with} \ \ 
A_i=\int_0^1 dx \, (u_i + \overline u_i -d_i - \overline d_i)_{\rm Sull}
\label{eqn:pi2nd}
\end{equation}
where $A_i$ are the $x$ integrated meson and baryon contributions to the quark
and anti-quark distributions in the nucleons.  These results, given in 
Table \ref{julichtable}, should be multiplied by the probabilities of finding
the meson-baryon states in the nucleon.
Although the nucleon ``core" satisfies the valence quark sum
$\int dx (u^v-d^v)=1/3$, its probability
is reduced by the normalization factor $Z$ due to
the presence of the virtual $MB$ states.
In this approach, no contribution from pions or rho mesons
enters the sum and the violation comes
from the normalization factor $Z$ and the baryon contributions
because $u+\overline u-d-\overline d$ vanishes
in the pion.

\begin{table}[hbtp]
\begin{center}
\begin{tabular}{ccccc}
\hline
$MB$         & $A_i^M$ & $A_i^B$ & $B_i^M$ & $B_i^B$ \\
\hline
$\pi N$      &    0    &  $-$1/3 &  $-$2/3 &    0    \\
$\pi\Delta$  &    0    &     5/3 &     1/3 &    0    \\
$\rho N$     &    0    &  $-$1/3 &  $-$2/3 &    0    \\
$\rho\Delta$ &    0    &     5/3 &     1/3 &    0    \\
$\omega N$   &    0    &     1   &     0   &    0    \\
$\sigma N$   &    0    &     1   &     0   &    0    \\
$\eta N$     &    0    &     1   &     0   &    0    \\
$K\Lambda$   &    1    &     0   &     0   &    0    \\
$K^*\Lambda$ &    1    &     0   &     0   &    0    \\
$K\Sigma$    & $-$1/3  &     4/3 &     0   &    0    \\
$K^*\Sigma$  & $-$1/3  &     4/3 &     0   &    0    \\
$KY^*$       & $-$1/3  &     4/3 &     0   &    0    \\
$K^*Y^*$     & $-$1/3  &     4/3 &     0   &    0    \\
\hline
\end{tabular}
\end{center}
\caption[]{Coefficients $A_i$ and $B_i$ in two different
descriptions \cite{JULICH}.  Reproduced from Ref.~\cite{Kumano} 
with permission from Elsevier Science.}
\label{julichtable}
\end{table}

The sum could be written in a different form to satisfy the valence quark sum
exactly.  Part of the meson and baryon contributions can be factored
into the valence sum, 1/3, and the deviation from 1/3 is identified
with the flavor asymmetry due to the Sullivan processes from the meson cloud,
$\int dx(\overline u-\overline d)_{\rm Sull}$
\cite{INDIANA,JULICH,KFS}.  The Gottfried integral then becomes
\begin{equation}
I_G= \frac{1}{3} \, (1+\sum_i B_i) \ , \ \ \ 
{\rm with} \ \ 
B_i=\int_0^1 dx \, (\overline u_i - \overline d_i)_{\rm Sull}
\ \ .
\label{eqn:pi1st}
\end{equation}
The coefficients $B_i$ in Eq.~(\ref{eqn:pi1st}) are also given in Table 
\ref{julichtable}.
The virtual $\pi B$ states contribute to the renormalization
of the valence-quark distributions. Therefore
the pionic renormalization contributions may be factored into the valence
quark sum factor in Eq. (\ref{eqn:pi1st})
\cite{INDIANA}.
Then the pion contributes to the deviation from the Gottfried
sum.
Because flavor symmetry is assumed in the $MB$ states, only
the pions and rho mesons contribute to the 
violation.  Thus the two different mesonic descriptions are 
equivalent.

Although there are slight model differences, the meson-cloud approach 
is successful
in explaining the major part of the NMC and E866 results.  The solid curve in
Fig.~\ref{dmumodel} 
shows a calculation with the meson-cloud model with $\pi NN$
and $\pi N \Delta$ contributions. A dipole form factor is used with $\Lambda_d
= 1$ GeV for the $\pi NN$ coupling and 0.8 GeV for the $\pi N \Delta$
coupling, corresponding to $\Lambda_m = 0.62$ GeV and 0.5 GeV respectively.  
Reasonably good agreement with the E866 $\overline d - \overline u$
data is found \cite{mcgmp}.

Even though meson cloud models, such as the calculation shown in
Fig.~\ref{dmumodel}, may successfully explain the difference $\overline d -
\overline u$, they usually have trouble reproducing the ratio $\overline
d/\overline u$.  An approach that has had relative success in obtaining
agreement with both the difference and the ratio is the inclusion of 
a perturbative
component in the meson cloud \cite{mc,brazil}.  The analysis of Ref.~\cite{mc},
based on the assumption that the proton is composed
of three dressed quarks or valons in the infinite momentum frame \cite{hwa}, is
discussed in more detail here.  One
of the valons may emit a gluon which subsequently decays into a $q \overline q$
pair.  This $q \overline q$ pair can then nonperturbatively rearrange itself
with the valons of the proton into a meson-baryon bound state.  The flavor
symmetric GRV distributions \cite{GRVold} are assumed to be the underlying
perturbative sea.

The ratio $\overline d/\overline u$ is computed assuming two components in the
sea, the nonperturbative meson cloud and a perturbative 
contribution,
\begin{eqnarray}
\frac{\overline d(x,Q^2)}{\overline u(x,Q^2)} = \frac{\overline d^{\rm
NP}(x,Q^2) + \overline q^{\rm P}(x,Q^2)}{\overline u^{\rm
NP}(x,Q^2) + \overline q^{\rm P}(x,Q^2)} \, \, . \label{mcdboub}
\end{eqnarray}
The difference $\overline d - \overline u$ is extracted from the $\overline d/
\overline u$ ratio according to Eq.~(\ref{e866diff}), as also done in the E866
analysis \cite{E866dy}, rather
than from the difference between the numerator and denominator of
Eq.~(\ref{mcdboub}).  The perturbative anti-quark distribution in
Eq.~(\ref{mcdboub}) is taken from the GRV LO symmetric sea \cite{GRVold}.  

The nonperturbative $\overline u$ and $\overline d$ distributions are
calculated in the meson cloud model assuming that the $ \overline d$ is
generated from $\pi^+ n$ and $\pi^+ \Delta^0$ fluctuations while the $\overline
u$ arises from $\pi^- \Delta^{++}$ fluctuations,
\begin{eqnarray}
\overline d^{\rm NP}(x,Q_V^2) & = & \int_x^1 \frac{dy}{y} \left[ P_{\pi^+ n}(y)
+ \frac{1}{6}P_{\pi^+ \Delta^0}(y) \right] \overline d_\pi \left( \frac{x}{y},
Q_V^2 \right) \label{mcdbnp} \\
\overline u^{\rm NP}(x,Q_V^2) & = & \int_x^1 \frac{dy}{y}
\frac{1}{2}P_{\pi^- \Delta^{++}}(y) \overline u_\pi \left( \frac{x}{y},
Q_V^2 \right) \label{mcubnp} \, \, .
\end{eqnarray}
The fluctuations containing $\pi^0$ mesons, {\it e.g.}\ 
$\pi^0 \Delta^+$, are not
included because flavor singlets such as $\pi^0 \sim d \overline d - u
\overline u$ are assumed to be more likely to annihilate than to recombine
with the valons in the proton.  The $\overline d_\pi$ and $\overline u_\pi$
valence distributions are taken from the GRV LO pion distributions
\cite{GRVpi}.  The pion probability densities from $\pi B$ states 
are calculated by the 
convolution of a recombination function associated with pion formation,
$R(x,y,z)$,
\begin{eqnarray}
R(x,y,z) = \gamma \frac{yz}{x} \delta(x-y-z) \, \, , \label{rxyzdef}
\end{eqnarray}
and a valon quark distribution function, $F(y,z)$,
\begin{eqnarray}
F(x,y) = \beta y v(y) z \overline q(z) (1 - y - z)^a \, \, , \label{fyzdef}
\end{eqnarray}
so that
\begin{eqnarray}
P_{\pi B}(x) = \int_0^1 \frac{dy}{y} \int_0^1 \frac{dz}{z} F(y,z) R(x,y,z) \,
\, . \label{pibconv}
\end{eqnarray}
In Eq.~(\ref{fyzdef}), the exponent $a$ is fixed by requiring that the pion and
baryon in the fluctuation have the same velocity, resulting in 
$a = 12.9$ for $\pi^+ n$ and 18 for the 
$\pi \Delta$ fluctuations.  The quark distribution is calculated
from the decays of the valon to gluons, $v \rightarrow v +g$ and gluon
splitting to $q \overline q$, $g \rightarrow q + \overline q$,
\begin{eqnarray}
q(x,Q_V^2) = \overline q(x,Q_V^2) = N \frac{\alpha_s^2(Q_V^2)}{4 \pi^2}
\int_x^1 \frac{dy}{y} P_{qg} \left( \frac{x}{y} \right) \int_y^1 \frac{dz}{z}
P_{gq} \left( \frac{y}{z} \right) v(z) \label{qvalon}
\end{eqnarray}
where $Q_V^2 = 0.64$ GeV$^2$ \cite{hwa,dashwa} and $P_{qg}$ and $P_{gq}$ 
are the valon and gluon
splitting functions for QCD evolution, as in Eqs.~(\ref{qdglap}) and
(\ref{gdglap}).  The valon distribution, independent of $Q^2$, is $v(x) =
(105/16) \sqrt{x} (1-x)^2$.  The normalization constants $N$, $\gamma$, and
$\beta$ are fixed from the difference
$\overline d - \overline u$ in Eq.~(\ref{e866diff}).

\begin{figure}[htbp]
\setlength{\epsfxsize=\textwidth}
\setlength{\epsfysize=0.6\textheight}
\centerline{\epsffile{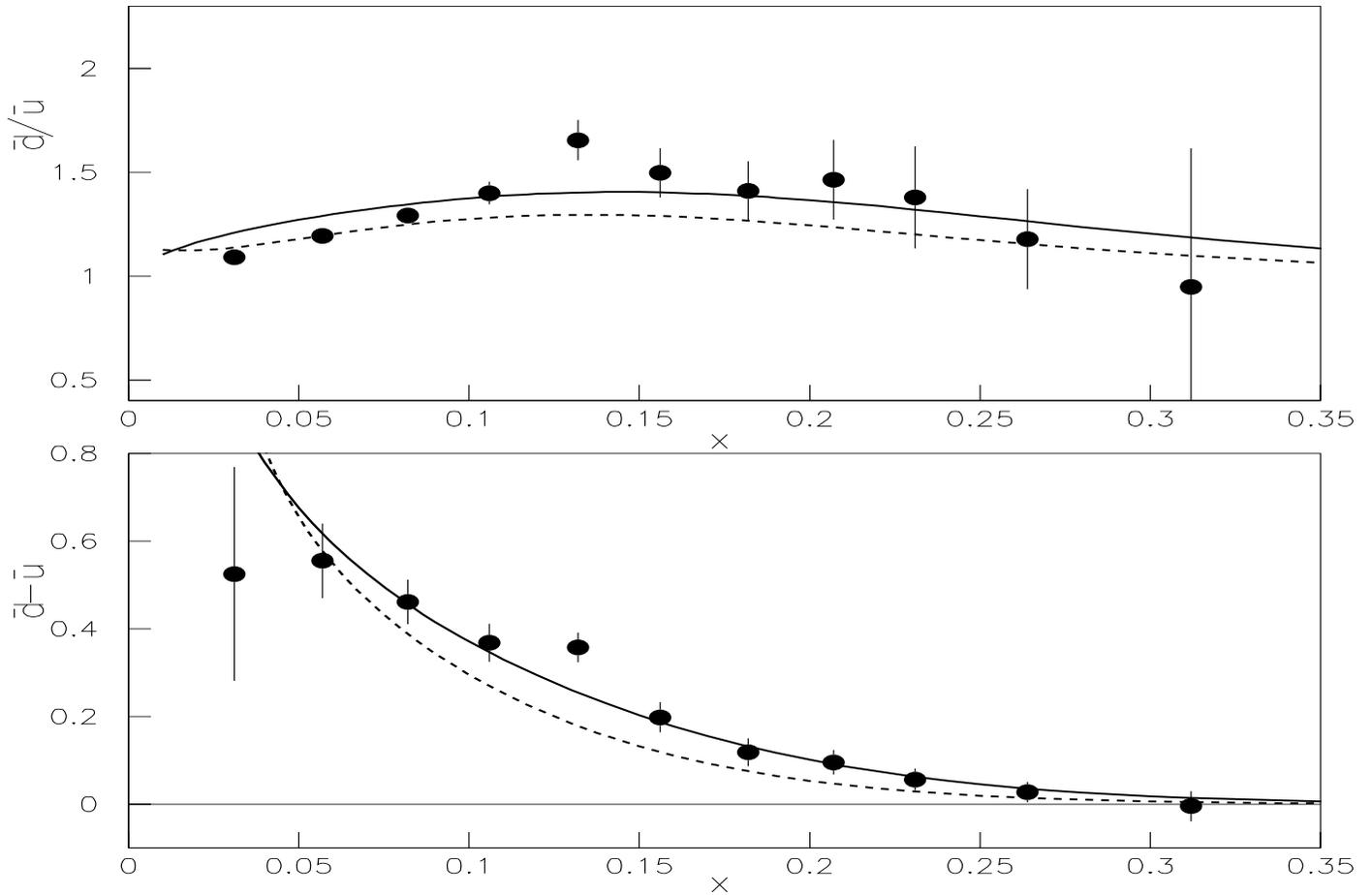}}
\caption[]{The E866 ratio $\overline d/\overline u$ and
difference $\overline d - \overline u$ are compared to calculations
at $Q^2 = 54$ GeV$^2$ \cite{E866dy} with the meson cloud model of
Ref.~\cite{mc}.   The solid curves show the $\overline u^{\rm NP}$ and
$\overline d^{\rm NP}$ distributions without $Q^2$ evolution (solid curves) and
with pseudo-evolution (dashed curves).  Reproduced from Ref.~\protect\cite{mc}.
Copyright 2000 by the American Physical Society.}
\label{mc1}
\end{figure}

The results for the ratio, Eq.~(\ref{mcdboub}), and the difference,
Eq.~(\ref{e866diff}), are shown in Fig.~\ref{mc1}.  The solid curve is the
result at the valon scale $Q^2 = 0.64$ GeV$^2$ while the dashed line  is a
pseudo-evolved result at $Q^2 = 54$ GeV$^2$ \cite{mc}.  Both calculations are
in reasonable agreement with the measurements due to the presence of the
perturbative component.  The calculation describes the ratio as well as the
difference because the valon model fixes the normalization of the perturbative
component while the meson cloud model shifts the contributions to $\overline d$
and $\overline u$ between the meson and baryon components.  Thus the meson
cloud model alone can only describe the difference.
Even better agreement is obtained if the valence
distribution in the pion in Eqs.~(\ref{mcdbnp}) and (\ref{mcubnp}) is
multiplied by an additional power of $x$ \cite{mc}.  This modified pion valence
distribution is not excluded
because the low $x$ pion parton distributions are not well known.

\subsubsection{Effects of Pauli blocking}

The Pauli exclusion principle forbids two fermions from having identical 
quantum numbers.  Thus, if a quark in the nucleon wavefunction has a 
certain set of quantum numbers, a $q$ in a virtual $q \overline q$ state cannot
be generated in the sea if it has the same quantum numbers as an already 
existing $q$.  The new $q$ is Pauli blocked.
Because the proton has two valence $u$ quarks and one valence $d$ quark, 
$u\overline u$ pair creation is more likely to be Pauli-blocked
than the $d\overline d$ pair, resulting in a difference between $\overline u$ 
and $\overline d$, as first suggested by Field and Feynman \cite{FF}.

Pauli blocking effects in the nucleon have been calculated in $1+1$
dimensions using the MIT bag model \cite{ADELAIDE,ST}.
In $1+1$ dimensions, there are three color states for each flavor.  In the 
ground state,
two of the three $u$-quark states and one of the three $d$-quark states
are occupied.  Only one more $u$ quark can be added to the
ground state, but two more $d$ quarks can be accommodated.  Since the 
additional $u$ or $d$ quarks in the ground state would be accompanied
by $\overline u$ and $\overline d$ anti-quarks, 
the expected sea quark asymmetry is 
rather large, $\overline d=2\overline u$.

No $3+1$ dimensional calculation exists at this time so that the Pauli
blocking effects are obtained from a naive counting estimate.
In the $3+1$ dimensional case, there are six states, three color and two 
spin, in the ground state.
There are four available ground states for $u$ quarks
and five states for $d$ quarks, so that $\overline d = 5 \overline u/4$. 
Because no valence anti-quarks exist in the bag, the sea quark contribution
comes from a quark being inserted, interacting in the bag, and then being
removed. Therefore, the $\overline d$ excess is related to the distribution
of a four-quark intermediate state $f_4(x)$
\begin{equation}
\int_0^1 dx \, [\overline d (x) - \overline u(x)]=\int_0^1 dx f_4(x)=1-P_2
\ \ \ ,
\end{equation}
where $P_2$ is the integral of a distribution associated with
a two-quark intermediate state.
Because the $\overline u$ and $\overline d$ distributions have not 
been calculated, the Pauli contributions were 
assumed to be $x^A (1-x)^B$ in Ref. \cite{ADELAIDE,MST99}.
The constant $A$ is chosen to match the small $x$ behavior
of the valence distributions used, and a large exponent, $B=7$, was chosen so
that Pauli blocking contributes only at small $x$. This value of $B$ is
consistent with counting rules for sea quarks \cite{brfar}.
The overall normalization
is not determined theoretically but $1-P_2$ is expected to be
$\approx 0.25$ because of 
the naive counting estimate in $3+1$ dimensions.

Pauli blocking effects could in principal, produce
the $\overline d$ over $\overline u$ excess.
Unfortunately, the qualitative $x$ dependence of the effect is only
calculated in $1+1$ dimensions.  It was suggested that a 10\% Pauli-blocking 
effect combined with
pionic contributions from the meson cloud could explain the difference
$F_2^{\gamma p}-F_2^{\gamma n}$ \cite{ADELAIDE} and the E866 
$\overline d/\overline u$ ratio and $\overline d - \overline u$ difference
relatively well \cite{brazil}.
However, if antisymmetrization of the quark wavefunctions is included,
the conclusion could change dramatically \cite{ST}.
The same $u$-valence excess which suppresses the $u\overline u$ pair creation
produces extra $u\overline u$ pairs
in the antisymmetrization of the second $u$.
This additional source of $\overline u$ could cause $\overline u > \overline
d$, in contradiction to the NMC and E866 results \cite{ST}.

\subsubsection{Chiral models}

Several different types of chiral models have been proposed to explain the 
Gottfried sum rule violation. 
In a chiral field theory with quarks, gluons, and Goldstone bosons (pions)
\cite{EHQ,KRE,CL,SBF}, 
flavor asymmetries arise from virtual photon interactions
with the pions.  On the other hand, 
in chiral soliton models \cite{SC,WAKA,BPG} a portion of the
isospin of the nucleon is carried by pions.  The Gottfried sum rule 
violation is then
determined from the moments of inertia of the nucleons and pions.
An extension of the linear sigma model to the Gottfried sum rule has also been
proposed \cite{WH,LI,SF}.
In this section, only the chiral field theory approach is briefly described.
For more details, see Ref.~\cite{Kumano}.

Because chiral symmetry is spontaneously broken, any description of low energy
hadron properties should include this symmetry
breaking.  The effective Lagrangian in
chiral field theory is \cite{Kumano}
\begin{equation}
{\mathcal L}= \overline\psi (iD_\mu+V_\mu) \gamma^\mu \psi
              +i g_{_A} \overline\psi A_\mu \gamma^\mu \gamma_5 \psi 
              +\cdot\cdot\cdot
\ \ \ ,
\end{equation}
where $\psi$ is the quark field and 
$D_\mu$ is the covariant derivative.
The vector and axial-vector currents are defined in terms of Goldstone bosons
so that
\begin{equation}
\left( \begin{array}{c} V_\mu \\ A_\mu \end{array} \right)
        = \frac{1}{2} (\xi^\dagger \partial_\mu \xi \pm
                       \xi \, \partial_\mu \xi^\dagger)
\ \ \ ,
\end{equation} where $\xi= e^{(i\Pi/f)}$ and the $+$-sign goes with the vector
current.  The meson matrix $\Pi = q \otimes
\overline q$ is
\be
\Pi=\frac{1}{\sqrt{2}}
\left( \begin{array}{ccc}
    \pi^0/\sqrt{2}+\eta/\sqrt{6}  & \pi^+ & K^+ \\
    \pi^- & -\pi^0/\sqrt{2}+\eta/\sqrt{6} & K^0 \\
    K^-   & \overline K^0     & -2\eta/\sqrt{6}
\end{array} \right)
\ \ \ .
\ee
Expanding $V_\mu$ and $A_\mu$ in powers of $\Pi/f$ gives
$V_\mu=O(\Pi/f)^2$ and $A_\mu=i\partial_\mu\Pi/f+O(\Pi/f)^2$.
Then the quark-boson interaction, proportional to $A_\mu$, becomes
${\mathcal L}_{\Pi q} = - (g_{_A}/f)
         \overline\psi\partial_\mu \Pi\gamma^\mu\gamma_5\psi$.

The Gottfried sum rule violation arises from $q \rightarrow q \pi$ splittings
where the final-state pion is massless.
The $u$ quark can split into either $u \rightarrow \pi^+ d \rightarrow u 
\overline d d$ or $u \rightarrow \pi^0 u \rightarrow u (\overline u u), \, u
(\overline d d)$.  Thus if $a$ is the splitting probability for $u \rightarrow
\pi^+ d$, 
\begin{equation}
u \rightarrow a\pi^+ +ad +\frac{a}{2}\pi^0 +\frac{a}{2} u = \frac{a}{4} \left[
7u + 5d + \overline u + 5 \overline d \right]
\ \ \ .
\end{equation}
Likewise,
\begin{equation}
d \rightarrow  a\pi^- +au +\frac{a}{2}\pi^0 +\frac{a}{2} d = \frac{a}{4} \left[
5u + 7d + 5 \overline u + \overline d \right]
\ \ \ .
\end{equation}
The probability $a$ is determined from the splitting function, $P_{\Pi q'
\leftarrow q}$, defined as
\begin{eqnarray}
P_{\Pi q'\leftarrow q} (z) = \frac{g_{_A}^2}{f^2} \, 
                             \frac{(m_q+m_{q'})^2}{32\pi^2} \, 
            z \int_{-\Lambda^2}^{t_{m}} dt \, 
                     \frac{(m_q-m_{q'})^2-t}{(t-M_\Pi^2)^2}
\ \ \ ,
\end{eqnarray}
where, as in the meson-cloud model, $t_{m}=m_q^2 z-m_{q'}^2z/(1-z)$, similar to
Eq. (\ref{eqn:fmb}), and 
$\Lambda\approx$1.169 GeV.  This large value of $\Lambda$ is assumed to be the
scale at which chiral symmetry is broken.
Integrating over $t$ for the process $u \rightarrow \pi^+ d$ gives
\begin{eqnarray}
a & = & \frac{g_{_A}^2 \, m_u^2}{8\pi^2 f^2} 
    \int_0^1 dz \, \theta (\Lambda^2-\tau(z)) \, z  
     \left \{ \ln \left [ \frac{\Lambda^2+M_\pi^2}{\tau(z)+M_\pi^2} \right ]
 \right.      \nonumber \\
  & + & \mbox{} \left. M_\pi^2 \left [ \frac{1}{\Lambda^2+M_\pi^2} -
                               \frac{1}{\tau(z)+M_\pi^2} \right ] \right \}
\end{eqnarray}
where $\theta (x) = 1$ when $x>0$ and zero otherwise.
Since $m_u \approx m_d$, $\tau(z)= -t_m(m_q = m_{q'}) = m_u^2 z^2/(1-z)$ 
With $\Lambda \approx 1.169$ GeV, the Gottfried integral becomes
$I_G = (1 - 2a)/3 = 0.278$.  Increasing $\Lambda$ decreases $I_G$.

The anti-quark distribution as a function of $x$ is \cite{EHQ,KRE}
\begin{equation}
\overline q_i (x) = \sum_{j,k,l}
               \left(  \delta_{jl} \delta_{ik} 
                 - \frac{\delta_{jk} \delta_{il}}{n_f} \right)^2
               \int_x^1 \frac{dy}{y} \int_{x/y}^1 \frac{dz}{z}
               \, \overline q_i^{\, (\Pi)} \left( \frac{x}{yz} \right) 
               \, P_{\Pi k\leftarrow j}(z)
               \, q_{v \, j}^{N}(y)
\label{eqn:cpt}
\end{equation}
where the indices $j$, $k$, and $l$ are summed over flavor.  The $\overline d
- \overline u$ difference 
$\overline d - \overline u$ is calculated at $Q^2 = 54$ GeV$^2$ using the
formulation of Szczurek {\it et al.} \cite{SBF} and compared with the E866
and HERMES data in Fig.~\ref{dmu}.  The difference in the chiral approach is
concentrated in the small $x$ region, even more so than in 
the meson-cloud model,
as seen in the dashed curve of Fig.~\ref{dmu}.  This concentration at low $x$
occurs because pions are
coupled to constituent quarks in the chiral model and the constituent quarks
carry a smaller fraction of the nucleon momentum than the virtual pions in the
meson cloud. 

\subsubsection{Charge Symmetry Violation}

Charge symmetry, Eq.~(\ref{pvsn}), 
is respected to a high degree of precision in nuclear physics with most  
low energy tests finding that it holds to
$\approx 1\%$ in reaction amplitudes \cite{Miller}. Thus, charge symmetry 
is generally assumed to be valid in strong interactions and all global 
analyses of parton distributions \cite{Lon98}.  

Experimental verification of charge symmetry is difficult for two reasons:
any violation effects are expected to be small and charge symmetry violation 
may mix with or be misinterpreted as flavor symmetry violation.  Thus,
Ma \cite{Ma1} suggested that the interpretation of the Gottfried sum rule
violation and the Drell-Yan ratio as flavor symmetry violation could instead
be the result of charge symmetry violation.  Given such ambiguities, 
experiments that could distinguish between charge and flavor symmetry 
violation are needed.  Some have already been proposed
\cite{Tim1}. 

Possible effects of charge 
symmetry violation have recently been examined by Boros
{\it et al.}\ \cite{AWT}.  They define charge symmetry violating ,CSV, 
distributions
\begin{eqnarray} 
\delta u(x)& =&  u_p(x) -d_n(x) \,\,\,\,\,\,\,\,\,\,
\delta d(x)  =   d_p(x) -u_n(x) \label{csvq} \\
\delta \overline u(x)& =& \overline u_p(x) - \overline d_n(x) 
\,\,\,\,\,\,\,\,\,\,
\delta \overline d(x)  =  \overline d_p(x) - \overline u_n(x) \, \, , 
\label{csvqb}
\end{eqnarray} 
which would disappear if charge symmetry holds.  The isoscalar structure
functions defined in Eqs.~(\ref{f2gamiso})-(\ref{f3wmiso}) are then modified
to include the violation \cite{AWT}
\begin{eqnarray}
  F_{2 \, {\rm CSV}}^{\gamma N_0}(x) & =& F_2^{\gamma N_0} - \frac{x}{18}[
 4(\delta d(x)+\delta \overline d(x)) + \delta u(x)+\delta 
   \overline u(x)] \label{f2gamcsv} \\
  F_{2 \, {\rm CSV}}^{W^+ N_0} (x) &=& F_2^{W^+ N_0}(x) -x[\delta u(x)+\delta 
\overline d(x)] \label{f2wpcsv} \\ 
  xF_{3 \, {\rm CSV}}^{W^+ N_0}(x) &=& xF_3^{W^+ N_0}(x) -x[\delta u(x)-\delta
\overline d(x)]  \label{f3wpcsv} \\ 
  F_{2 \, {\rm CSV}}^{W^- N_0} (x) &=& F_2^{W^- N_0}(x) -x[\delta d(x)+\delta
\overline u(x)] \label{f2wmcsv} \\
  xF_{3 \, {\rm CSV}}^{W^- N_0}(x) &=& xF_3^{W^- N_0}(x) -x[\delta d(x)-\delta
\overline u(x)]  \label{f3wmcsv} 
\end{eqnarray}

Likewise, one can form a charge ratio relating the neutrino structure function
to the charged lepton structure function similar to Eq.~(\ref{f2diff})
\cite{AWT} 
\begin{eqnarray}
 R_c (x) & \equiv  & \frac{F_2^{\gamma N_0}(x)}{\frac{5}{18}
 F_2^{W^+ N_0}(x) - \frac{x}{6}( s(x) +\overline s(x))} \nonumber\\ 
&\approx & 1 - \frac{s(x) -\overline s(x)}{\overline{Q}(x)} + 
 \frac{4\delta u(x) - \delta \overline u(x) - 4 \delta d(x) 
+\delta \overline d(x)}{5 \overline{Q}(x)}   
\label{rc}
\end{eqnarray}
where $\overline Q(x) \equiv \sum_{q=u,d,s} (q(x)+\overline q(x)) - 
\frac{3}{5}(s(x)+\overline s(x))$.  The denominator of Eq.~(\ref{rc})
was expanded retaining only
leading terms in the small quantities $s(x) - \overline s(x)$ and $\delta
q(x)$. If $R_c(x) \ne 1$ at any $x$, then either $s(x) \ne \overline s(x)$
or charge symmetry is violated.  The CCFR collaboration compared their neutrino
structure function $F_2^{W^+ {\rm Fe}}$ \cite{CCFRNLO} with 
$F_2^{\gamma {\rm D}}$ measured by the
NMC collaboration \cite{NMC} and found that for $x < 0.1$, $R_c \approx 0.9$ in
the region of $Q^2$ overlap between the two experiments.  When nuclear 
shadowing corrections were applied to the targets, $R_c$
remained inconsistent with unity.
It therefore appears that in this region, $s(x) \ne \overline s(x)$ or charge
symmetry is violated.

\begin{figure}[htbp]
\setlength{\epsfxsize=\textwidth}
\setlength{\epsfysize=0.5\textheight}
\centerline{\epsffile{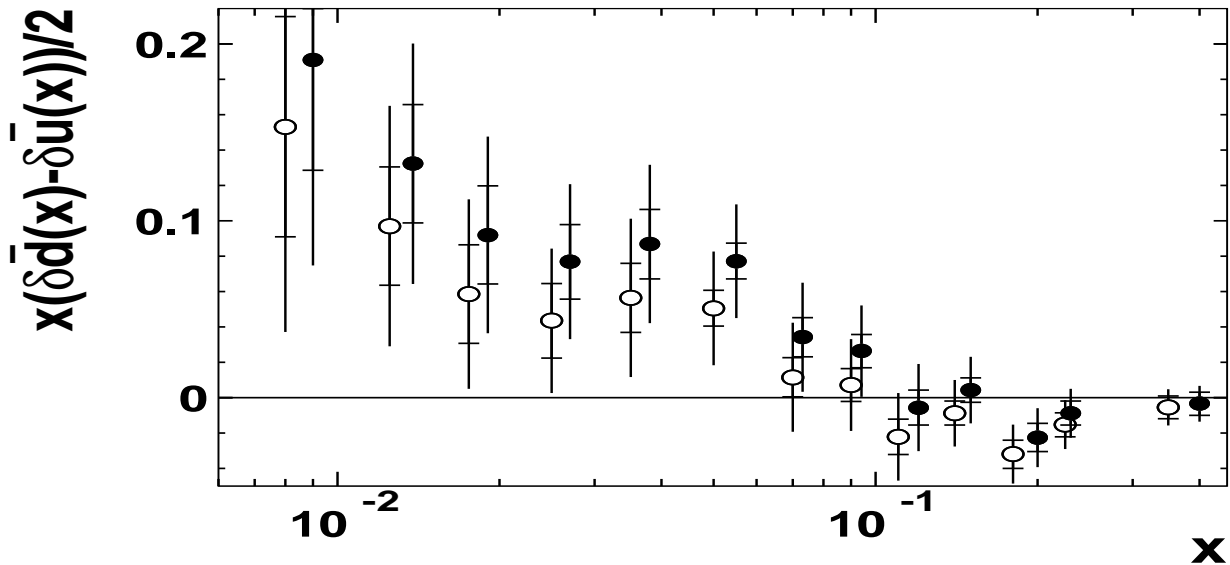}}
\caption[]{Degree of charge symmetry violation obtained from
Eq.~(\protect\ref{f2sdiff}) assuming that $s(x) = \overline s(x)$ (open
circles) and $\overline s(x) = 0$ (filled circles).  Statistical and systematic
errors have been added in quadrature.  Reproduced from
Ref.~\cite{AWT}. Copyright 1998 by the American Physical Society.} 
\label{csvio}
\end{figure}

The CCFR data allow $s(x)$ and $\overline s(x)$ to be determined
independently.  Such an analysis by the CCFR collaboration \cite{CCFRNLO}
will be discussed in more detail in Section
3.2.1.  Here, the CCFR dimuon data were used to set limits on charge symmetry
violation effects \cite{AWT}.  
Dimuon production from $\nu_\mu$ and $\overline \nu_\mu$
interactions with the strange nucleon sea is defined as
\begin{equation} 
 x s^{\mu\mu}(x) = \frac{1}{2}\, x \,[s(x) + \overline s(x)] + \frac{1}{2}  
       (2\beta^\prime -1 )\, x\, [s(x) - \overline s(x)] \, \, ,
\label{smumu}
\end{equation} 
where experimentally
$\beta' = N_\nu^{\mu\mu}/( N_\nu^{\mu\mu} + N_{\overline \nu}^{\mu\mu})
\approx \beta$ in Eq.~(\ref{f2diff}).  Combining Eq.~(\ref{smumu}) with
Eq.~(\ref{f2diff}), one can simultaneously solve for $s(x)$ and 
$\overline s(x)$.  The anti-strange distribution determined in this fashion is
negative and thus unphysical \cite{AWT}.  If charge symmetry is assumed to be
violated, the difference between Eqs.~(\ref{f2diff}) and (\ref{smumu}) can be
used to estimate the size of the violation \cite{AWT}
\begin{eqnarray} 
\lefteqn{\frac{5}{6} F_2^{W^\pm N_0}(x) -  3 F_2^{\gamma N_0}(x)
 -x s^{\mu\mu}(x) =  \frac{x(2\beta -1)}{3}[ s(x) -\overline{s}(x)]} 
\nonumber \\ 
&  & \mbox{} + \frac{x}{6} \, [ 
  (5\beta -1)(\delta d(x)-\delta u(x))+ 
  (4-5\beta )(\delta\overline d(x) -\delta\overline u(x))] \nonumber \\ 
  &\approx & {x(2\beta -1)\over 3}[s(x) 
 -\overline{s}(x)] + \frac{1}{2} \, x \, [\delta \overline d(x) -\delta 
\overline u(x)] \, \, . 
\label{f2sdiff} 
\end{eqnarray} 
In the last line, it is assumed that $\beta = \beta'$ and 
$\delta q^v = \delta q
- \delta \overline q \approx 0$.  The assumption $\delta q^v \approx 0$ means
that charge symmetry violation should
manifest itself most strongly in the sea.  Fig.~\ref{csvio} show the
magnitude of the effect obtained for the two extremes $s(x) = \overline s(x)$
and $\overline s(x) = 0$ \cite{AWT}.  
The result is surprisingly large.
Theoretically, $\delta\overline d(x) \approx 
-\delta\overline u(x)$ \cite{Ben98} which has the effect that $\overline u_p +
\overline d_p = \overline u_n + \overline d_n$.  The results of
Fig.~\ref{csvio} suggest a charge symmetry violation of $\approx 25$\% at low
$x$.  Such large effects, if they exist, 
would necessarily have to be incorporated into future
fits of parton distribution functions.

\begin{figure}[htbp]
\setlength{\epsfxsize=\textwidth}
\setlength{\epsfysize=0.5\textheight}
\centerline{\epsffile{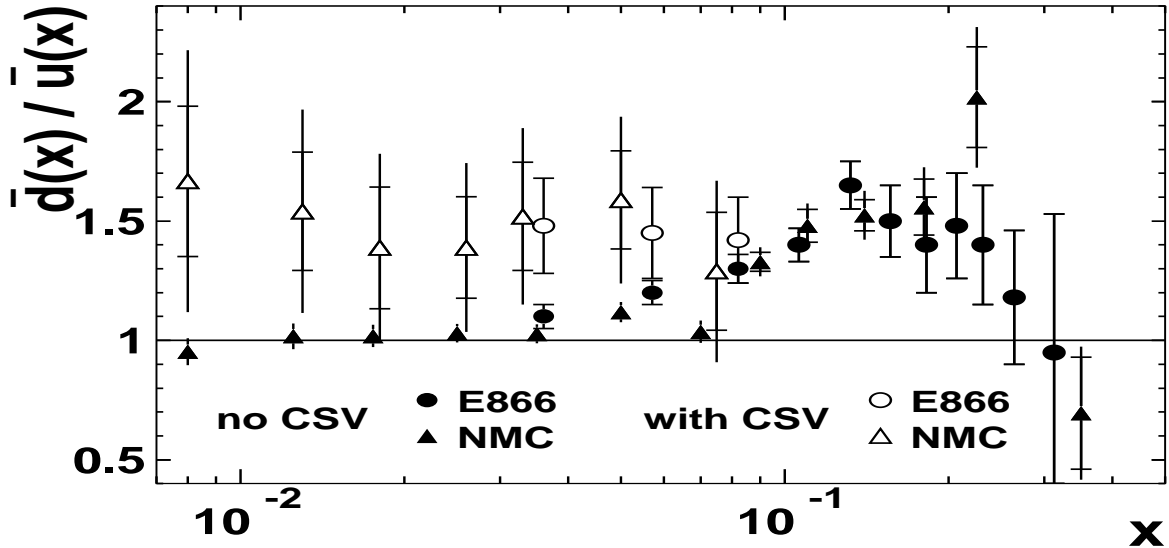}}
\caption[]{The ratio $\overline d/\overline u$ as a function of $x$ for the NMC
\cite{NMC94} (triangles) and E866 \cite{E866dy} (circles) data.  The solid
points do not assume any charge symmetry violation while the open points
include charge symmetry violation at low $x$ as parameterized in
Fig.~\ref{csvio}.  Reproduced from Ref.~\cite{BLT}.  Copyright 1999 by the 
American Physical Society. } 
\label{BLTCSV}
\end{figure}

This rather large charge symmetry violation at low $x$ would be mixed with
flavor asymmetry in the DIS and Drell-Yan studies of the Gottfried sum rule
violation.  In this case, with $-\delta \overline u \approx \delta \overline
d$, Eq.~(\ref{eqn: F2P-M}) would be rewritten as
\begin{equation}
\frac{1}{x} \left( F_2^{\gamma p}(x) - F_2^{\gamma n}(x) \right) =  
                \frac{1}{3} \, [ u^v(x) - d^v(x) ]
              + \frac{2}{3} \, [ \overline u(x) - \overline d(x) + \delta
              \overline d ]
\ .
\label{gsrcsv}
\end{equation}
Since measurements of $I_G$ suggest that the second term on the right hand side
of Eq.~(\ref{gsrcsv}) is negative and $\delta \overline d$ is positive in
Fig.~\ref{csvio}, charge symmetry violation would still require a large SU(2)
flavor asymmetry, even larger than that suggested without charge symmetry
violation.  Note that the magnitude of $I_G$ would not change, only the
interpretation of the result if charge symmetry violation mixes with flavor
asymmetry \cite{BLT}.  The $\overline d/\overline u$ ratio with and without
charge symmetry violating effects is shown in Fig.~\ref{BLTCSV}.  The open
points show how the flavor asymmetry would need to be enhanced for $I_G$ to
remain the same after charge symmetry violation is accounted for.  This would
indeed require a substantial modification of the low $x$ $\overline d$ and
$\overline u$ distributions in the global analyses of parton distribution
functions. 

\begin{figure}[htbp]
\setlength{\epsfxsize=\textwidth}
\setlength{\epsfysize=0.5\textheight}
\centerline{\epsffile{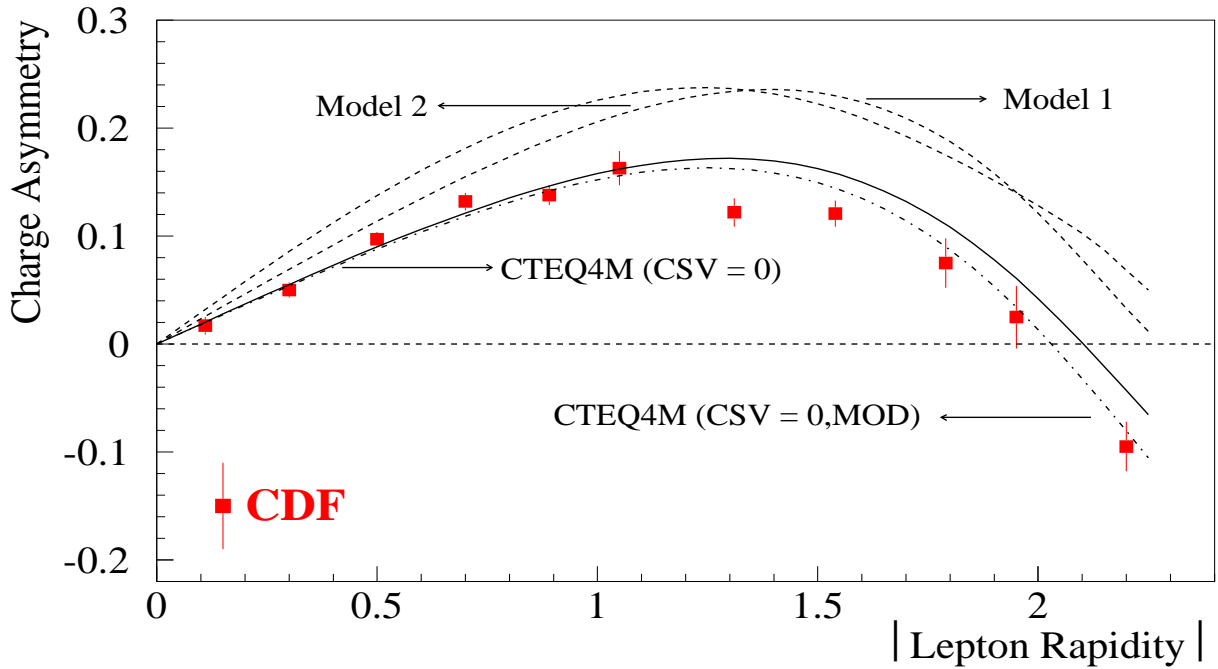}}
\caption[]{The CDF $W^\pm$ asymmetry as a function of rapidity \cite{CDFdat}
compared to calculations without charge symmetry violation, labeled CTEQ4M
(CSV=0) and CTEQ4M (CSV=0,MOD), and with the effect of Fig.~\ref{csvio} using
the two parameterizations of Bodek {\it et al.}\ \cite{Bodek}, labeled Model 1
and Model 2.  Reproduced from Ref.~\cite{Bodek}.  Copyright 1999 by the
American Physical Society. } 
\label{cdfcsv}
\end{figure}

Another test of charge symmetry violation can be made with 
$W^\pm$ production at
hadron colliders.  Boros {\it et al.}\ \cite{BLT} suggest comparing $W^+$ and
$W^-$ production in $p {\rm D}$ 
collisions at the Relativistic Heavy Ion Collider
\cite{RHIC} at Brookhaven National Laboratory, BNL, 
and the Large Hadron Collider
\cite{LHC} at CERN.  However, as recently pointed out by Bodek {\it et al.}\
\cite{Bodek}, the DIS and Drell-Yan measurements discussed here cannot extract
charge symmetry violating effects alone.  Measurements of $d_p/u_p$ or
$d_n/u_n$ would isolate these effects more definitively.  
At the Fermilab Tevatron, $W^+$ and
$W^-$ are produced by $u \overline d \rightarrow W^+$ and $d \overline u 
\rightarrow W^-$ respectively.  The $u$ quarks carry more of the
proton momentum than
the $d$ quarks, see Figs.~\ref{cteqpdf}-\ref{grvpdf}, so that the $W^+$
follows the incoming proton while the $W^-$ follows the anti-proton, creating a
charge asymmetry between the two in rapidity.  This asymmetry, measured through
the decay leptons, is directly sensitive to the $d_p/u_p$ ratio \cite{Bodek}.
Bodek {\it et al.}\ parameterized the charge symmetry violation of
Fig.~\ref{csvio} in two different ways.  In the first, the average of the
$\overline d$ and $\overline u$ distributions is unchanged by charge symmetry 
violation while
in the second, the entire violation 
is placed on the $\overline d$, leading to a
shift in the total sea quark distribution.  Both parameterizations are compared
to the Tevatron CDF data at 1.8 TeV \cite{CDFdat} in Fig.~\ref{cdfcsv}.  The
large charge symmetry violation suggested in Refs.~\cite{AWT,BLT} 
overpredicts the
asymmetry for $y>0$ while calculations with parton distribution functions 
assuming no charge symmetry violation agree with the data.  
Bodek {\it et al.}\ proposed that even though there is no evidence for charge
symmetry violation in the CDF data, the large violation proposed by Boros
{\it et al.}\ \cite{AWT,BLT} can be preserved if charge
symmetry is violated only in the neutron parton distributions.  In
this case, the neutron sea would be larger than the proton sea.  Thus, while
the results of Ref.~\cite{Bodek} weaken the case for charge symmetry violation,
they cannot eliminate it until better measurements can be made.  Note however
that the treatment of $F_3^{W^\pm N}$ by Boros {\it et al.}\ \cite{AWT,BLT} is
at leading order \cite{Bodek}.  A treatment beyond leading order might
eliminate the need for charge and flavor symmetry violations in the neutron.

\subsubsection{Light sea summary}

Experiments have clearly shown that not only is $\overline u \neq \overline d$
but also that the difference is a strong function of $x$ \cite{NMC94,E866dy}.
A number of models have been proposed to describe this difference, including
meson clouds, Pauli blocking, chiral models, and charge symmetry violation.
All but the last assume that $\overline u \neq \overline d$ is due to SU(2)
flavor asymmetry.  Meson cloud models have generally had the most success
describing the difference $\overline d - \overline u$ but, 
with the exception of Ref.~\cite{mc}, 
generally have not had as much success with the ratio $\overline
d/\overline u$.  Chiral models tend to concentrate the difference at low $x$,
underestimating $\overline d - \overline u$ as a function of $x$.  The proposed
charge symmetry violation would also require a large flavor asymmetry 
\cite{AWT,BLT} but
relies on the accuracy of the CCFR strange quark measurements \cite{CCFRNLO}.

\subsection{The strange sea}

Possible manifestations of nonperturbative effects in the strange sea
are now discussed along with some phenomenological explanations.  The CCFR data
which might imply a difference between the $s$ and $\overline s$ distributions
\cite{CCFRNLO} are discussed first.  Three experiments which indirectly measure
strange form factors, two low-energy elastic $ep$ scattering determinations 
of the parity-violating asymmetry \cite{SAMPLE,HAPPEX} 
and neutral current $\nu$ and a
$\overline \nu$ scattering experiment 
on nucleons \cite{Ahrens87}, are also introduced.

\subsubsection{Experiments}

The CCFR collaboration has performed LO and NLO evaluations 
of dilepton production in neutrino scattering off nucleons
to more directly study the strange quark distributions
\cite{CCFRLO,CCFRNLO}.  Although, as mentioned earlier, they found $s(x) = 
\overline s(x)$ within their uncertainties, the result is not inconsistent
with $s(x) \ne \overline s(x)$ in some $x$ regions, as is now discussed.

Pairs of opposite sign muons can be produced via charm production in $\nu_\mu$
and $\overline \nu_\mu$ scattering off nucleons:
\begin{eqnarray}
\nu_\mu\; + \; N \;
 \longrightarrow \; \mu ^{-} \; \!\! & + & \;
\! c \; +\;  X \label{nunscat} \\
& & \; \!\!\hookrightarrow \;  \mu ^{+} \; + \; \nu_\mu \nonumber \\
\overline\nu_\mu\; + \; N \;
 \longrightarrow \; \mu ^{+} \; \!\! & + & \; \! \overline c \; +\; X
 \label{nubnscat} \\
& & \; \!\! \hookrightarrow \;  \mu ^{-} \; + \; \overline\nu_\mu \, \, . 
\nonumber
\end{eqnarray}
In the first case, the neutrino interacts with an $s$ or $d$ quark, $\nu + s,d
\rightarrow c + \mu^-$, to produce a charm quark
which then decays semileptonically to a $\mu^+$.  
Because the $d$-induced channel is Cabibbo suppressed relative
to that of the strange quark, this process can provide a determination of the
strange sea.  Likewise, the anti-neutrino-induced interaction is an independent
measurement of the anti-strange distribution.

The non-strange quark and anti-quark components of the sea 
were assumed to be symmetric so that $\overline u(x) = u_s(x)$ and 
$\overline d(x) = d_s(x)$.
An isoscalar correction was also applied assuming 
$\overline u(x) = \overline d(x)$.
The strange sea was allowed to have a different 
magnitude and shape than the non-strange sea.
Shape parameters were determined from two fits, both starting from the same set
of proton parton distributions at initial scale $\mu_0^2 = 1$ GeV$^2$
\cite{distribs}. 
In the first fit, $s(x) = \overline s(x)$, while in the second $s(x)$ and
$\overline s(x)$ were determined independently.
The $x$-integrated strange fraction of the sea is set by the parameter 
\begin{equation}
\kappa= \frac{\int_0^1 dx \,[xs(x,\mu^2)+x\overline s(x,\mu^2)]}{\int_0^1 dx \,
[x\overline u(x,\mu^2)+x\overline d(x,\mu^2)]} \, \, , \label{kapdef}
\end{equation}
where $\kappa=1$ indicates an SU(3) flavor symmetric sea.
The shape of the strange quark distribution is
related to that of the non-strange sea by a shape parameter $\delta$.
If $\delta=0$, the strange sea would have the same $x$
dependence as the non-strange sea. 

Assuming that $s(x) = \overline s(x)$, the sea quark
distributions were parameterized as
\begin{eqnarray}
xs(x,\mu^2) & = & A_s(1-x)^\delta \left[\frac{x\overline u(x,\mu^2) + 
x\overline d(x,\mu^2)}{2}\right]  \label{ssymfit} \\
x\overline q(x,\mu^2) & = & (2 + A_s(1-x)^\delta ) 
\left[\frac{x\overline u(x,\mu^2) + 
x\overline d(x,\mu^2)}{2}\right] \label{qbsymfit} \, \, .
\end{eqnarray}
The normalization $A_s$ in Eq.~(\ref{ssymfit}) is defined in terms of 
$\kappa$ and $\delta$.  With the requirement $s(x) = \overline s(x)$, the
parameters are
\begin{eqnarray}
\kappa&=&0.477 \;^{+\; 0.051}_{-\; 0.050} \;^{-\; 0.017}_{+\; 0.036}
\label{kapses} \\
\delta& =& -0.02 \;^{+\; 0.66}_{-\; 0.60} \;^{+\; 0.08}_{-\; 0.20}
\label{alfses} \\
m_c & = & 1.70 \pm 0.19 \pm 0.02 \; {\rm GeV} \, . \label{mcses}
\end{eqnarray}
Note that the charm quark mass enters the fit through the original charm
production reaction in Eqs.~(\ref{nunscat}) and (\ref{nubnscat}).
\begin{figure}[htbp]
\setlength{\epsfxsize=\textwidth}
\setlength{\epsfysize=0.3\textheight}
\centerline{\epsffile{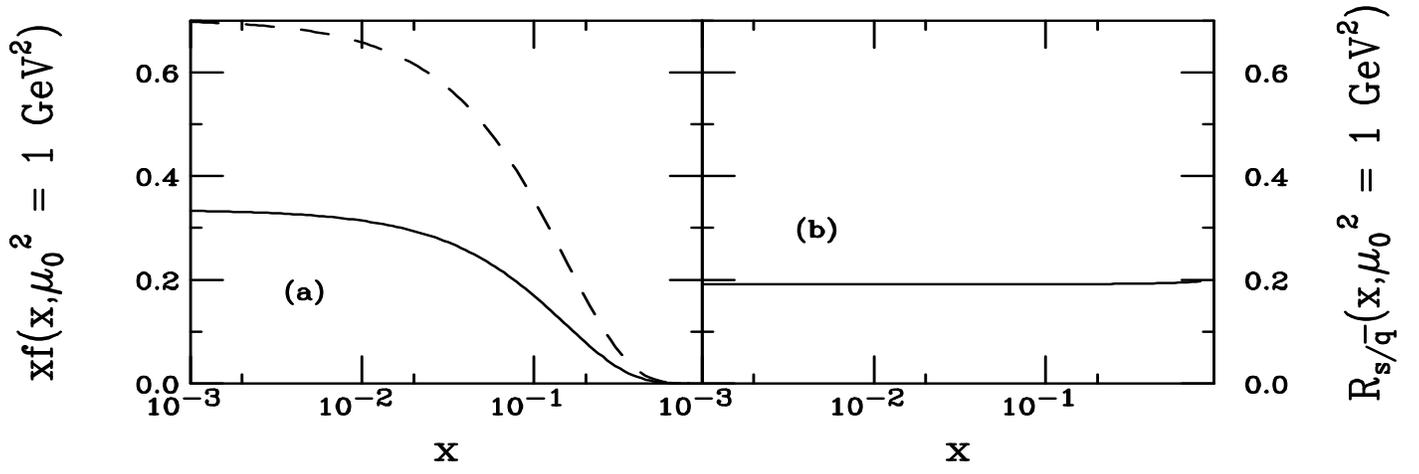}}
\caption[]{The CCFR $x\overline q$ and $xs$ distributions in the $s(x) =
\overline s(x)$ fit are compared at the
initial scale 1 GeV$^2$.  In (a) the individual distributions are shown with
the strange quark distribution (solid curve) multiplied by a factor of 2.5.
In (b), the ratio $xs/x \overline q$ is shown. } 
\label{ccfrssb}
\end{figure}

The strange quark, Eq.~(\ref{ssymfit}), and the total anti-quark,
Eq.~(\ref{qbsymfit}), $x \overline q = x \overline u + x \overline d + x
\overline s$, distributions in this fit are shown in Fig.~\ref{ccfrssb}
at the initial scale $\mu_0^2 = 1$ GeV$^2$.
The strange quark distribution is scaled up by a factor of 2.5 in
Fig.~\ref{ccfrssb}(a) to show its similarity to the anti-quark distribution.
Since $\delta \ne 0$ would indicate 
a shape difference between $x\overline
q(x)$ and $xs(x)$, the fit value $\delta=-0.02\;^{+\;0.66}_{-\;0.60} $ 
indicates no shape difference at NLO. In a previous leading order fit, they 
found that the strange quark distribution was softer than the total 
quark sea by $(1-x)^{2.5\pm 0.7}$.
The difference between the LO and NLO fits may be attributable to a softer
$x\overline q(x)$ at NLO than LO.  This similarity is highlighted in
Fig.~\ref{ccfrssb}(b) where the ratio of the strange quark distribution to $x
\overline q$ is shown.  The ratio is essentially constant over all $x$.

\begin{figure}[htbp]
\setlength{\epsfxsize=\textwidth}
\setlength{\epsfysize=0.3\textheight}
\centerline{\epsffile{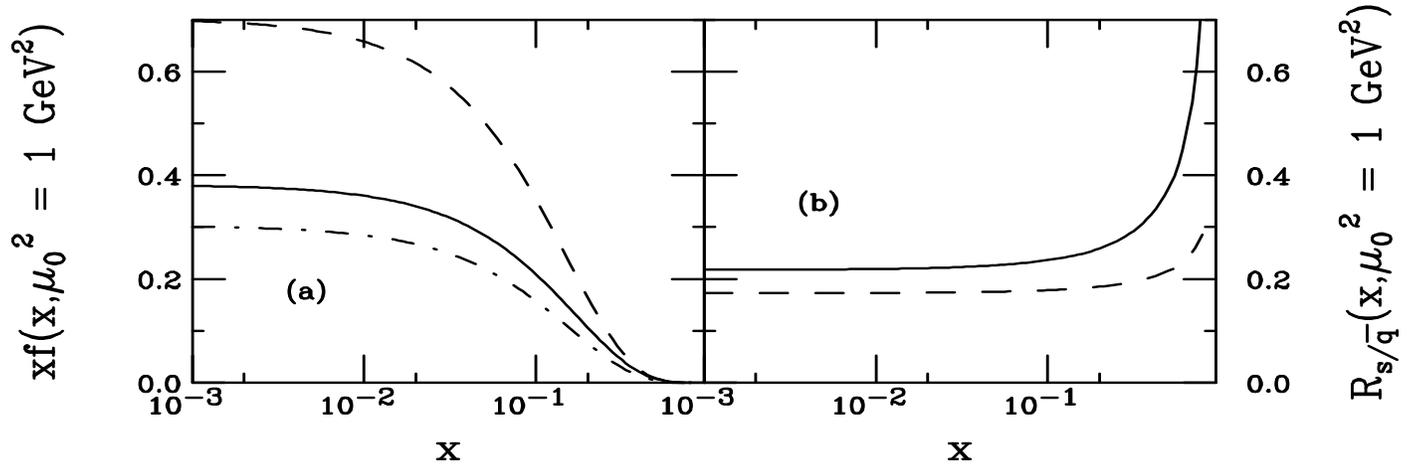}}
\caption[]{The CCFR $x\overline q$, $xs$, and $x \overline s$ distributions 
in the $s(x) \neq \overline s(x)$ fit are compared at the
initial scale 1 GeV$^2$.  In (a) the individual distributions are shown with
the strange quark (solid curve) and anti-strange quark (dot-dashed curve)
multiplied by a factor of 2.5.
In (b), the ratios $xs/x \overline q$ (solid curve) and $x \overline s/x
\overline q$ (dashed curve) are shown. } 
\label{ccfrssbdiff}
\end{figure}

If one allows the momentum distributions of the 
$s$ and $\overline s$ quarks to be different, the sea quark distributions can
be parameterized by:
\begin{eqnarray}
 xs(x,\mu^2) & = & A_s(1-x)^\delta 
\left[\frac{x\overline u(x,\mu^2) 
+x\overline d(x,\mu^2)}{2}\right] \label{sasymfit} \\
 x\overline s(x,\mu^2) & = & A_s^\prime(1-x)^{\delta - \Delta \delta}
\left[\frac{x\overline u(x,\mu^2) 
+x\overline d(x,\mu^2)}{2}\right] \label{sbasymfit} \\
x\overline q(x,\mu^2) & = & (2 + \frac{1}{2}(A_s(1-x)^\delta \nonumber \\ 
&  & \mbox{} + A_s^\prime(1-x)^{\delta - \Delta \delta}))
\left[ \frac{x\overline u(x,\mu^2) + 
x\overline d(x,\mu^2)}{2} \right] \, \, . \label{qbasymfit}
\end{eqnarray}
The fit required the number of $s$ and $\overline s$ quarks to be equal:
\begin{equation}
\int_0^1 dx \, s(x,\mu^2) =  \int_0^1 dx \, \overline s(x,\mu^2) \, \, . 
\label{seqsb}
\end{equation}
The normalization factors $A_s$ and $A_s^\prime$ are defined 
in terms of $\kappa$, $\delta$ and
$\Delta \delta$.
In this case they found
\begin{eqnarray}
 \kappa & = & 0.536 \pm 0.030 \pm 0.036 
 \;^{-\; 0.064}_{+\; 0.098} \pm 0.009 \label{kapsnes} \\
 \delta & = & -0.78 \pm 0.40 \pm 0.56 \pm 0.98 \pm 0.50 \label{alfsnes} \\
 \Delta\delta & = & -0.46 \pm 0.42 \pm 0.36 \pm 0.65 \pm 0.17 \label{dalfsnes}
 \\
 m_c & = & 1.66 \pm 0.16 \pm 0.07 \;^{+\; 0.04}_{-\; 0.01}
\pm 0.01 \; {\rm GeV} \, \, . \label{mcsnes}
\end{eqnarray}
In this analysis, the total
strange quark content, given by $\kappa$, is slightly higher and the shape of
the $s$ quark distribution is different from the non-strange sea.  Compare
Eqs.~(\ref{kapsnes}) and (\ref{alfsnes}) with
the values of $\kappa$ and $\delta$ in Eqs.~(\ref{kapses}) and (\ref{alfses}).
The charm quark mass parameters in the two fits are very similar. 
The uncertainty in $\Delta\delta$ is large enough for the
$s$ and $ \overline s$ momentum distributions to be consistent with a 
difference limited to $-1.9 < \Delta\delta < 1.0$ at the 90\% confidence level.
The resulting $s(x)$ and $\overline s(x)$ distributions in
Eqs.~(\ref{sasymfit}) and (\ref{sbasymfit}) are compared to the $\overline q$
distribution, Eq.~(\ref{qbasymfit}), in 
Fig.~\ref{ccfrssbdiff} at the initial scale $\mu_0^2 = 1$ GeV$^2$.  Both the
$s$ and $\overline s$ distributions are multiplied by a factor of 2.5 to
facilitate comparison with the larger $\overline q$ distribution.  The strange
quark distribution is larger than the anti-strange distribution and
deviates more from the total anti-quark distribution than does the anti-strange
distribution. 

It is important to note that the starting distributions \cite{distribs} used in
the fits assume flat sea and gluon distributions as $x \rightarrow 0$ and a 
flavor symmetric light quark sea.  Both these assumptions have had to be
modified recently.  If the analysis was repeated with newer starting
distributions which take these new developments into account, perhaps the
results would change.  In particular, the discrepancy between these 
data and the muon-induced DIS data which led Boros {\it et al.}\ to argue 
for a large charge 
symmetry violation \cite{AWT} could be reduced and the difference, if any,
between $s(x)$ and $\overline s(x)$ could be more clearly delineated.

\begin{figure}[htbp]
\setlength{\epsfxsize=\textwidth}
\setlength{\epsfysize=0.5\textheight}
\centerline{\epsffile{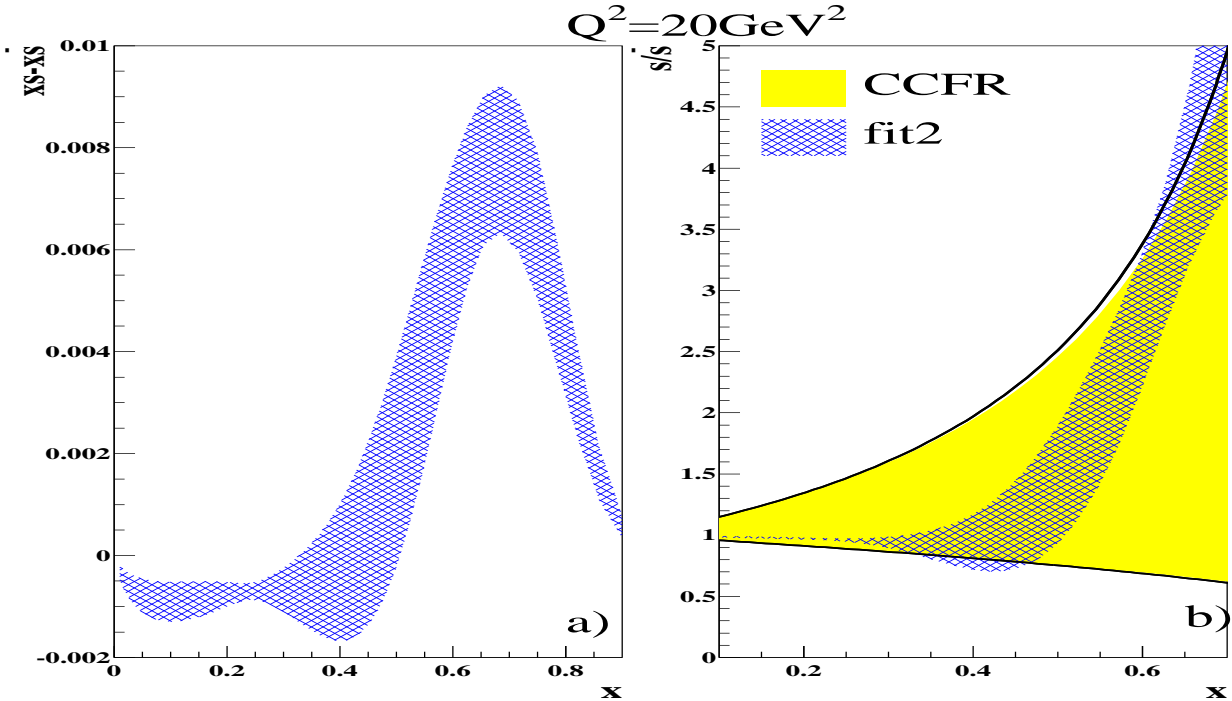}}
\caption[]{Results of a fit to all DIS data allowing $s(x) \neq \overline s(x)$
\cite{BPZ}.  The results for the difference $xs(x) - x \overline s(x)$ is shown
in (a) while the ratio $s(x)/\overline s(x)$ of the fit is compared to the CCFR
result at $Q^2 = 20$ GeV$^2$ in (b).  Reproduced from Ref.~\cite{BPZ} with
permission from Springer-Verlag.} 
\label{BPZ17}
\end{figure}

A step in this direction was recently taken by Barone {\it et al.} \cite{BPZ}
in an analysis of DIS data that emphasized the $\nu$ and $\overline \nu$
induced cross section data.  The CCFR data were 
not used in their global fit but
the fitted strange quark distribution was shown to agree with their data.
Barone {\it et al.}\ also allowed $s(x) \neq \overline s(x)$ in one of their
fits to all DIS data but did not constrain the shapes to be proportional to
$0.5 [\overline u(x) + \overline d(x)]$ as in Eqs.~(\ref{sasymfit}) and
(\ref{sbasymfit}).  Since the DIS data included in the Barone {\it et al.}\ fit
was more balanced between $\nu$ and $\overline \nu$ induced events with better
statistics, tighter constraints on the difference between $s(x)$ and $\overline
s(x)$ were set.  They note that the assumption $s(x) \neq \overline s(x)$ gave
a better overall fit to the data with $x s(x)$ harder than $x \overline s(x)$.
The difference is largest at $x> 0.4$, as seen in Fig.~\ref{BPZ17} \cite{BPZ}.

Strange form factors have been inferred from the parity violating asymmetry in
electron-proton elastic scattering
\be
A & = & \frac{\sigma_R - \sigma_L}{\sigma_R + \sigma_L} \nonumber \\
  & = & \frac{-G_F Q^2}{\pi \alpha \sqrt{2}} \frac{\epsilon G_E^\gamma 
G_E^Z + \tau G_M^\gamma G_M^Z - 0.5(1 - 4 \sin^2 \theta_W) \epsilon'
G_M^\gamma F_A^Z}{\epsilon(G_E^\gamma)^2 + \tau(G_M^\gamma)^2} \, \, ,
\label{rlasymm}
\ee
where $\tau = Q^2/4m^2$, $\epsilon = 1/(1 + 2(1 + \tau)\tan^2(\theta/2))$,
$\epsilon' = \sqrt{\tau(1 + \tau)(1 - \epsilon^2)}$, and $\theta$ is the 
scattering angle.
This asymmetry between right and left handed incident electrons is due to the
interference of the electromagnetic and neutral weak amplitudes.  The
electromagnetic electric and magnetic form factors are $G_E^\gamma$ and
$G_M^\gamma$ while the neutral weak electric, magnetic, and axial form factors
are $G_E^Z$, $G_M^Z$ and $F_A^Z$.  
To leading order, the proton weak magnetic form factor is
\be
G_M^Z = \frac{1}{4} \left[ G_M^{\gamma p}(1 - 4 \sin^2 \theta_W) - 
G_M^{\gamma n} - G_M^s \right] \, \, . \label{gmz}
\ee
The electroweak radiative corrections to Eq.~(\ref{gmz}) have been calculated
\cite{4Sample}.  The weak mixing angle $\theta_W$ has been determined with
high precision and $G_M^{\gamma p}$ and $G_M^{\gamma n}$ are known.

The strange magnetic form factor has been determined by the SAMPLE 
collaboration \cite{SAMPLE} at backward angles.  A 200 MeV polarized
electron beam is directed on a liquid hydrogen target.  The polarized electron
source is a GaAs photoemission crystal.  The photoemission is stimulated by a
laser beam incident on the crystal.  The laser beam is circularly polarized and
the electron beam helicity is reversed by changing the voltage to the cell 
which polarizes the laser beam, reversing the circular polarization of the 
light.  The helicity is chosen at random for ten consecutive pulses and the
complement helicities are used for the next ten pulses.  The scattered 
electrons are detected by a large Cerenkov detector which accepts momentum
transfers of $Q^2 \sim 0.1$ GeV$^2$.  The asymmetry is found for pairs of
pulses separated by 1/60 s to minimize systematic uncertainties.  Each ``pulse
pair'' asymmetry is equivalent to a measurement of the parity violating
asymmetry, Eq.~(\ref{rlasymm}).

The SAMPLE collaboration determined $G_M^Z$ from $A$.  
The strange magnetic form factor $G_M^s$ can be obtained from $A$ 
since the backward scattering angles and low $Q^2$ cause the second term in the
numerator of Eq.~(\ref{rlasymm}) to dominate the asymmetry.  They found
\be
A & = & -6.34 \pm 1.45 \pm 0.53 \, {\rm ppm}\, \, , \label{rlasamp} \\
G_M^Z & = & 0.34 \pm 0.09 \pm 0.04 \pm 0.05 \, {\rm n.m.} \label{gmzsamp} 
\ee
at $Q^2  = 0.1$ GeV$^2$.  The last uncertainty on $G_M^Z$ 
is due to theoretical 
uncertainties in the form factors.  If $G_M^s = 0$, then $G_M^Z = 0.40$ n.m.
\cite{SAMPLE}.  Thus the difference between the expected and the measured 
$G_M^Z$ corresponds to \cite{SAMPLE}
\be
G_M^s (Q^2 = 0.1 \, 
{\rm GeV}^2) = 0.23 \pm 0.37 \pm 0.15 \pm 0.19 \, {\rm n.m.}
\, \, ,
\label{SAMPLEres}
\ee
consistent with zero.

The HAPPEX collaboration performed a similar measurement of $A$ at a higher
energy, with 3.356 GeV electrons, and the scattered electrons were measured at
more forward angles $\langle \theta_{\rm lab} \rangle \sim \pm 12.3^\circ$
\cite{HAPPEX}.  The polarized source was a GaAs photocathode excited by a
circularly polarized laser.  The helicity was set every 33.3 ms and structured
as pairs of consecutive 33.3 ms periods of opposite helicity.  The presence of
false asymmetry was ruled out by inserting a half-wave plate in the laser
beam.  The correlation between the half-wave plate and the sign of the
asymmetry was an unambiguous signal of parity violation.  The measured
asymmetry was 
\be
A = -14.5 \pm 2.0 \pm 1.1 \, {\rm ppm}
\label{rlahapp}
\ee 
at $Q^2 = 0.48$ GeV$^2$.  Since $\tau$ and $\epsilon$ are of the same order at
forward angles, $G_E^s$ and $G_M^s$ cannot be separated from each other.  Thus 
the combination of the strange electric and magnetic form
factors was reported \cite{HAPPEX}
\be
G_E^s + \frac{\tau}{\epsilon} \frac{G_M^\gamma}{G_E^\gamma} G_M^s = 0.023 \pm
0.034 \pm 0.022 \pm 0.026(\delta G_E^{\gamma n}) \, {\rm n.m.} \, .
\label{gesgmscombo}
\ee
The last uncertainty is an estimate of the theoretical uncertainty on
$G_E^{\gamma n}$.  
A new run should improve the precision of Eq.~(\ref{gesgmscombo}) by
a factor of two \cite{Armst}.

Both the SAMPLE and HAPPEX results are consistent with strong
suppression of the strange sea at low energies \cite{GI}.  To separate $G_E^s$
from $G_M^s$ at fixed $Q^2$, a number of scattering angles and thus beam
energies are needed \cite{Armst}.  An isoscalar target would also help separate
$G_E^s$ and $G_M^s$ \cite{HAPPEX}.
Additional measurements of weak currents and strange form factors are planned
at Mainz \cite{13Sample} and the Thomas Jefferson Nation Accelerator
Facility, TJNAF \cite{14Sample}.

Information about the strange form factors of the
nucleon can also be obtained from scattering of $\nu$ and
$\overline \nu$ on nucleons.  The BNL experiment E734 \cite{Ahrens87} performed
such an investigation using wideband neutrino and anti-neutrino beams on a 
$^{12}$C target with an average energy of 1.3 GeV and momentum transfer
$0.5 < Q^2 < 1$ GeV$^2$.  The majority, 80\%, of the data were due to 
quasielastic proton knockout from the carbon nuclei, $\nu n \rightarrow \mu^- 
p$ and $\overline \nu p \rightarrow \mu^+ n$, charged current processes,
and the remaining 20\% arose from
$\nu$ and $\overline \nu$ elastic scattering on free protons, $\nu p 
\rightarrow \nu p$ and $\overline \nu p \rightarrow \overline \nu p$, governed
by neutral current interactions.  Since the quasielastic interactions do not
involve strange quarks, the strange form factors only appear in neutral current
processes. 

Experimental differential cross sections are obtained by a convolution of the 
expected differential distribution with the neutrino/anti-neutrino energy 
spectrum, $\phi_\nu(\epsilon_\nu)$, so that 
\begin{eqnarray}
\langle \frac{d\sigma_\nu}{dQ^2}\rangle =
\frac{1}{\Phi_\nu}\int_{\epsilon_{\rm min}}^{\epsilon_{\rm max}}
d\epsilon_\nu \frac{d\sigma_\nu}{dQ^2} \phi_\nu (\epsilon_\nu)
\label{dsdq2nu}
\end{eqnarray}
where $\epsilon_{\rm min} = 0.2$ GeV and $\epsilon_{\rm max} = 5$ GeV.
In Eq.~(\ref{dsdq2nu}), $\nu$ represents either neutrinos or anti-neutrinos.  
The total neutrino flux is $\Phi_\nu$.  In the elastic scattering events,
$Q^2$ is directly obtained from the kinetic energy of the scattered proton,
$T_p$, in 
the laboratory frame, $Q^2 = 2m_p T_p$.  In the quasielastic knockout
events, an effective momentum transfer for $\nu$ and $\overline \nu$ scattering
off free nucleons was obtained by correcting for the binding energy and Fermi
motion of the struck nucleon.  The total cross sections, $\langle \sigma_\nu
\rangle$, were obtained by integrating Eq.~(\ref{dsdq2nu}) over $Q^2$ in 
the range of the data.  The cross section ratios were \cite{Ahrens87}:
\begin{eqnarray}
R &=& 
\frac{\langle \sigma_{\overline{\nu} p\rightarrow 
\overline{\nu} p} \rangle }
{\langle \sigma_{\nu p\rightarrow
\nu p} \rangle } = 0.302 \pm 0.019 \pm 0.037 \label{ratel} \\
R_\nu &=& 
\frac{\langle \sigma_{\nu p\rightarrow \nu p} \rangle }
{\langle \sigma_{\nu n\rightarrow \mu^- p} \rangle } = 
0.153 \pm 0.007 \pm 0.017 \label{ratnu} \\
R_{\overline{\nu}} &=& 
\frac{\langle \sigma_{\overline{\nu} p\rightarrow 
\overline{\nu} p} \rangle }
{\langle \sigma_{\overline{\nu} p\rightarrow
\mu^+ n} \rangle} = 0.218 \pm 0.012 \pm 0.023 \label{ratnub} 
\end{eqnarray}
where $R$ is the elastic $\overline \nu/\nu$ scattering ratio while $R_\nu$
and $R_{\overline \nu}$ are the ratios of elastic scattering to quasielastic
knockout cross sections after the correction to ``free scattering''.
Thus $R_\nu$ and $R_{\overline \nu}$ are ratios of charged to neutral current
processes while $R$ involves only neutral current scattering.

In the following sections, models of the interactions discussed here are
introduced and compared to the data.

\subsubsection{$s$/$\overline s$ asymmetry}

The CCFR NLO evaluation of the strange quark distributions did not exclude the
possibility that the strange and anti-strange quark distributions are different.
The analysis indicated that this asymmetry is small, on the order of 
$(1-x)^{-0.46}$ \cite{CCFRNLO}, as suggested by Eqs.~(\ref{sasymfit}), 
(\ref{sbasymfit}), and (\ref{dalfsnes}).  There is actually no reason why the
sea quark and anti-quark distributions must be the same besides phenomenological
prejudices as long as their total numbers are identical.  Nothing in the QCD
Lagrangian explicitly relates the sea quark and anti-quark distributions in the
nucleon.  Charge conjugation symmetry only tells us that the quark distribution
in a proton is the same as the anti-quark distribution in an anti-proton 
\cite{jitang}.

Ji and Tang showed that for every sea quark interaction, the corresponding
sea anti-quark interaction may be obtained by changing the direction of
quark propagation in the Feynman diagram \cite{jitang}.  However, different
color factors are associated with the quark line direction, changing the
relative interaction strengths.  They found that while the quark and anti-quark
diagrams are of the same order in $1/N_c$ ($N_c$ is the number of colors)
the numerical coefficients are not constrained  to be identical.

A difference between the strange and anti-strange quark distributions can arise
naturally in the meson cloud-type models introduced in Section 3.1.2.  Since
the proton can fluctuate into a virtual kaon-hyperon intermediate state, the
strange anti-quark distribution in the kaon is different from the strange quark
distribution in the hyperon \cite{SigTom}.  The results of meson cloud model
calculations \cite{pnndb,cm,bma,MelMalmcm} 
will be discussed in this section followed by a
calculation of the strange and anti-strange distributions based on the locality
of the strange sea \cite{jitang}.

\begin{figure}[htbp]
\setlength{\epsfxsize=\textwidth}
\setlength{\epsfysize=0.5\textheight}
\centerline{\epsffile{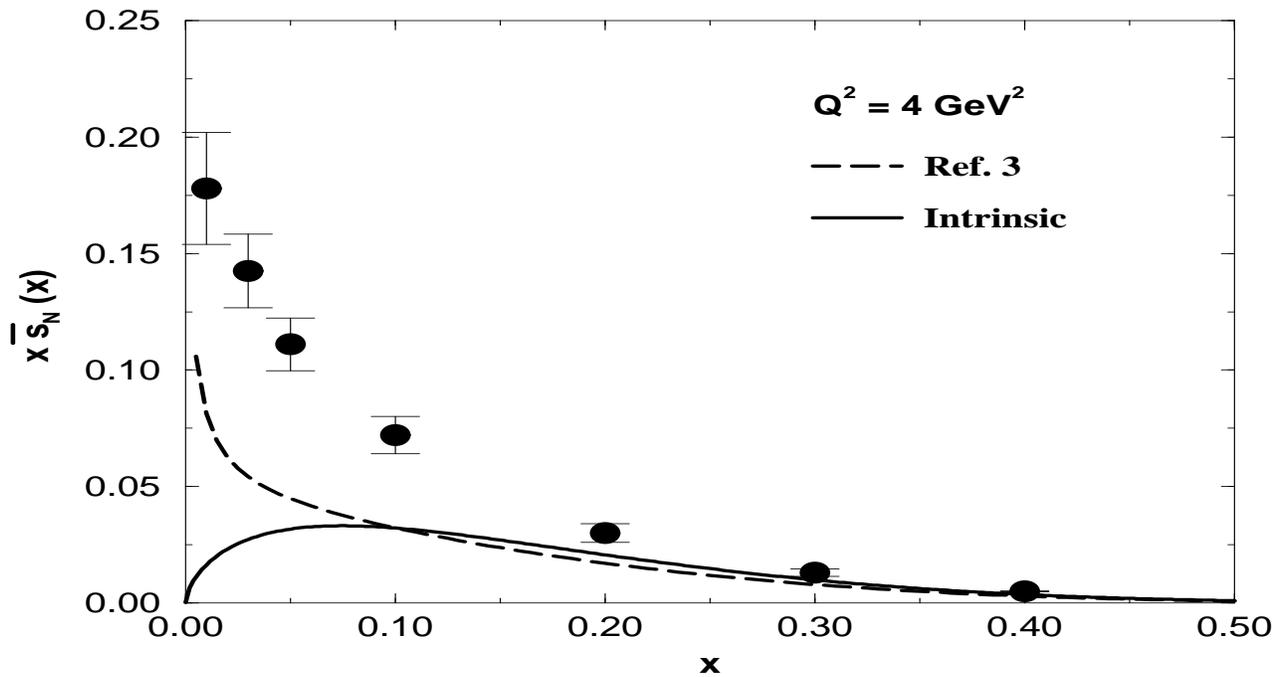}}
\caption[]{The intrinsic $x \overline s_N$ distribution calculated with the
meson cloud model (solid curve).  
The result is compared to the CCFR \cite{CCFRNLO} data
and the model calculation of \cite{KFS} (dashed curve), labeled Ref.\ 3
on the figure.  Reproduced from Ref.~\cite{pnndb} with permission from
World Scientific. } 
\label{ismcm}
\end{figure}

The first model, by Paiva {\it et al.}\ \cite{pnndb}, 
follows directly from the one outlined in 
Eqs.~(\ref{fmtot})-(\ref{eqn:CONV1}).  If the kaon is assumed to be the 
intermediate meson state and the $\Lambda$, $\Sigma$, and $\Sigma^*$ are
the intermediate hyperon states, the strange anti-quark distribution in the
nucleon is
\begin{equation}
x {\overline s}_N (x,Q^2) =\sum_{Y} \tau_Y \int_{x}^{1}
dy\, f_{KY} (y)\, \frac{x}{y} \, {\overline s}_K^{v}(\frac{x}{y},Q^2)\; .  
\label{xsbnmcm}
\end{equation} 
following Eq.~(\ref{eqn:CONV1}) where $\tau_Y$ are the kaon-hyperon spin-flavor
SU(6) Clebsch-Gordon coefficients.  The kaon distribution in the nucleon cloud,
$f_{KY}(y)$, is calculated as in Eq.~(\ref{eqn:fmb}) with $M \rightarrow K$
and $B \rightarrow Y$.  An exponential form factor, Eq.~(\ref{eqn:exponf}),
was used with $\Lambda_e = 1.2$ GeV, corresponding to $\Lambda_m = 0.94$ GeV. 
The strange valence quark distribution in
the kaon was assumed to be identical to the pion valence quark distributions.
In Ref.~\cite{pnndb} the SMRS \cite{SMRS} pion structure function was used to
obtain $x \overline s_K^v(x,Q^2)$.  
Paiva {\it et al.}\ refer to the distribution
in Eq.~(\ref{xsbnmcm}) as the intrinsic anti-strange quark distribution, after
the intrinsic charm model of Brodsky {\it et al.}\ \cite{intc}.  Their
intrinsic distribution is compared to the CCFR data and another meson-cloud
model calculation \cite{KFS} based on the GRV pion distribution functions
\cite{GRVpi} in Fig.~\ref{ismcm}.
The shape of the intrinsic distribution is valence-like, as expected in such
models \cite{intc}.  The GRV pion-based result, labeled Ref.\ 3 on 
Fig.~\ref{ismcm}, has the opposite behavior at low $x$.  While both 
calculations agree with the CCFR data for $x > 0.3$, neither can describe the
low $x$ behavior of the data.

Christiansen and Magnin \cite{cm} developed their approach in the context of 
the valon model \cite{hwa}, described in Section 3.1.2.  In this case, 
the gluon emitted by the valon can decay into an
$s \overline s$ pair.  This three valon + $s \overline s$ state can then 
rearrange itself to form a kaon-hyperon bound state.  They point out that the
strange
meson and baryon distributions inside the nucleon are not independent but
are constrained by the requirements of zero strangeness in the nucleon and
momentum conservation of the meson-baryon state,
\begin{eqnarray}
\int_0^1 dx \left[P_B(x) - P_M(x) \right] & = & 0 \label{szero} \\
\int_0^1 dx \left[xP_B(x) + xP_M(x) \right] & = & 1 \,\, . \label{momcons}
\label{sconds}
\end{eqnarray}
Both conditions are satisfied if $P_M(x) = P_B(1-x)$ where $P_M(x)$ is the 
strange meson distribution in the proton, corresponding to $f_{KY}$ in 
Eq.~(\ref{xsbnmcm}).  Note that in this model, no explicit form factor is
required since the valon model is defined on the light cone.  Thus, the 
conditions in Eqs.~(\ref{szero}) and (\ref{momcons}) can be satisfied.

They use the recombination model \cite{dashwa} in Eq.~(\ref{pibconv}) with
$a=1$ in Eq.~(\ref{fyzdef}) to predict the rearrangement of 
the valons and the strange partons in the proton so that
\begin{eqnarray}
P_M (x) = \int_0^x \frac{dy}{y} \int_0^{x-y}\frac{dz}{z} F(y,z) R(x,y,z) \,
\, .
\label{pmdist}
\end{eqnarray}
The normalizations of $F(y,z)$ and $R(x,y,z)$ are
fixed by assuming the probability of $s \overline s$ pair production
by a valon is $\sim 4-10$\% \cite{cm}.  

\begin{figure}[htbp]
\setlength{\epsfxsize=\textwidth}
\setlength{\epsfysize=0.65\textheight}
\centerline{\epsffile{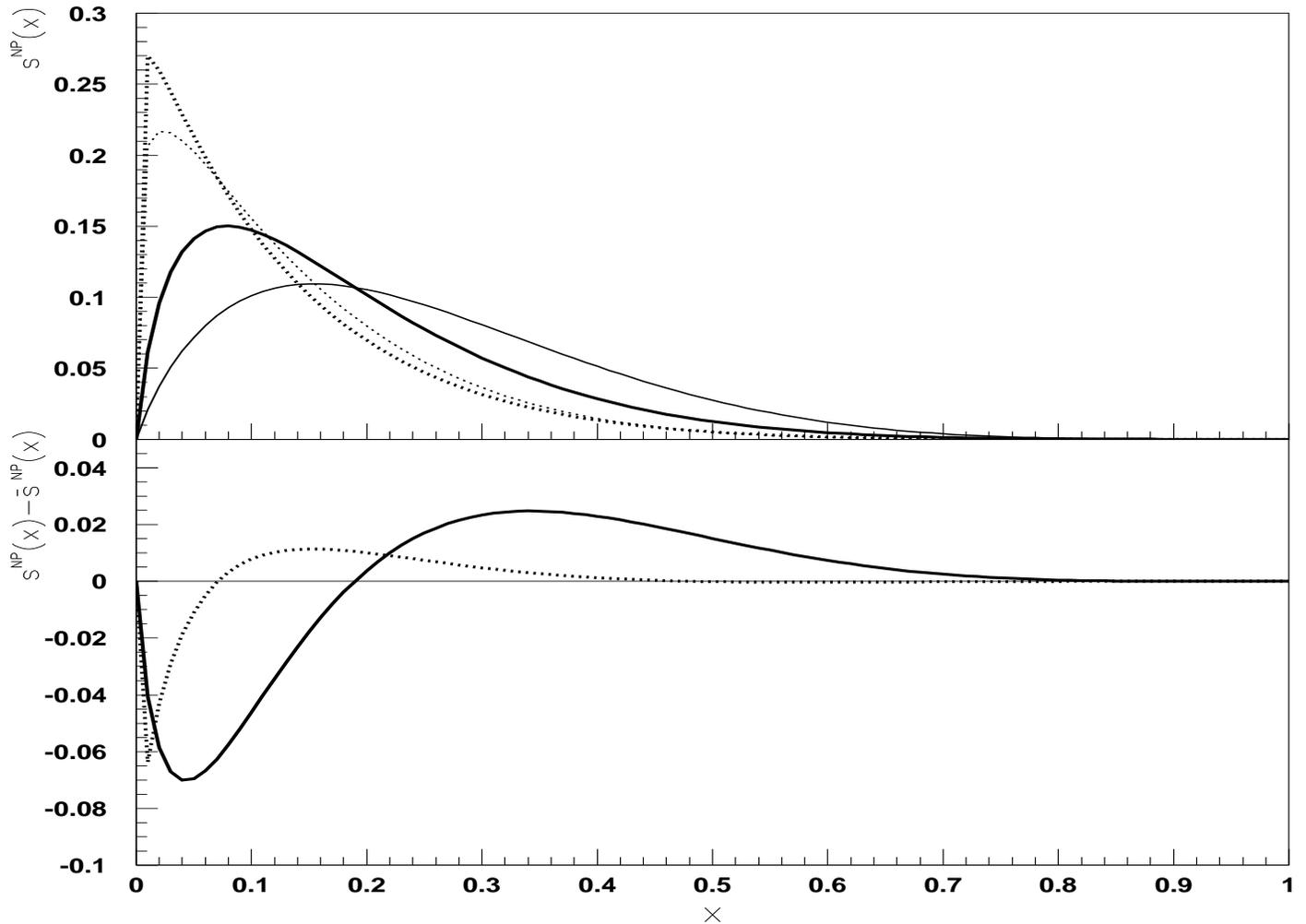}}
\caption[]{The upper plot shows the nonperturbative strange (thin lines) and
anti-strange (thick lines) distributions in the nucleon.  The solid curves
are calculated assuming the schematic kaon and hyperon distributions of 
Eq.~(\ref{approx}) while the dotted curves are calculated with 
Eq.~(\ref{shadrs}).  The lower plot gives the difference $s^{\rm NP}(x) -
\overline s^{\rm NP}(x)$ obtained with Eq.~(\ref{approx}) (solid curve) and 
Eq.~(\ref{shadrs}) (dotted curve).  Reproduced from Ref.~\cite{cm} with
permission from Elsevier Science. } 
\label{cmmcm}
\end{figure}

The nonperturbative strange and anti-strange quark densities
in the nucleon are
\begin{eqnarray}
s^{\rm NP}(x) & = & \int^1_x \frac{dy}{y} P_B(y)\ s_{B} \left( \frac{x}{y} 
\right) \label{xsnpcm} \\
\overline{s}^{\rm NP}(x) & = & \int^1_x \frac{dy}{y} P_M(y)\ 
\overline{s}_{M} \left( \frac{x}{y} \right) \,\, .
\label{xsbnpcm}
\end{eqnarray}
Two different approximations of $s_B(x)$ and $\overline s_M(x)$ were tested,
\be
 \overline s_M(x) & = & 6x(1-x) \,\,\,\,\,\,\,\,\,\,\,\,\, 
\,\,\,\,\,\,\,\,\,\,\,
 s_B(x) = 12x (1-x)^2
\label{approx} \\
 \overline s_M(x) & = & q^\pi(x) (1-x)^{0.18} \,\,\,\,\,\,\,\,
 s_B(x) = \frac{1}{2} u^v_p(x)  \label{shadrs} \,\, .
\ee
The strange and anti-strange densities in the proton, 
Eqs.~(\ref{xsnpcm}) and (\ref{xsbnpcm}), are shown in Fig.~\ref{cmmcm} along 
with the difference $s^{\rm NP}(x) - \overline s^{\rm NP} (x)$.  
Although they do not
compare the calculations directly to data, it is clear that the 
kaon and hyperon distributions in Eq.~(\ref{shadrs}) agree better with the
trends of the CCFR data since the differences between the $s$ and $\overline s$
distributions are small except at low $x$ where the valon model is expected
to break down.
An ``intrinsic''-type meson cloud model 
calculation was performed by Brodsky and Ma
\cite{bma} with results similar to those shown in the solid curve in
Fig.~\ref{cmmcm}. A light-cone calculation of $s(x) - \overline s(x)$ by
Melnitchouk and Malheiro shows the opposite behavior: $s(x) - \overline s(x)$
is positive at low $x$ and negative at large $x$ \cite{MelMalmcm}.

Ji and Tang discussed the implications of the CCFR data on the ``locality'' of
the strange sea \cite{jitang}.  They define 
two limits on the locality.  If the 
$s$ and $ \overline s$ are bound in pairs or if they move independently of each
other but have similar interactions with other partons, the $s$ and $\overline
s$ have essentially the same spatial and momentum wavefunctions.  This limit
corresponds to small locality.  If, instead, the quarks and anti-quarks move
independently of each other with different interactions with other
partons, their
distributions should be different, corresponding to large locality.  The 
locality in coordinate space, the strangeness radius, is
\begin{equation}
\langle r^2_s \rangle = \int d{\vec r} \; r^2 \,
(|s(\vec r)|^2-|\overline s (\vec r)|^2) \, \, .
\label{sradius}
\end{equation}
The corresponding locality in momentum space is \cite{jitang}
\begin{equation}
  L_s = \int_0^1 dx \; |s(x)-\overline s (x)| \, \, .
\label{slmom}
\end{equation}
A small $L_s$ or $\langle r^2_s \rangle$ is equivalent to a large locality
while a large $L_s$ or $\langle r^2_s \rangle$ represents a small 
locality.

They concluded that the standard meson cloud picture of the $s$ and $\overline
s$ distributions, as in Eq.~(\ref{xsbnmcm}) \cite{KFS,pnndb}, predicts a
larger locality than the CCFR data can accommodate, see Fig.~\ref{cmmcm}.
They attempted to  model the $s$ and $\overline s$ distributions in a manner
that could correlate coordinate and momentum space localities.  One obvious
way of doing this is to assume that $g \rightarrow s \overline s$ is the
dominant $s \overline s$ production mechanism at scale $\mu^2 = 1$ GeV$^2$.
This perturbative production is regulated by introducing effective $s$ and
$\overline s$ masses which could arise from mean-field interactions between
the $s$ and $\overline s$ with other components of the sea.  They allow
$m_s \neq m_{\overline s}$ to account for differences between the $s$ and 
$ \overline s$ interactions in the nucleon.  The larger the effective mass, the
harder the momentum distributions.  They calculate 
the strange quark distribution
in the gluon by \cite{jitang}
\begin{eqnarray}
f_s(x,\mu^2) & = & \frac{\alpha_s}{4\pi} \int_0^{\mu^2}
dk_{\perp}^2  [(k_{\perp}^2+ m_{\overline s}^2)
+(1-x)(m_s^2- m_{\overline s}^2)]^{-2}
\left\{ 2x(1-x)m_s m_{\overline s} \right. \nonumber \\
&   & \mbox{} \left. + (k_{\perp}^2+m_{\overline s}^2)[x^2+(x-1)^2] +(1-x)^2(
m_s^2- m_{\overline s}^2)\right\}\ 
\label{x}
\end{eqnarray}
where $\alpha_s = 0.5$ and $f_{\overline s}(x,\mu^2) = f_s(1-x,\mu^2)$.

\begin{figure}[htbp]
\setlength{\epsfxsize=\textwidth}
\setlength{\epsfysize=0.4\textheight}
\centerline{\epsffile{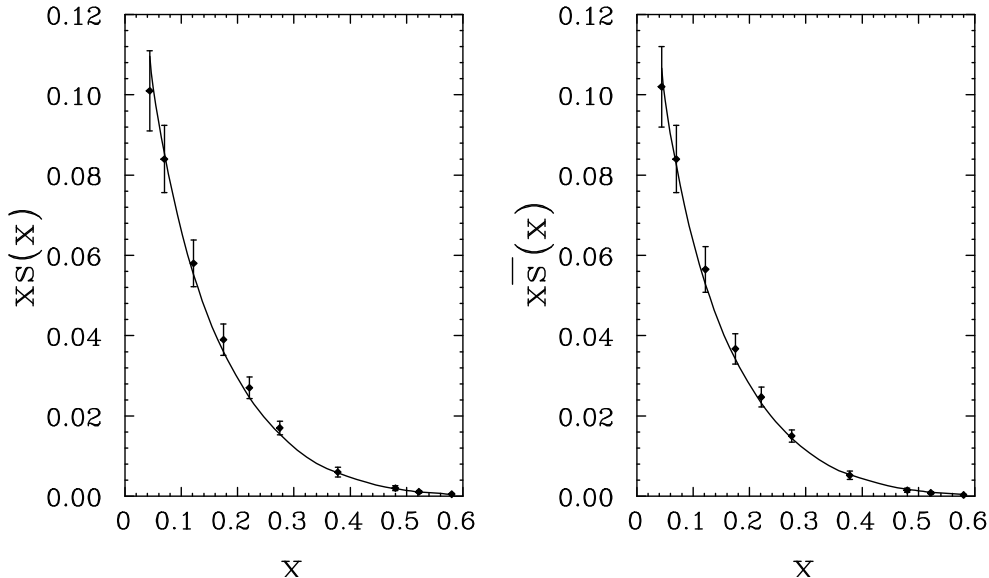}}
\caption[]{Comparison of the strange and anti-strange quark distributions 
calculated with Eq.~(\ref{sdistjt}) with the CCFR data \cite{CCFRNLO}.
The effective masses $m_s = 260$ MeV and $m_{\overline s} = 220$ MeV are used.
Reproduced from Ref.~\cite{jitang} with permission from Elsevier Science. } 
\label{jtfit}
\end{figure}

The strange quark distribution in the nucleon is obtained by convoluting 
$f_s(x,\mu^2)$ with the nucleon gluon distribution,
\begin{equation}
\label{sdistjt}
xs(x,m_s,m_{\overline s})=\int_x^1
{dy\over y} ~xf_s({x\over y},m_s, m_{\overline s})~G(y ,\mu^2=1 \,
{\rm GeV}^2) \ .
\end{equation}
They fit Eq.~(\ref{sdistjt}) and the corresponding anti-strange distribution to
the CCFR $s$ and $\overline s$ results by adjusting $m_s$ and $m_{\overline 
s}$.  With the CTEQ3 gluon distribution \cite{cteq3}, they found
$m_s=260 \pm 70$ MeV and $\overline m_s=220 \pm 70 $ MeV.  Note that the lower
limit of $m_{\overline s}$ is correlated with the upper limit of $m_s$.
Within the range of uncertainties, the effective masses are equivalent and the
agreement with the data is good, as seen in Fig.~\ref{jtfit}.

Comparing the meson-cloud type models of Figs.~\ref{ismcm} and \ref{cmmcm}
with the gluon splitting picture of Fig.~\ref{jtfit}, it appears that the
$s$ and $\overline s$ distributions are more strongly correlated than allowed
by the meson cloud model unless rather hard valence distributions are assumed
for the valence quark distributions in the strange mesons and baryons, such
as in
the dotted curves of Fig.~\ref{cmmcm}.  The data thus seem to imply that the
$s$ and $\overline s$ quarks undergo similar interactions and their 
distributions have a high degree of locality.

The rather significant correlation between the $s$ and $\overline s$ 
distributions suggests that the strangeness radius, Eq.~(\ref{sradius}), is
small.  The strangeness radius can be extracted by measuring the strange quark
contribution to the nucleon elastic form factor, as done by SAMPLE 
\cite{SAMPLE}.  Both Paiva {\it et al.}\ \cite{pnndb} and Ji and Tang 
\cite{jitang} provided estimates of the strangeness radius within their models.

In the intrinsic strangeness calculation of Paiva {\it et al.}\, the integral
over $\overline s_N$ is equivalent to the probability of finding an intrinsic
$s \overline s$ pair in the proton wavefunction.  They obtain $P_{\rm IS}
= \int dx \, \overline s_N(x) = 0.12$.  The probability is related to the 
strangeness radius by
\begin{equation}
P_{\rm IS} = {|\langle r_s^2 \rangle |\over | \langle r_p^2 \rangle |}\; ,
\label{pis}
\end{equation}
where, using $r_p = 0.72$ fm and the integrated value of $P_{\rm IS}$, 
they find \cite{pnndb}
\be
| \langle r_s^2 \rangle | = 0.0622 ~{\rm fm}^2 \, \, .
\ee
They also note that if they only considered $K \Lambda$ components in the
cloud, the strangeness radius would be reduced by more than a factor of two.
However, they can only predict the absolute value of $| \langle r_s^2 \rangle 
|$, not its sign.

Ji and Tang translate their small locality in momentum space into a coordinate
space locality by dimensional analysis \cite{jitang}
\be
| \langle r_s^2 \rangle | \sim {(m_s-\overline m_s)\over \mu^3} 
\le 0.005 ~{\rm fm }^2 \ ,
\ee
considerably smaller than the prediction of Paiva {\it et al.}\ \cite{pnndb}
as well as other earlier theoretical predictions \cite{GI,jitang} 
although the result is similar
to that of Ref.~\cite{MelMalmcm}.

\subsubsection{Strange form factors}

The strange form factors have been studied phenomenologically in the context of
the E734 $\nu$ and $\overline \nu$ scattering experiment \cite{Ahrens87}.  
The neutral current cross
section is sensitive to the electric, magnetic, and axial strange form factors
of the nucleon while the charged current interactions do not access the strange
quarks in the nucleon sea at low $Q^2$.  

The neutral current elastic scattering cross sections
are \cite{Wanda}
\begin{eqnarray}
\frac{d\sigma_{\nu p \rightarrow \nu p}}{d Q^2} &=& 
\frac{G_F^2}{2\pi}\left\{
C_M\left[\epsilon_W G_M^{\gamma p}- G_M^{\gamma n} -G_M^s\right]^2
\right. 
\nonumber\\
& & \mbox{} + C_E\left[ \epsilon_W G_E^{\gamma p} -G_E^{\gamma n} -
G_E^s\right]^2 +C_A(F_A-F_A^s)^2
\nonumber \\
& &\left. \mbox{}
+ C_{AM} (F_A-F_A^s)\left(\epsilon_W G_M^{\gamma p} -G_M^{\gamma n} -
G_M^s\right) \right\} \label{ncnu} \\
\frac{d\sigma_{\overline \nu p \rightarrow \overline \nu p}}{d 
Q^2} &=& 
\frac{G_F^2}{2\pi}\left\{
C_M\left[\epsilon_W G_M^{\gamma p}- G_M^{\gamma n} -G_M^s\right]^2
\right. 
\nonumber\\
& & \mbox{} + C_E\left[ \epsilon_W G_E^{\gamma p} -G_E^{\gamma n} -
G_E^s\right]^2 +C_A(F_A-F_A^s)^2
\nonumber \\
& &\left. 
\mbox{} - C_{AM} (F_A-F_A^s)\left(\epsilon_W G_M^{\gamma p} -
G_M^{\gamma n} -G_M^s\right) \right\} \label{ncnubar}
\end{eqnarray}
where $\epsilon_W = 1 - 4 \sin^2 \theta_W$ and $G_F$ is the Fermi constant.  
The $Q^2$ dependence of the form factors has been suppressed in
Eqs.~(\ref{ncnu}) and (\ref{ncnubar}).  The kinematic quantities
$C_M$, $C_E$, $C_A$, and $C_{AM}$ above are defined as
\begin{eqnarray}
C_M&=&\frac{1}{2}y^2
\\
C_E&=&1-y-\frac{My}{2E_\nu}
\\
C_A&=& 1-y+ \frac{1}{2}y^2 +\frac{My}{2E_\nu}
\\
C_{AM}&=& 2y\left(1-\frac{1}{2}y\right)
\end{eqnarray}
where $y = Q^2/(2m_p E_\nu)$ and $E_\nu$ is the incident neutrino energy.
The neutral current cross sections include terms independent of the strange
form factors as well as terms with linear and quadratic dependencies on
$G_M^s$, $G_E^s$, and $F_A^s$.

The corresponding charged current cross sections are \cite{Wanda}
\begin{eqnarray}
\frac{d\sigma_{\nu n \rightarrow \mu^- p}}{d Q^2}
&=& \frac{G_F^2}{2\pi}|V_{ud}|^2\left\{
C_M\left(G_M^{\gamma p} -G_M^{\gamma n}\right)^2 +C_E\left(G_E^{\gamma p} -
G_E^{\gamma n}\right)^2
\right.
\nonumber\\
& &\left. 
\mbox{} + C_AF_A^2 + C_{AM} F_A \left(G_M^{\gamma p} -G_M^{\gamma
n}\right)\right\} \label{ccnu} \\
\frac{d\sigma_{\overline \nu p \rightarrow \mu^+ n}}{d Q^2}
&=& \frac{G_F^2}{2\pi}|V_{ud}|^2\left\{
C_M\left(G_M^{\gamma p} -G_M^{\gamma n}\right)^2 +C_E\left(G_E^{\gamma p} -
G_E^{\gamma n}\right)^2 \right.
\nonumber\\
& &\left. \mbox{} 
+ C_AF_A^2 - C_{AM} F_A \left(G_M^{\gamma p} -G_M^{\gamma n}\right)\right\}
\label{ccnubar}
\end{eqnarray}
where $|V_{ud}|^2$ is the square of the Cabibbo-Kobayashi-Maskawa matrix
element mixing up and down quarks.  The $Q^2$ dependence of the form factors
has again been suppressed in Eqs.~(\ref{ccnu}) and (\ref{ccnubar}).
It is necessary to check whether the equivalent free scattering cross section
obtained by E734 for quasielastic knockout is appropriate to use to extract 
the relative strengths of the strange magnetic form factors from the E734 
charged current data.
Alberico {\it et al}.\ showed that while including final-state
interactions between the ejected nucleon and the residual nucleus suggests
that only half of the $^{12}$C reactions correspond to elastic proton knockout,
the ratios, Eqs.~(\ref{ratel})-(\ref{ratnub}) are affected at the percent level
for neutrino energies greater than 1 GeV \cite{Alber}.  Since E734 reported
ratios, Eqs.~(\ref{ratel})-(\ref{ratnub}), only elastic scattering of 
neutrinos on free protons were considered.

Integrating the cross sections over $Q^2$, it is possible to compare these
results with the ratios reported by E734.  The neutral current ratio $R$, 
Eq.~(\ref{ratel}), is formed by the ratio of Eq.~(\ref{ncnubar}) to 
Eq.~(\ref{ncnu}).  The charged current ratios $R_\nu$, Eq.~(\ref{ratnu}), and
$R_{\overline \nu}$, Eq.~(\ref{ratnub}), are the ratios of Eq.~(\ref{ncnu}) to
Eq.~(\ref{ccnu}) and Eq.~(\ref{ncnubar}) to Eq.~(\ref{ccnubar}) respectively.

By forming the asymmetry of neutral current
to charged current interactions, defined as the ratio of the difference between
Eqs.~(\ref{ncnu}) and (\ref{ncnubar}) to the difference between
Eqs.~(\ref{ccnu}) and (\ref{ccnubar}), 
\begin{eqnarray}
{\cal A}(Q^2) = \frac{ (d\sigma_{\nu p\rightarrow \nu p}/dQ^2) -
(d\sigma_{{\overline\nu} p\rightarrow {\overline\nu} p}/dQ^2) }
{(d\sigma_{\nu n\rightarrow \mu^- p}/dQ^2) -
(d\sigma_{{\overline\nu} p\rightarrow \mu^+ n}/dQ^2) } \, \, ,
\label{asymmdif}
\end{eqnarray}
model independent information on the
axial, $F_A^s$, and magnetic, $G_M^s$, strange form factors of the nucleon
can be obtained \cite{Alber}.
The differences in the cross sections can be written in terms of the form 
factors so that the asymmetry becomes
\begin{eqnarray}
{\cal A}(Q^2) = \frac{1}{4|V_{ud}|^2}
\left(1-\frac{F_A^s}{F_A}\right)
\left(1-2\sin^2\theta_W
\frac{G_M^{\gamma p}}{G_M^3} -\frac{G_M^s}{2G_M^3}\right)\, ,
\label{asymff}
\end{eqnarray}
where $F_A$ is the charged current axial form factor and $G_M^3=
0.5(G_M^{\gamma p}- G_M^{\gamma n})$ is
the charged current isovector magnetic form factor of the nucleon.
Note that ${\cal A}$ is independent of the electric form factors.

To compare the calculations to the neutrino energy and momentum transfer
averaged BNL E734 ratios, in  Eqs.~(\ref{ratel})-(\ref{ratnub}), they 
integrated over the neutrino energy spectrum, Eq.~(\ref{dsdq2nu}), and $Q^2$
to obtain the integral asymmetry
\begin{eqnarray}
\langle {\cal A} \rangle & = & \frac{
\langle \sigma_{\nu p\rightarrow
\nu p} \rangle - \langle \sigma_{
\overline{\nu} p \rightarrow \overline{\nu} p} \rangle }
{\langle \sigma_{\nu n\rightarrow
\mu^- p} \rangle - \langle \sigma_{
\overline{\nu} p \rightarrow \mu^+ n} \rangle }
\label{asymm} \\
    & = &
\frac{R_\nu(1-R)}{1-RR_\nu/R_{\overline{\nu}}} 
\label{asym2} \\
& = & 0.136 \pm 0.008 \pm 0.019 
\label{asymexp}
\end{eqnarray}
from the data, Eqs.~(\ref{ratel})-(\ref{ratnub}).

Dipole parameterizations of the non-strange and strange form
factors were assumed.  The strange axial and vector form factors are
\cite{Alber} 
\be
F_A^s(Q^2) & = &  \frac{g_A^s M_A^4}{(M_A^2 + Q^2)^2} \label{saxial} \\
G_M^s(Q^2) & = &  \frac{\mu_s M_V^4}{(M_V^2 + Q^2)^2} \label{smagnet} \\
G_E^s(Q^2) & = & \frac{\rho_s Q^2}{4m_p^2} \frac{M_V^4}{(M_V^2 + Q^2)^2} 
\label{select}
\ee
where the strange couplings $g_A^s$, $\mu_s$ and $\rho_s$ are left 
as free parameters.  The vector and axial cutoff masses are $M_V = 0.84$ GeV
and $M_A = 1.032$ GeV respectively.
Other $Q^2$ parameterizations of the form factors stronger
than the dipole would reduce the overall effect of strangeness at high $Q^2$
but in the range $0.5 < Q^2 < 1$ GeV$^2$ changing the $Q^2$ dependence has a
very weak effect on the results.

\begin{figure}[htb]
\setlength{\epsfxsize=\textwidth}
\setlength{\epsfysize=0.5\textheight}
\centerline{\epsffile{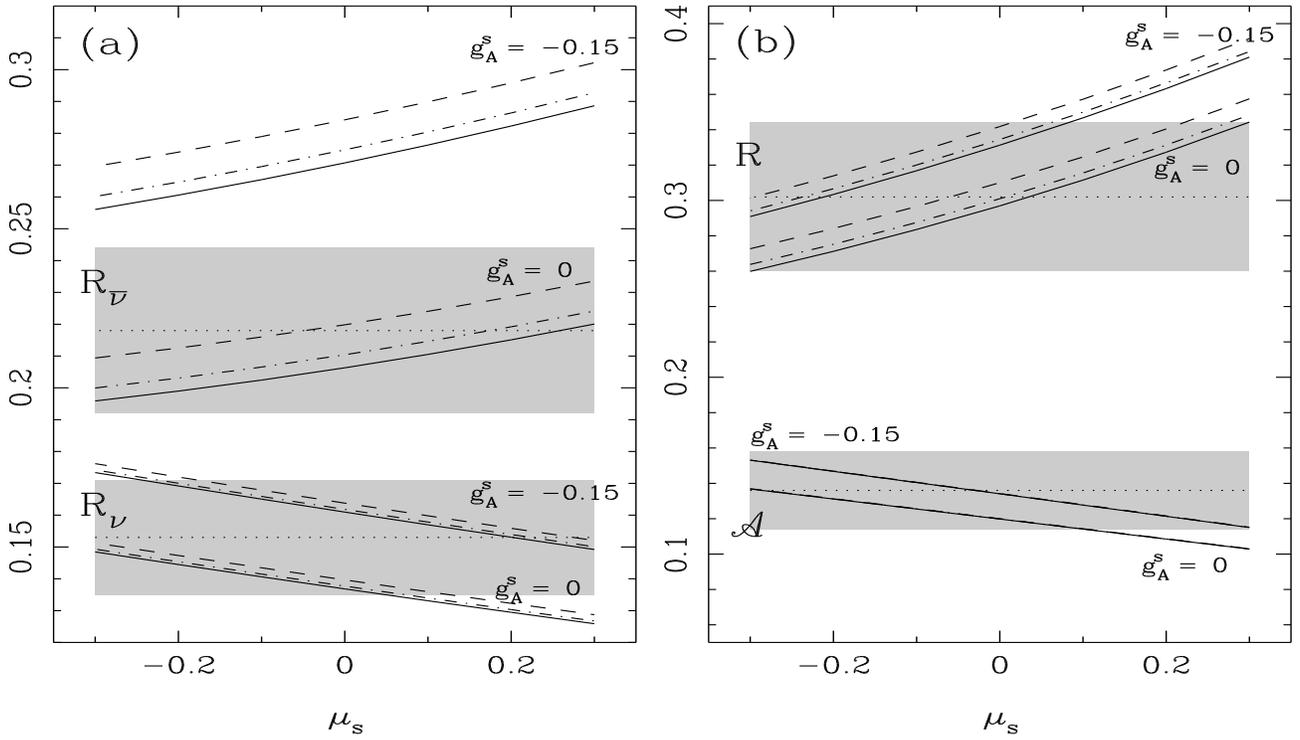}}
\caption[]{The ratios $R_\nu$, $R_{\overline \nu}$ (a), and $R$ and the
asymmetry ${\cal A}$
(b) are calculated as a function of $\mu_s$ and compared to the E734 data 
(shaded areas).  All the calculations correspond to
elastic scattering.  Two choices of the axial coupling $g_A^s$ are shown:
0 and $-0.15$.  For each value of $g_A^s$, three values of $\rho_s$ are chosen:
$-2$ (dot-dashed), 0 (solid), and 2 (dashed).  Reproduced from 
Ref.~\cite{Alber} with permission from Elsevier Science.
}
\label{alber2}
\end{figure}

The ratios $R_\nu$, $R_{\overline \nu}$, and $R$ and the asymmetry ${\cal A}$
are shown in  Fig.~\ref{alber2} along with several choices of $g_A^s$,
$\mu_s$, and $\rho_s$.  
Note that ${\cal A}$ is independent of $\rho_s$.  The
ratios and asymmetry are given as a function of $\mu_s$ for $g_A^s = 0$ and 
$-0.15$ with $\rho_s = -2$, 0, and 2.  Changing $g_A^s$ 
influences the ratios more than does changing $\rho_s$, especially for 
$R_{\overline \nu}$.  In fact, the sensitivity of $R_{\overline \nu}$ to
$g_A^s$ excludes $g_A^s \leq -0.15$.  The calculations of $R_\nu$ and 
${\cal A}$ are most consistent with the E734 data when $g_A^s$ is close to 
zero and $\mu_s$ is negative.  Varying the axial
cutoff $M_A$ by $\pm 3.5$\% only
has a significant effect on $R_{\overline \nu}$.  Lowering $M_A$ increases
the allowed range of $g_A^s$ if $\mu_s = \rho_s = 0$ while increasing $M_A$
restricts $g_A^s$ to values close to zero.  

Although the results obtained by 
the SAMPLE collaboration, Eq.~(\ref{SAMPLEres}), suggest that $\mu_s$ should 
be positive, the large uncertainties cannot rule out $\mu_s \leq 0$.  Thus
so far the E734, SAMPLE, and HAPPEX results are consistent with each other.  As
pointed out in Ref.~\cite{Alber}, if the parity-odd asymmetry measurements
are improved, then the more stringent bounds on $\mu_s$ can place better
constraints on $g_A^s$ and $\rho_s$.  The strange form factors have also been
computed in a light-cone approach \cite{MelMalsff} and compared to the HAPPEX
data.  In this case, Melnitchouk and Malheiro calculated 
the combination of $G_E^s$ and $G_M^s$ in
Eq.~(\ref{gesgmscombo}) and found that the strange electric and magnetic form
factors were small and positive, consistent with the experiments. 

\subsubsection{Strange sea summary}

The difference between the $s$
and $\overline s$ distributions determined by the CCFR collaboration
\cite{CCFRNLO} has been studied in the context of meson cloud models
\cite{pnndb,cm,bma,MelMalmcm} 
and has been modeled by gluon splitting into $s$ and $\overline
s$ quarks with different effective masses \cite{jitang}.  Meson cloud models 
tend to produce larger differences between $s(x)$ and $\overline s(x)$ than can
be accommodated by the data, leading Ji and Tang to suggest that the 
strange sea
is highly localized \cite{jitang}.  However, further measurements are needed
with analyses based on more recent parton distribution functions.  A reanalysis
of the neutrino data is underway which could provide some answers \cite{Bodek}.

Information on strange form factors in the nucleon has been obtained by the
SAMPLE \cite{SAMPLE}, HAPPEX \cite{HAPPEX}, and E734 \cite{Ahrens87} 
collaborations.  The results
are consistent with each other and show that the strange form factors
could be small and positive but are also consistent with zero \cite{Alber}.

\subsection{Heavy quark contributions to the sea}

\begin{figure}[htb]
\setlength{\epsfxsize=\textwidth}
\setlength{\epsfysize=0.3\textheight}
\centerline{\epsffile{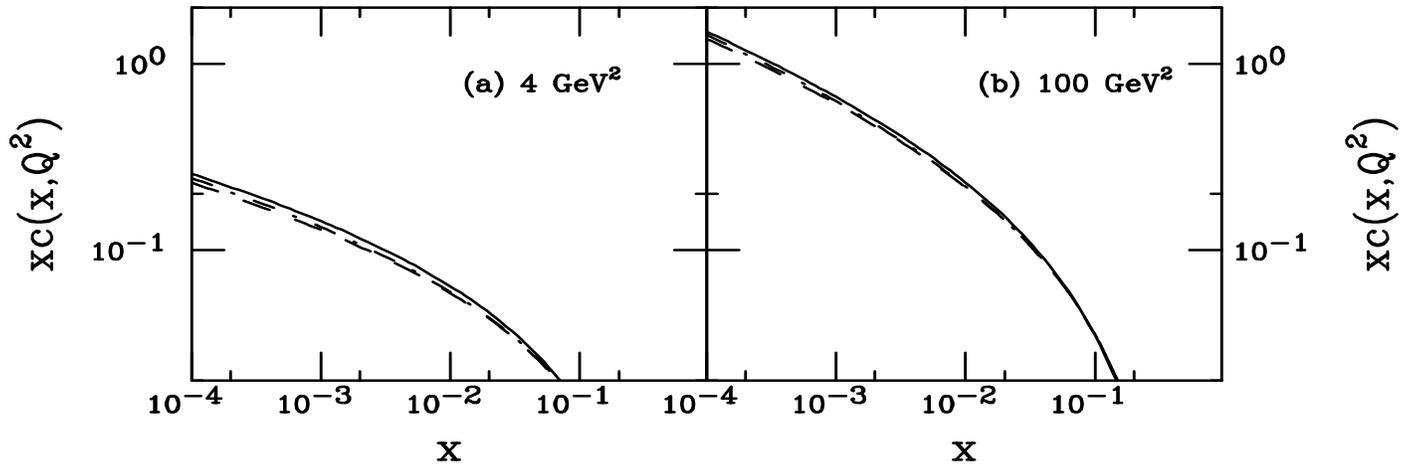}}
\caption[]{Charm quark parton distribution in the Variable Flavor Number 
scheme,
CTEQ 5M (solid), the Fixed Flavor Number scheme, CTEQ 5FF (dashed), and the 
interpolating scheme of Ref.~\cite{alvazis}, CTEQ 5HQ (dot-dashed).}
\label{cteqcharm}
\end{figure}

Quarks more massive than the strange quark, {\it e.g.}\ charm and bottom
quarks, 
have also been included in global analysis of parton distribution functions.
(See Ref.~\cite{cteq5} for a recent discussion and further references.)  Early 
treatments of heavy quark parton distributions assumed that below a scale
$Q^2 \sim m_Q^2$, there are $n_f$ flavors in the perturbative sea and in the
calculation of $\alpha_s$.  When $Q^2 > m_Q^2$, $n_f + 1$ massless flavors
are included in the perturbative sea and the running coupling constant. 
Thus above $Q^2 = m_Q^2$, the heavy quark is treated as massless 
\cite{mrst}.  This
scheme is now generally referred to as the Variable Flavor Number scheme.
At the opposite extreme, the heavy quark is never treated as part of the 
nucleon sea but is produced perturbatively through photon-gluon fusion.  The
number of flavors remains fixed, regardless of $Q^2$.  This treatment, 
used in the GRV 94 parton densities \cite{grv94}, is 
referred to as the Fixed Flavor Number scheme.  Obviously neither of these
schemes is correct at all $Q^2$.  The Fixed Flavor Number scheme is good near 
threshold but cannot incorporate large logarithms at $Q^2 \gg m_Q^2$ while
the heavy quark should not be treated as massless near $Q^2 \sim m_Q^2$, as in
the Variable Flavor Number
scheme.  Interpolating schemes, such as the one used by
CTEQ \cite{cteq5,alvazis}, have been introduced which reproduce relevant
features of both schemes.  The heavy quarks are essentially as produced by
photon-gluon fusion when $Q^2 \sim m_Q^2$ and behave as massless quarks when
$Q^2 \gg m_Q^2$.  All three schemes have been included in the recent
global analysis by the CTEQ collaboration.  The charm quark distribution from
three of their fits is shown in Fig.~\ref{cteqcharm}.

Charm DIS data are discussed first.  The EMC large $x$ charm data \cite{emcc}
implied some nonperturbative source of charm production.  These data have not 
been followed up with subsequent large $x$ charm measurements.
More recent HERA data \cite{zeuscharm,h1charm} suggest that
only perturbative production is needed at low $x$.  Models of nonperturbative
charm production are then introduced which could be important at large $x$.

\subsubsection{Current experimental evidence}

\begin{figure}[htb]
\setlength{\epsfxsize=\textwidth}
\setlength{\epsfysize=0.5\textheight}
\centerline{\epsffile{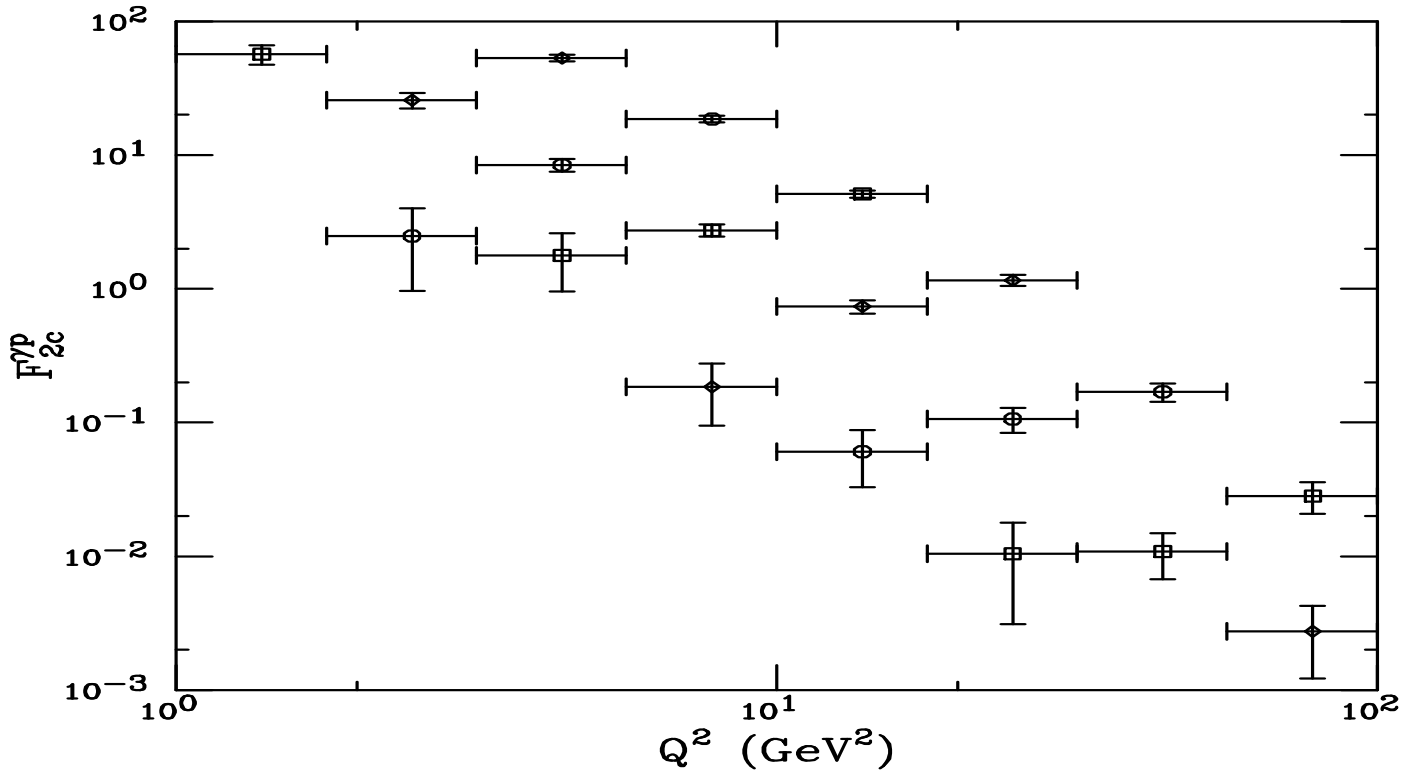}}
\caption[]{The charm structure function $F_{2\, c}^{\gamma p}$ as a function of
$Q^2$ measured by the European Muon Collaboration \protect\cite{emcc}. 
From top to bottom the values of
the $x$ bins are:  0.0044, 0.0078, 0.014, 0.025, 0.044, 0.078, 0.14, 0.25,
and 0.44. To separate the data, $F_{2\, c}^{\gamma p}$ in each $x$ bin is
scaled by a factor of 4 from the next higher $x$ bin.  Therefore, only the
highest $x$ data (lowest points) have the correct scale.  Only the statistical
uncertainty is shown.}
\label{emccharm}
\end{figure}

The European Muon Collaboration, EMC, \cite{emcc} at CERN first measured the
charm structure function, $F_{2 \, c}^{\gamma p}$.
In the relatively large $x$ region of the EMC measurement, charm only 
contributed $\approx 1$\% to the total structure function $F_2^{\gamma p}$.
Charm was identified by dimuon events, where one muon was consistent with charm
meson decay, and trimuon events, which were consistent with a $\mu^+ \mu^-$
pair produced by the semileptonic decays of a pair of charm mesons.  
In both cases, the additional final-state muon was the scattered projectile
muon.  They
compared their results to a leading order photon-gluon fusion calculation.  In
their analysis, they assumed a simple gluon momentum distribution, $xg(x) =
3(1-x)^5$ and used the $c \overline c$ pair mass in the scale of the strong
coupling constant, $\mu^2 = Q^2 + m_{c \overline c}^2$.  They included $\sim
83$\% of the $c \overline c$ cross section below the $D \overline D$ threshold
in the total charm cross section.  
Their data, which has never been incorporated into global analyses of parton 
distributions, are shown in Fig.~\ref{emccharm} as a function of $Q^2$
with uncertainties on both $x$ and $Q^2$ given.  A nonnegligible $Q^2$
dependence is clearly indicated.

\begin{figure}[htb]
\setlength{\epsfxsize=\textwidth}
\setlength{\epsfysize=0.5\textheight}
\centerline{\epsffile{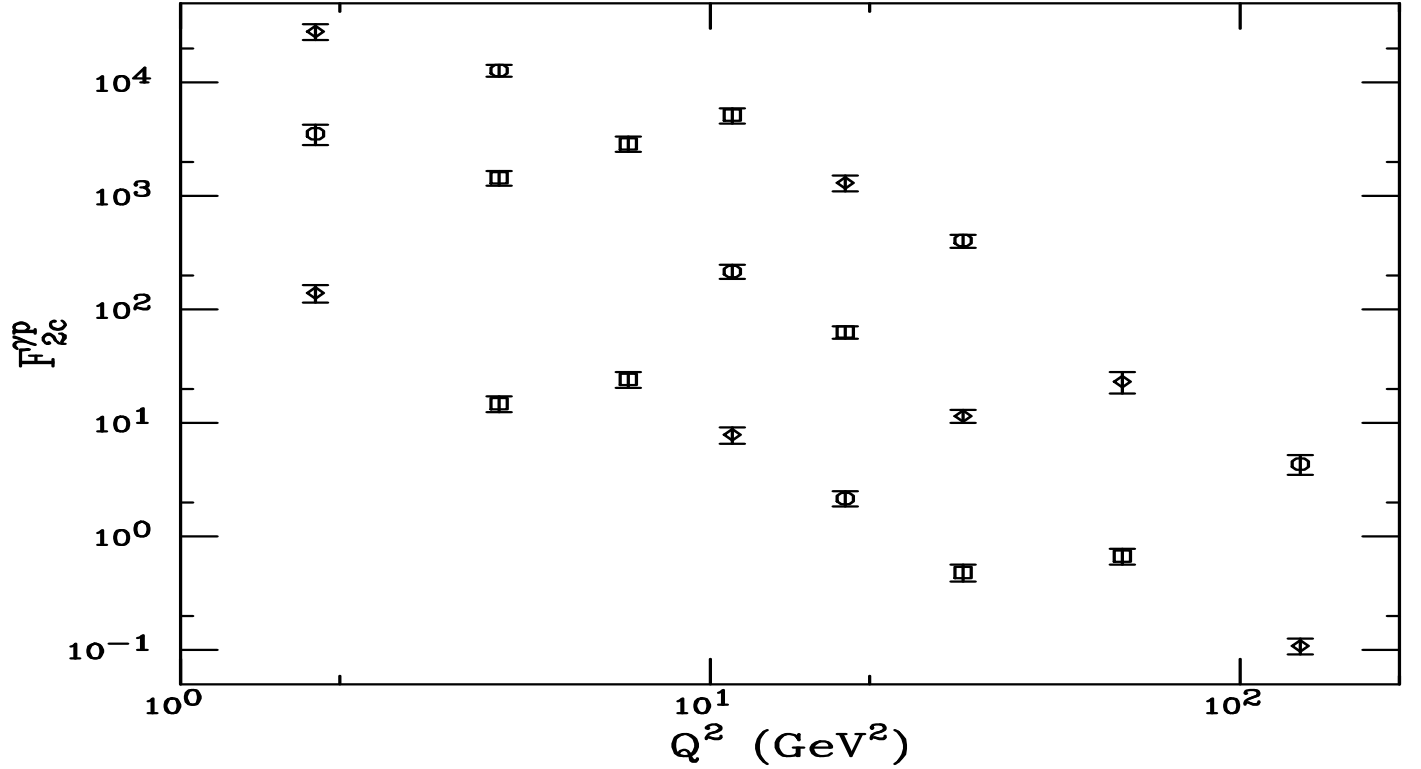}}
\caption[]{The charm structure function $F_{2\, c}^{\gamma p}$ as a function of
$Q^2$ measured by ZEUS \protect\cite{zeuscharm}. 
From top to bottom the values of
the $x$ bins are:  0.00005, 0.00013, 0.0003, 0.0005, 0.0008, 0.0012, 0.002,
0.004, 0.008, and 0.02. 
To separate the data, $F_{2\, c}^{\gamma p}$ in each $x$ bin is
scaled by a factor of 4 from the next higher $x$ bin.  Therefore, only the
highest $x$ data (lowest points) have the correct scale.  Only the statistical
uncertainty is shown.}
\label{heracharm}
\end{figure}

More recently, the charm structure function has been measured to very 
low $x$ at HERA \cite{zeuscharm,h1charm}.  
At these low $x$ values, $F_{2 \, c}^{\gamma p}/
F_2^{\gamma p} \sim 0.25$ at $Q^2 \sim 11$ GeV$^2$, 
a surprisingly large fraction.  The charm structure function is measured in the
reactions $e^+ p \rightarrow e^+ D^{* \pm} X$ by studying the decay channels
$D^{* +}(2010) \rightarrow D^0(1864) \pi_s^+ + \, {\rm charge \, conjugate}$ 
where
$\pi_s^+$ is a slow $\pi^+$.  The $D^0$ and $\overline D^0$ from $D^{* \pm}$
decays subsequently decay into $K (n \pi)$ final states,
$D^0 \rightarrow K^- \pi^+$ or $K^- \pi^+ \pi^- \pi^+$ ($+$ charge
conjugates) resulting in $D^{* \pm}$ decays with two or four pions in the final
state.  Once the final states have been reconstructed, the $c \overline c$
cross section is obtained from the $D^{* \pm}$ production cross section,
extrapolating to the full kinematical phase space, and employing the
hadronization fraction $F(c \rightarrow D^{* +})$ 
to obtain the total charm cross
section.  The extrapolations necessary to go to
full phase space were significant, resulting in a factor of ten correction
for the four-pion final state at low $Q^2$ and a factor of
four at high $Q^2$, and
neglect charm production by processes other than photon-gluon fusion.  The $c
\overline c$ production cross section is then
\begin{eqnarray}
\frac{d^2 \sigma^{c \overline c}(x,Q^2)}{dx dQ^2} = \frac{2 \pi \alpha^2}{xQ^2}
\left\{ \left[1 + (1-y)^2 \right] F_{2 c}^{\gamma p}(x,Q^2) - y^2 F_{L
c}^{\gamma p}(x,Q^2) \right\} \, \, ,
\label{sigccb}
\end{eqnarray}
where the contribution from the second term is negligible.
Equation~(\ref{sigccb}) can be obtained from Eq.~(\ref{dissigpart}) for the
virtual photon by changing integration
variables from $y$ to $Q^2$ and taking the $m_h/E
\ll 1$ limit.  The hadronization fraction of charm to $D^*$, 
assuming that final
state production of charm hadrons is the same in $e^+ e^-$ annihilation and
DIS, is $F(c \rightarrow D^*) = 0.222 \pm 0.014 \pm 0.014$ \cite{OPAL}.  They
assume $m_c = 1.4$ GeV and the renormalization and factorization  scales used
in the photon-gluon fusion production of charm at NLO are $\mu = \sqrt{4m_c^2 +
Q^2}$.  Recall that the EMC data were analyzed with $\mu = \sqrt{m^2_{c 
\overline c} + Q^2}$.  The
data from the two different decay modes are combined in the final data set and
the charm structure function is extracted from Eq.~(\ref{sigccb}).
The ZEUS data are shown as a function of $Q^2$
in Fig.~\ref{heracharm}.  The structure function shows significant $Q^2$
dependence as well as a steep rise as $x$ decreases with $Q^2$ fixed.
The ZEUS data and the low $x$ EMC data
are consistent with each other and with perturbative generation of the charm
sea.  

Since H1 has a smaller charm sample, they only present the ratio
$F_{2 \, c}^{\gamma p}/F_2^{\gamma p}$ averaged over the kinematic range
$10 < Q^2 < 100$ GeV$^2$ and $0.0008 < x < 0.008$.  They found 
$F_{2 \, c}^{\gamma p}/F_2^{\gamma p} = 0.237 \pm 0.021
\;_{-\; 0.039}^{+\; 0.043}$, an order of magnitude greater than at larger $x$.
Their result is compatible with the ZEUS result in the same kinematic interval.

\subsubsection{Nonperturbative heavy quarks}

The inclusion of heavy quarks in the perturbative generation of the sea 
and global analyses of the parton distributions has
already been discussed.  However, the large $x$ and $Q^2$ EMC charm structure
function data hinted that these results were inconsistent with production of 
charm by photon-gluon fusion.  This led to the speculation that charm could be
generated in the sea by a higher-twist mechanism.  Two nonperturbative models
of charm in the sea are discussed here, intrinsic charm \cite{intc} and
the meson cloud model \cite{pnndb,SMT}.

\subsubsection{Intrinsic charm}

The QCD wavefunction of a hadron can be represented as a
superposition of quark and gluon Fock states. For example, at fixed
light-cone time, a hadron wavefunction can be
expanded as a sum over the complete basis of free quark and gluon
states: $\vert \Psi_h \rangle = \sum_m \vert m \rangle \,
\psi_{m/h}(x_i, k_{T,i})$ where the color-singlet
states, $\vert m \rangle$, represent the fluctuations in the hadron
wavefunction with the Fock components $\vert
q_1 q_2 q_3 \rangle$, $\vert q_1 q_2 q_3 g \rangle$, $\vert
q_1 q_2 q_3 c \overline c \rangle$, {\it etc}. The boost-invariant
light-cone wavefunctions, $\psi_{m/h}(x_i, k_{T,i})$
are functions of the
relative momentum coordinates $x_i = k_i^+/P^+$ and $k_{T,i}$.
Momentum conservation demands $\sum_{i=1}^n x_i = 1$ and
$\sum_{i=1}^n \vec{k}_{T,i}=0$ where $n$ is the number of partons
in state $\vert m \rangle$.
For example, intrinsic charm, IC,
fluctuations \cite{intc} can be liberated by a soft interaction which breaks
the coherence of the Fock state \cite{BHMT}
provided the system is probed during
the characteristic time that such fluctuations exist.

Microscopically, the intrinsic heavy quark Fock component in the
proton wavefunction, $|u u d c \overline c \rangle$, is
generated by virtual interactions such as $g g \rightarrow Q
\overline Q$ where the gluons couple to two or more
valence quarks. The probability for $c \overline c$ fluctuations to
exist in a hadron is higher twist since it scales as $1/m_c^2$
relative to the extrinsic, EC,
leading-twist production by photon-gluon fusion \cite{VB}.

The dominant Fock state configurations are not far off
shell and thus have minimal invariant mass, $M^2 = \sum_i^n \widehat{m}_i^2/
x_i$ where  $\widehat{m}_i^2 = m_i^2 + \langle
\vec k_{T, i}^2 \rangle$ is the square of the average transverse mass
of parton $i$.
The general form of the Fock state wavefunction appropriate to any frame at
fixed light-cone time is \be
\Psi(x_i, \vec k_{\perp i}) = \frac{\Gamma(x_i, \vec k_{\perp i}) }{m_h^2 -
M^2 } \, \,  \ee where $\Gamma$ is a
vertex function, expected to be a slowly-varying,
decreasing function of $m_h^2 - M^2$.
The particle distributions are then
controlled by the light-cone energy denominator and  phase space.
This form for the higher Fock components is applicable to an
arbitrary number of light and heavy partons.
Intrinsic $c \overline c$ Fock components with minimum invariant
mass correspond to configurations with equal rapidity constituents.
Thus, unlike extrinsic heavy quarks generated from a single parton, intrinsic
heavy quarks carry a larger fraction of the parent momentum
than the light quarks in the state \cite{intc}.

The parton distributions
reflect the underlying shape of the Fock state wavefunction.
Assuming it is sufficient to
use $\langle k_T^2 \rangle$ for the transverse momentum,
the probability distribution as a function of $x$ in 
a general $n$--particle intrinsic
$c \overline c$ Fock state is
\be
\label{icprobtot}
\frac{dP_{\rm IC}}{dx_i \cdots dx_n} =  N_n 
%\alpha_s^4(M_{c \overline c})
\ \frac{\delta(1-\sum_{i=1}^n x_i)}{(m_h^2 - \sum_{i=1}^n
(\widehat{m}_i^2/x_i)
)^2} \, \, ,
\ee
where $N_n$ normalizes the $n$-particle Fock state probability.  

The intrinsic charm structure
function at leading order, LO, and next-to-leading order, NLO, 
is now described.  At LO in the
heavy quark limit, $\widehat{m}_c$, $\widehat{m}_{\overline c} \gg m_h$,
$\widehat{m}_q$,
\be
\frac{dP_{\rm IC}}{dx_i \cdots dx_n} & = & N_n 
%\alpha_s^4(M_{c \overline c}) 
\frac{x_c x_{\overline c}}{(x_c + x_{\overline c})^2}
\ \delta(1-\sum_{i=1}^n x_i) \nonumber \\
F_{2 \, c}^{\rm IC \, LO}(x) & = & 
\frac{8}{9} xc(x) = \frac{8}{9}x \int dx_1 \cdots
dx_{\overline c} \frac{dP_{\rm IC}}{dx_i \cdots dx_{\overline c}
dx_c} \nonumber \\
& = & \frac{8}{9} x \frac{1}{2} N_5 x^2 [\frac{1}{3}(1-x)(1 +
10x + x^2) + 2x(1+x)\ln x] \, \, .
\label{massless}  \ee
If $n=5$, the minimal intrinsic charm Fock
state of the proton, $|uudc \overline c \rangle$, is assumed and
$P_{\rm IC} = 1$\% \cite{intc}, then $N_5 = 36$.  
In what follows, a 1\% intrinsic
charm contribution is assumed unless otherwise noted.

Hoffmann and Moore \cite{hm} incorporated mass effects into the
analysis.  They first included threshold effects by introducing a scaling
variable $\xi = 2ax/ [1 + (1 + 4 x^2m_p^2/Q^2)^{1/2}]$ with $a = [1+ (1 
+ 4m_c^2/Q^2)^{1/2}]/2$.   The
$c \overline c$ mass threshold requires $\xi \leq \gamma < 1$ where
$\gamma = 2a\hat x [1 + (1 + 4 \hat x^2 m_p^2/Q^2)^{1/2}]^{-1}$ and 
$\hat x = Q^2/[Q^2 + 4
m_c^2 - m_p^2]$.
Then Eq.~(\ref{massless}) is replaced by
\be \label{xiscale}
F_{2 \, c}^{\rm IC \, LO}(x,Q^2,m_c^2) = \frac{8}{9} \xi c(\xi,\gamma) \, ,
\ee
with $c(z,\gamma)=c(z)-zc(\gamma)/\gamma$ for $z \leq \gamma$ and zero
otherwise.   The final LO result can be written more generally as in 
Eq.\ (18) of Ref.~\cite{hm}, 
\be
\label{fulllo}
F_{2 \, c}^{\rm IC \, LO}(x,Q^2,m_c^2) = \frac{8x^2 Q^3}{9(Q^2 
+ 4 x^2m_p^2)^{3/2}}
\left[ \frac{(Q^2 +4m_c^2)}{\xi Q^2} c(\xi, \gamma)  
+ \hat g (\xi,\gamma) \right] 
\ee
where
\be
\hat g(\xi,\gamma) = \frac{6xm_p^2}{(Q^2 + 4x^2m_p^2)}
\int_\xi^\gamma\, \frac{dt}{t} c(t,\gamma)
\left( 1 - \frac{m_c^2}{m_p^2 t^2} \right)
\left[1 + \frac{2 x t m_p^2}{Q^2} + \frac{2x m_c^2}{tQ^2}\right] \,. \nonumber
\ee
The NLO IC component of the structure function is given by
\be
\label{nloic}
F_{2 \, c}^{\rm IC \, NLO} 
(x, Q^2, m_c^2) = \frac{8}{9} \xi \int_{\xi / \gamma}^1 \frac{dz}{z}
c(\xi / z, \gamma) \sigma_2^{(1)} (z, \lambda) \,.
\ee
The lowest order cross section is normalized to
$\sigma^{(0)}_2(z,\lambda) = \delta(1-z)$.
The NLO QCD corrections
to the IC contribution are given in Eq.\ (51) of Ref.~\cite{hm}.
The IC results at NLO are the sum of the kinematically
corrected LO result, Eq.~(\ref{fulllo}), and the full NLO correction.\\
\begin{figure}[htb]
\setlength{\epsfxsize=0.95\textwidth}
\setlength{\epsfysize=0.4\textheight}
\centerline{\epsffile{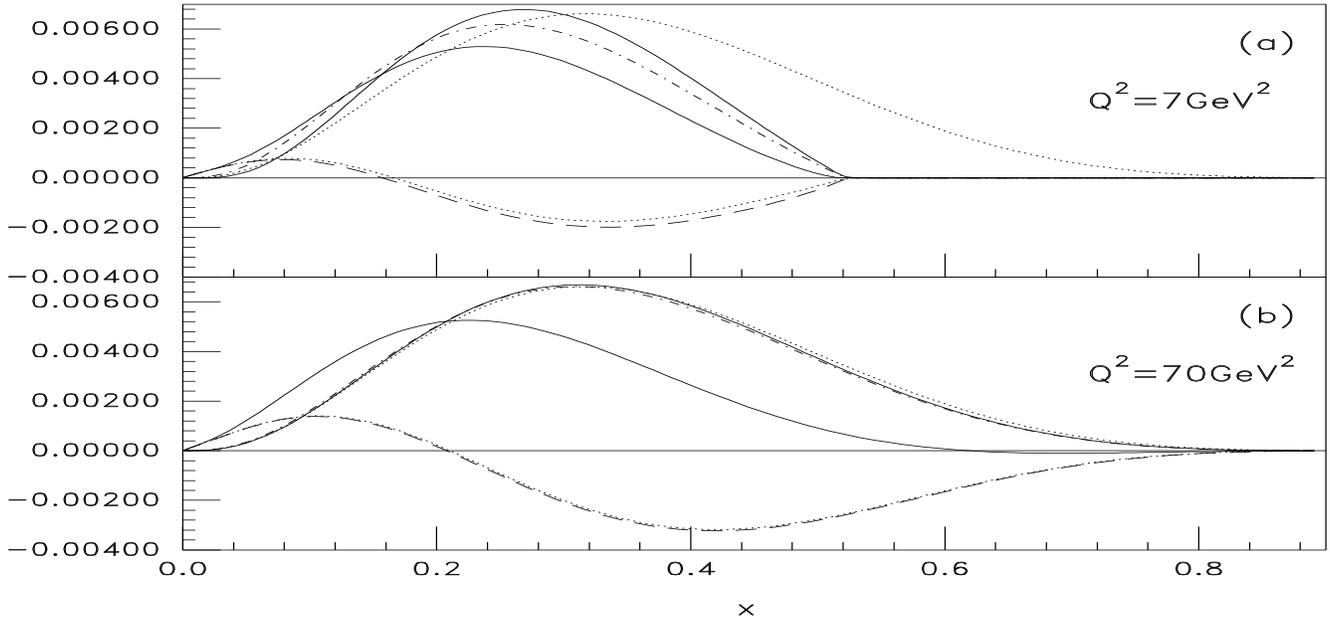}}
\caption[]{(a) The contributions to the structure function
$F_{2 \, c}^{\rm IC}(x,Q^2,m_c^2 )$
at $Q^2 = 7$ GeV$^2$: the massless result, Eq.~(\ref{massless}),
(upper dotted line);
the $\xi$-scaling result, Eq.~(\ref{xiscale}), (dot-dashed line);
and the kinematically corrected formula, Eq.~(\ref{fulllo}), (top solid line).
Also shown are the NLO corrections given by Eq.~(\ref{nloic})
with the leading-log result, Eq.~(54) in \cite{hm}, (dashed line),
and the full result, Eq.~(51) in \cite{hm}, (lower dotted line).
The sum of Eqs.~(\ref{fulllo}) and (\ref{nloic}) using the full result,
Eq.~(51) in \cite{hm},
represents the total IC contribution to $F_{2 \, c}^{\rm IC \, 
NLO}(x,Q^2,m_c^2)$
(lower solid line). (b) Same as part (a) for $Q^2=70$ GeV$^2$. 
Reproduced from Ref.~\cite{hsv} with permission from Elsevier Science.}
\label{iclonlo}
\end{figure}

Figure \ref{iclonlo} shows the LO and NLO IC contributions
to $F_2(x,Q^2,m_c^2)$ for $Q^2 = 7$ and 70 GeV$^2$.  At LO
the massless, $Q^2$ independent, result,
Eq.~(\ref{massless}), the $\xi$ scaling formula, Eq.~(\ref{xiscale}),
and the full calculation, Eq.~(\ref{fulllo}), show
substantial differences at low $Q^2$ while the three are nearly
indistinguishable when $Q^2$ is large.  The
NLO corrections calculated using Eq.~(\ref{nloic}) are also shown.
The leading-logarithmic approximation, Eq.~(54) in \cite{hm},
and the complete result, Eq.~(51) in \cite{hm}, differ at small $Q^2$ but
are almost identical at large $Q^2$.
The total NLO IC results are
actually negative at large $x$ and $Q^2$,
indicating that even higher order terms are needed in the perturbative
expansion.

The EMC data have been analyzed with a combination of photon-gluon fusion and
intrinsic charm, first by Hoffmann and Moore \cite{hm} and later by Harris {\it
et al.} \cite{hsv}.  
In both cases, the intrinsic charm calculation was carried to
next-to-leading order.  While only leading order
photon-gluon fusion was 
available for the first study \cite{hm}, in the more recent analysis, the 
photon-gluon fusion
calculation was carried to next-to-leading order as well \cite{hsv}.
Photon-gluon fusion is used to generate charm perturbatively because at the
measured $Q^2$ of the EMC experiment, the mass of the charm quark is 
not small.

To NLO \cite{LRSvN12}, the extrinsic charm structure function
$F_{2 \, c}^{\gamma p}$ is
calculated from the inclusive virtual photon-induced reaction at the parton
level, $\gamma^\ast  + a_1 \rightarrow  c + \overline c + a_2$
where $a_1$ and $a_2$ are massless partons.  In an inclusive calculation,
the $\overline c$ and $a_2$
momenta are integrated over.
The charm 
structure function is the sum of the longitudinal and transverse virtual
photon total cross sections, $F_{2 \, c}^{\gamma p} \propto 
(Q^2/4 \pi^2 \alpha)(\sigma_T +
\sigma_L)$.  At NLO, the extrinsic charm structure function is then
\begin{eqnarray}
\label{fhad}
\lefteqn{F_{2 c}^{\gamma p}(x,Q^2,m^2_c) = 
\frac{Q^2 \alpha_s(\mu^2)}{4\pi^2 m^2_c} \!\!
\int_{\xi_0}^1 \frac{d\xi}{\xi} \!\! \left[ \,e_c^2 f_g(\xi,\mu^2)
 c^{(0)}_{2,g} \, \right] } \nonumber \\
& & \!\!\!\!\!\!\!\!\!\!\!\! + \, \frac{Q^2 \alpha_s^2(\mu^2)}{\pi m^2_c} \!\!
\int_{\xi_0}^1 \frac{d\xi}{\xi} \!\! \left\{ e_c^2
f_g(\xi,\mu^2) \!\!
\left( c^{(1)}_{2,g} + \overline c^{(1)}_{2,g} \ln \frac{\mu^2}{m^2_c} \right)
\right. \nonumber \\
& &  \!\!\!\!\!\!\!\!\!\!\!\! + \sum_{i=q,\overline q}
f_i (\xi,\mu^2) \!\! \left. \left[ e_c^2
\left( c^{(1)}_{2,i} + \overline c^{(1)}_{2,i} \ln \frac{\mu^2}{m^2_c} \right)
+ e^2_i \, d^{(1)}_{2,i} + e_c\, e_i \, o^{(1)}_{2,i} \,
\Large\right] \Large\right\} \, .
\end{eqnarray}
where $\xi_0 = x(4m^2_c+Q^2)/Q^2$.  The quark charges are in units of $e$.
The parton momentum distributions in the proton are denoted by
$f_i(\xi,\mu^2)$ where $\mu$,
the mass factorization scale, has been set equal to the
renormalization scale in $\alpha_s$.  The scale
independent parton coefficient functions,
$c^{(l)}_{2,i}$, $l=0,1$, $\overline c^{(1)}_{2,i}$,
$d^{(1)}_{2,i}$ and
$o^{(1)}_{2,i}$ originate from the coupling of the virtual photon to the
partons.
The functions
$c^{(l)}_{2,i}$ and $ \overline c^{(1)}_{2,i}$ represent the
virtual photon-charm quark coupling, thus appearing for both charged parton
and gluon-induced reactions.  The virtual photon-light quark
coupling gives rise to $d^{(1)}_{2,i}$ while $o^{(1)}_{2,i}$ is an interference
term proportional to $e_c e_i$.

\begin{figure}[htb]
\setlength{\epsfxsize=0.95\textwidth}
\setlength{\epsfysize=0.4\textheight}
\centerline{\epsffile{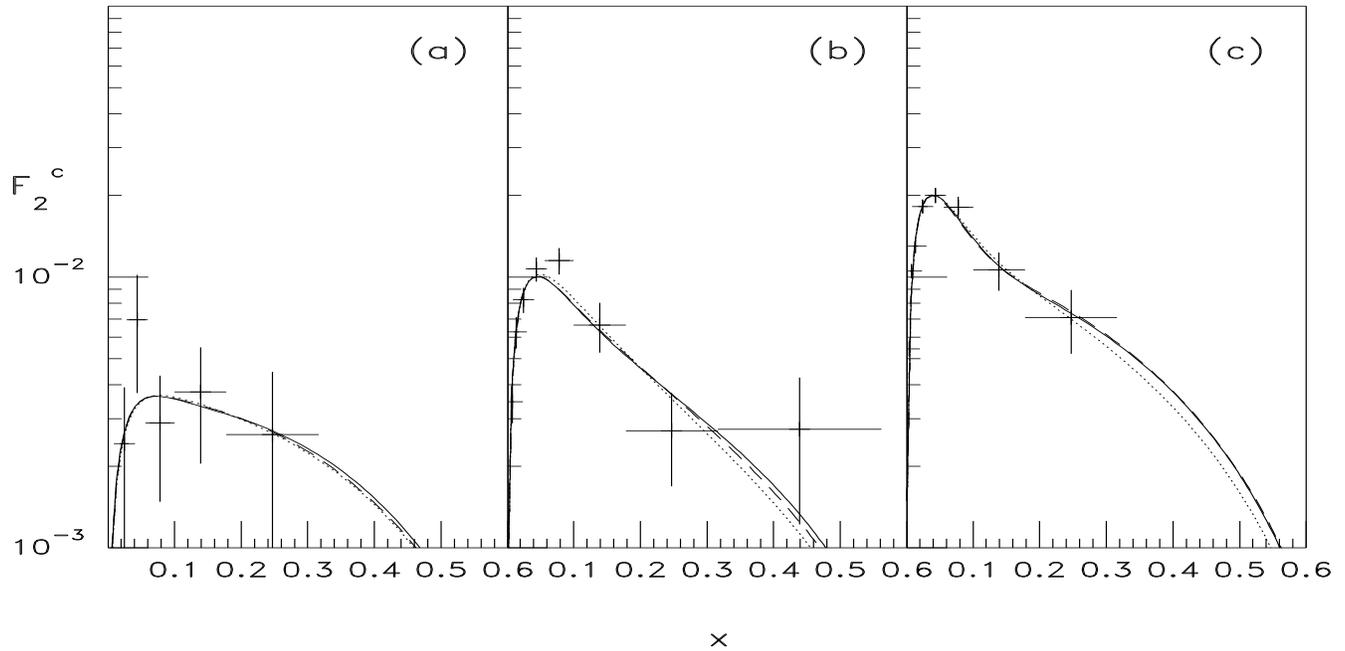}}
\caption[]{(a) The EMC data for the structure function $F_{2\, 
c}(x,Q^2,m_c^2)$ at
$\bar{\nu} = 53$ GeV are plotted as a function of $x$ together with
the fit from Eq.~(\ref{f2ccombo}) with $\epsilon$ and
$\delta$ taken from Table~\ref{ecicfit}. The
CTEQ3 (solid), MRS G (dashed) and GRV 94 HO (dotted) parton densities are
shown. (b) Same as (a) for $\bar{\nu} = 95$ GeV.
(c) Same as (a) for $\bar{\nu} = 168$ GeV. Reproduced from Ref.~\cite{hsv}
with permission from Elsevier Science.}
\label{emcsum}
\end{figure}

The EMC data \cite{pdg} with $\overline{\nu} = \overline{Q^2} / 2m_p 
\overline{x} = 53,\, 95, \,$ and
168 GeV were fit by a sum of the photon-gluon fusion 
and IC components \cite{hsv}.
The normalization
of both the IC and EC components are free parameters so that
\be \label{f2ccombo}
F_{2 \,c}(x,\mu^2,m_c^2) = \epsilon  F_{2 \, c}^{\gamma p}(x,\mu^2,m_c^2)
                   + \delta  F_{2 \, c}^{\rm IC}(x,\mu^2,m_c^2) \, \, ,
\ee
with the scale $\mu = \sqrt{m_{c \overline c}^2 + Q^2}$.  Since $\mu(
m_{c \overline c}^2)$, an exclusive $c \overline c$ production calculation at
NLO \cite{Brian} was carried out to more precisely match the EMC analysis.
The parameter $\epsilon$, typically larger than unity, can
be considered an estimate of the
NNLO contribution to photon-gluon fusion.
Since a 1\% normalization of the IC component is already assumed, the
fitted $\delta$ is the fraction of this normalization.
Several different parton distribution functions were used which produced
essentially the same results.
The results are presented in Table~\ref{ecicfit}.
\begin{table}
\begin{footnotesize}
\begin{center}
\begin{tabular}{ccccccc}
\hline
\multicolumn{1}{c}{} & \multicolumn{2}{c}{$\bar{\nu}=53$ GeV} &
\multicolumn{2}{c}{$\bar{\nu}=95$ GeV} & \multicolumn{2}{c}{$\bar{\nu}=168$
GeV} \\
\hline
PDF & $\epsilon$ & $\delta$ & $\epsilon$ & $\delta$ & $\epsilon$ & $\delta$ \\
\hline \hline
CTEQ3 \cite{cteq3} 
& 1.0 $\pm$ 0.6 & 0.4 $\pm$ 0.6 & 1.2 $\pm$ 0.1 & 0.4 $\pm$ 0.3
& 1.3 $\pm$ 0.1 & 0.9 $\pm$ 0.5 \\
MRS G \cite{MRSG} 
& 1.0 $\pm$ 0.7 & 0.3 $\pm$ 0.6 & 1.4 $\pm$ 0.2 & 0.3 $\pm$ 0.3
& 1.5 $\pm$ 0.1 & 0.8 $\pm$ 0.5 \\
GRV 94 \cite{grv94}
& 1.2 $\pm$ 0.8 & 0.3 $\pm$ 0.6 & 1.4 $\pm$ 0.2 & 0.3 $\pm$ 0.3
& 1.5 $\pm$ 0.1 & 0.9 $\pm$ 0.5 \\
\hline
\end{tabular}
\end{center}
\end{footnotesize}
\caption[]{Results of the least squares fit of the photon-gluon fusion 
and IC contributions to the
EMC data according to Eq.~(\ref{f2ccombo}). Adapted from \cite{hsv}.}
\label{ecicfit}
\end{table}
The errors in the table correspond to a fit at the
$95\%$ confidence level.
The final results for the combined model of Eq.~(\ref{f2ccombo})
are shown in Fig.~\ref{emcsum}
for the CTEQ3, MRS G, and GRV 94 parton densities.
Given the quality of the data, no
statement can be made about the intrinsic charm content when
$\bar{\nu}=53$ and $95 \: {\rm GeV}$. However,
with $\bar{\nu}=168 \: {\rm GeV}$
an intrinsic charm contribution of $(0.86\pm0.60)\%$ is indicated.
These results are consistent with those of the original analysis by Hoffman
and Moore \cite{hm}.\\

Only the intrinsic charm contribution to the charm structure function has been
discussed here.  There are many other applications of intrinsic charm in
charm hadron production.  See {\it e.g.}\ \cite{VB,VBH1,VBH2,VBlam,GutVogt}
for more details.

\subsubsection{Meson cloud models}

\begin{figure}[htb]
\setlength{\epsfxsize=\textwidth}
\setlength{\epsfysize=0.4\textheight}
\centerline{\epsffile{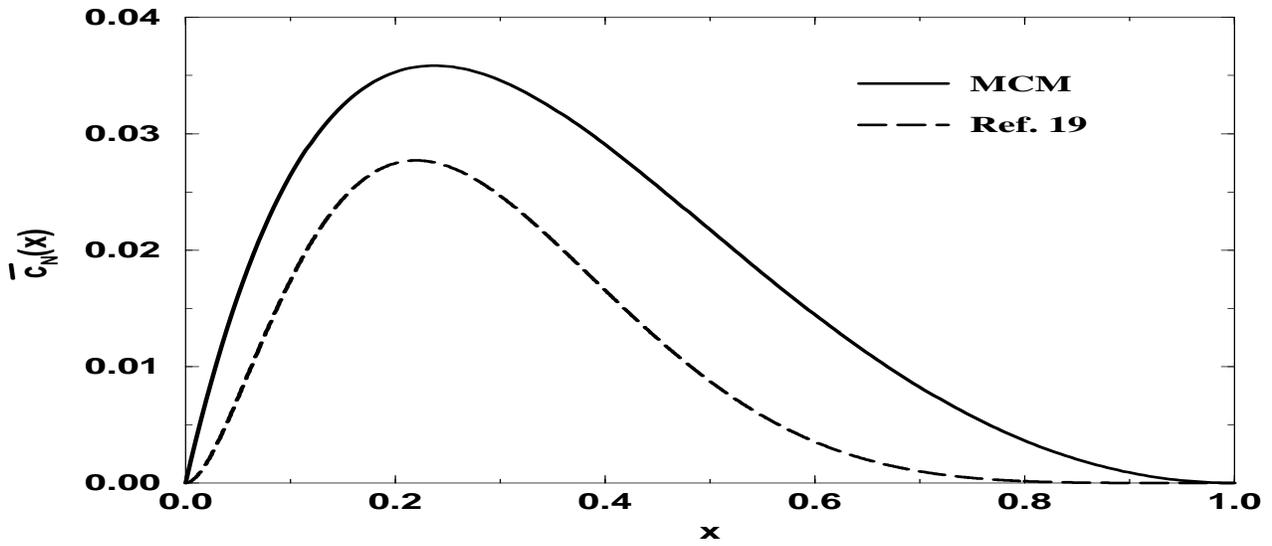}}
\caption[]{The $\overline c_N$ distribution calculated with the meson cloud
model, Eq.~(\ref{cn}), (solid curve) compared to the intrinsic $\overline c_N$
distribution of Eq.~(\ref{massless}) (dashed curve) labeled Ref.~19.  
Reproduced from Ref.~\cite{pnndb} with permission from World Scientific.}
\label{2pnndb}
\end{figure}

Paiva {\it et al.} have also calculated an ``intrinsic'' charm component of the
nucleon sea within the context of the meson cloud model \cite{pnndb}.  They
assumed that the nucleon can fluctuate into $\overline D \Lambda_c$.  The
$\overline c$ distribution in the nucleon is then
\begin{equation}
x {\overline c}_N (x) = \int_{x}^{1}
dy\, f_{\overline D} (y)\, \frac{x}{y} \, {\overline c}_{\overline D} 
(\frac{x}{y})\; .  
\label{cn}
\end{equation}                       
where
\begin{equation}
f_{\overline D} (y) = \frac{g^2_{ \overline D N\Lambda_c}}{16 \pi^2} \, y \, 
\int_{-\infty}^{t_{\rm max}}dt \, \frac{[-t+(m_{\Lambda_c}-m_N)^2]}{[t-
m_{\overline D}^2]^2}\,
F^2 (t)\; ,
\label{fdbar}
\end{equation}
and $t_{\rm max} = m^2_N y- m^2_{\Lambda_c} y/(1-y)$.  In this case they 
chose a monopole form factor with $\Lambda_m = 1.2$ GeV.  The coupling 
constant was assumed to be $g_{\overline D N \Lambda_c} = -3.795$.  From heavy
quark effective theories \cite{man}, the $\overline c$ distribution in the 
$\overline D$ is expected to be hard because in the bound state, the $\overline
c$ exchanges momenta much less than $m_c$.  They make the extreme assumption
that the entire $\overline D$ momenta is carried by the charm quark,
$\overline c_{\overline D} = x \delta(x-y)$.  The resulting $\overline c$
distribution is compared to the intrinsic
charm quark distribution in the heavy quark
limit, Eq.~(\ref{massless}), with $x \overline c_N = xc_N = 9F_{2 \, c}^{\rm IC
\, LO}/8$ in Fig.~\ref{2pnndb}.

More recently Steffens {\it et al.}\ investigated all 
the charm structure function
data with two variants of intrinsic charm \cite{SMT}.  The first was that of 
Eq.~(\ref{massless}), called IC1 in their paper, while the second was a meson
cloud model, IC2.  In the second approach, the $\overline c$ distribution is
obtained from the light cone distribution of $\overline D^0$ mesons in the
nucleon,
\be
\overline c^{\rm IC2}(x) & \approx & f_{\overline D}(x) \label{cbaric2} \\
& = & \frac{1}{16\pi^2}
\int_0^\infty dk_\perp^2 \frac{g^2(x,k_\perp^2)}{x(1-x)(s_{\overline D 
\Lambda_c}-m_N^2)^2} \frac{k_\perp^2 + (m_{\Lambda_c} - (1-x)m_N)^2}{1-x}
\nonumber \, \, .
\ee
A hard charm momentum distribution was assumed in the $\overline D$, similar to
that of Ref.~\cite{pnndb}.
The vertex function $g^2(x,k_\perp^2)$ is parameterized as $g^2 =
 g_0^2(\Lambda^2 + m_N^2)/(\Lambda^2 + s_{\overline D \Lambda_c})$ where
$ s_{\overline D \Lambda_c}$ is the square of the center of mass energy of
the $\overline D \Lambda_c$ system and $g_0^2$ the coupling constant at
$ s_{\overline D \Lambda_c} = m_N^2$.  For an intrinsic charm probability of
1\%, $\Lambda \approx 2.2$ GeV.  The charm distribution is then
\be
c^{\rm IC2}(x) \approx \frac{3}{2} f_{\Lambda_c} \left(\frac{3x}{2} \right)
\label{cic2}
\ee
where the charm distribution in the $\Lambda_c$ is assumed to be $c_{\Lambda_c}
\sim \delta(x - 2/3)$ and $f_{\Lambda_c}(x) = f_{\overline D}(1-x)$.

\begin{figure}
\begin{center}
\begin{picture}(120,150)(90,0)
\epsfig{figure=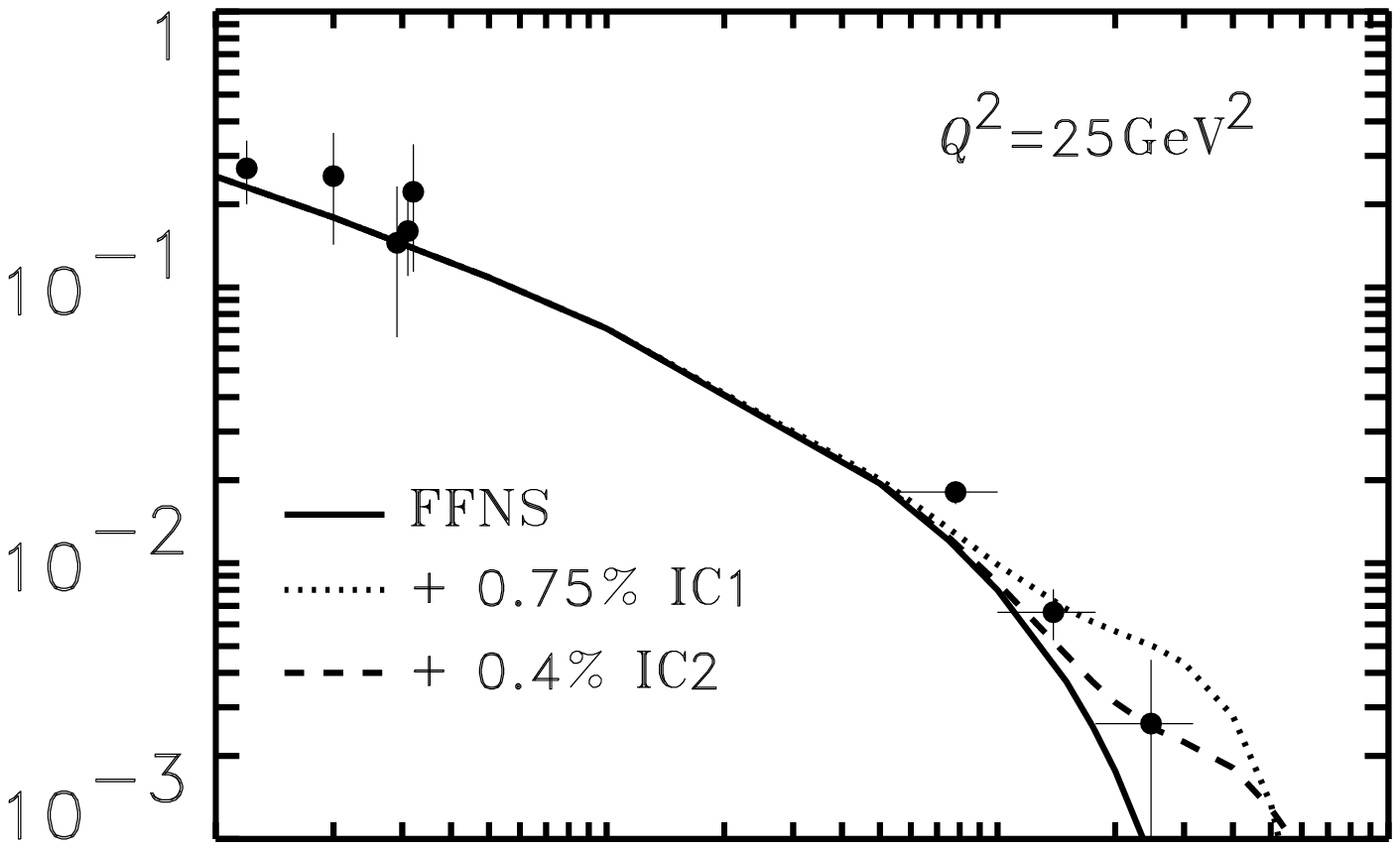,height=7cm}
\put(-279,-135){\epsfig{figure=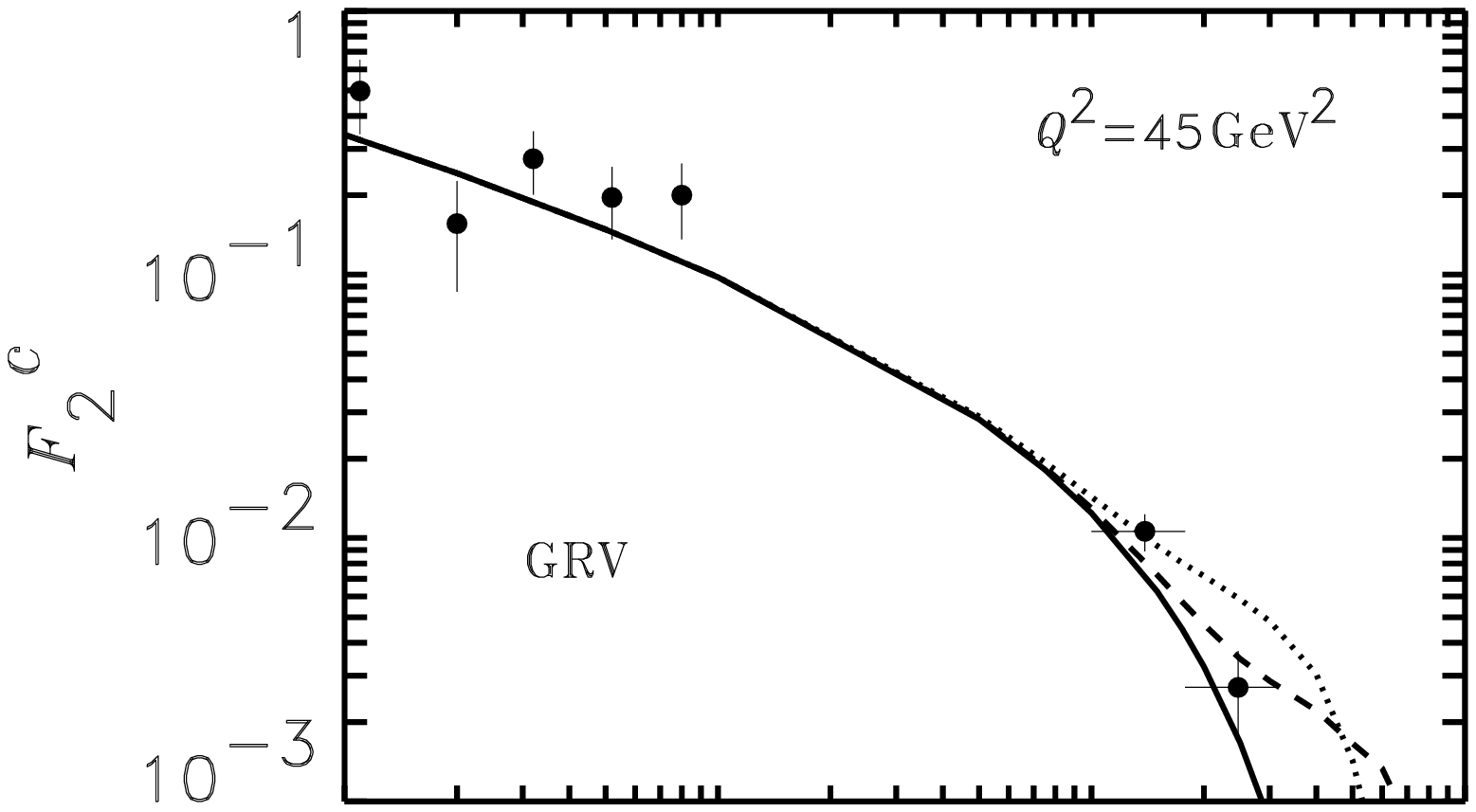,height=7cm}}
\put(-279,-270){\epsfig{figure=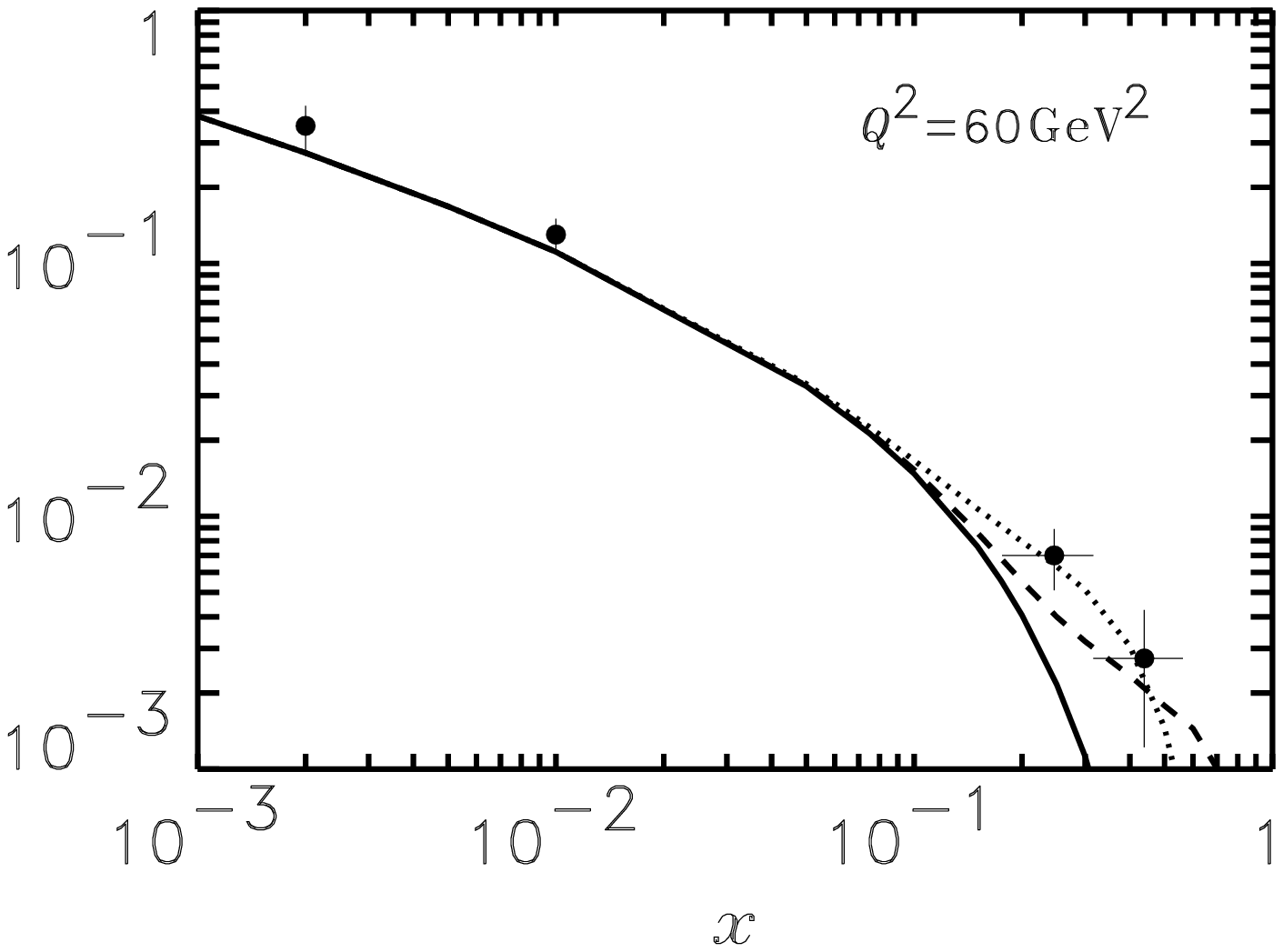,height=7cm}}
\end{picture}
\end{center}
\vspace*{10cm}
\caption[]{The charm structure function calculated in the fixed flavor number
scheme (FFNS) for three $Q^2$ bins
with the intrinsic charm models IC1, Eq.~(\ref{massless}), 
and IC2, Eqs.~(\ref{cbaric2}) and (\ref{cic2}).  The IC1 result
is normalized to 0.75\% while IC2 is normalized to 0.4\%.  The most recent
GRV parton distributions \cite{GRV98}
are used for the light quark and gluon densities.  Reproduced from 
Ref.~\cite{SMT} with permission from Springer-Verlag.}
\label{9smt}
\end{figure}

Using several different schemes to calculate the extrinsic charm sea, 
Steffens {\it et al.}\
compared their two component model to the combined ZEUS and EMC data.  As
expected from the agreement of the low $x$ ZEUS data to the perturbative
generation of the charm sea, the ZEUS data show no sign of intrinsic charm.
However, the EMC data still indicate a need for intrinsic charm at large
$x$ and $Q^2$, as found in Ref.~\cite{hsv}.  The best results, shown in
Fig.~\ref{9smt}, are obtained in the fixed flavor scheme, as outlined at the
beginning of this section.  They find $P_{\rm IC} \sim 0.75$\% for the
standard massless intrinsic charm at leading order, their IC1, and $P_{\rm IC}
\sim 0.4$\% with their meson cloud approach, IC2 \cite{SMT}.  They claim that
the data indicate a slight preference for model IC2.  However, unlike the 
results shown in the previous section, they did not include the NLO corrections
to $F_{2 \, c}^{\rm IC}$ which reduce the large $x$ structure function 
relative to Eq.~(\ref{massless}), as seen in Fig.~\ref{iclonlo}.
Clearly, more data are needed at large $x$ and $Q^2$ to finally resolve the 
situation.

\subsubsection{Heavy sea summary}

Considerable progress has been made in the understanding of the charm component
in the nucleon sea, particularly at small $x$.  Treating the heavy quark as
massless at large $Q^2$ and massive at $Q^2 \sim m_Q^2$ put the heavy quark
contribution to the sea on a firmer theoretical footing.  Charm data at HERA
\cite{zeuscharm,h1charm} are in good agreement with perturbative production of
charm.  However, the large $x$ and $Q^2$ EMC data \cite{emcc} still do not
fit into this scheme.  Intrinsic charm on the level of $\sim 1$\% can explain
this lower energy data \cite{intc,hsv}.  Whether the $c$ and $\overline c$
distributions are identical \cite{intc} or differ as in an {\it e.g.}\ 
$\overline D\Lambda_c$ cloud \cite{SMT} is unclear.  More data in this
interesting region are needed.  

In addition to the reconstructed charm structure function data at HERA, 
deep-inelastic scattering
data with significant statistics at high $Q^2$ and high $x$
are now available~\cite{h1hixQ,zeushixQ} which could also shed some light on 
charm production processes.  
There is an excess in the data beyond the predictions based
on standard QCD parton distribution functions. 
These data have been compared to calculations with intrinsic charm that enhance
the standard model charm cross section \cite{rvgunic,mtic}
in the $x$ and $Q^2$ regions of interest to the HERA excess.  It was
found that intrinsic charm could account for a significant enhancement over
the standard model predictions in charged current scattering.  More data are
needed to confirm the effect and to clarify the role of charm in this region.

A nonperturbative bottom sea component in the nucleon has not been discussed
here but would be a plausible extension of the models.  See Ref.~\cite{VB} for
some predictions of intrinsic bottom effects.

\section{Summary}

There is only room in this review for discussion of a limited subset of all
experimental results and theoretical progress towards understanding the data.
For more in-depth discussion of the work described here as well as other 
results, the reader is encouraged to consult the appropriate references.

As discussed in this review, studies of the nucleon sea have revealed a rich
structure compared to early evaluations of parton distribution functions with 
an SU(3) symmetric sea, {\it e.g.}\ Ref.~\cite{DO}.  Of these studies, the 
Gottfried sum rule 
violation, which showed that $\overline d > \overline u$, is the best
documented experimentally.  Interpretation of the NMC and E866 data as an
indication of a cloud of virtual mesons and baryons surrounding the nucleon
has proved fruitful.

Fewer data are available so far for the more massive components of the sea
such as strangeness and charm.  While the CCFR data suggest that a difference
between the $s$ and $\overline s$ distributions cannot be excluded, the
distributions appear to be more strongly correlated than a kaon-hyperon
cloud can accommodate.  The high $x$ and $Q^2$ EMC data still indicate the
presence of a nonperturbative intrinsic charm component of the nucleon sea
but the exact form of such a component is not fixed.  In both cases, more
data are needed to resolve both the origin and shape of the strange and charm
contributions to the sea, particularly at large $x$ and $Q^2$. 

A relatively short time ago, there was a strong bias that the nucleon sea was
flavor symmetric and that the quark and anti-quark distributions should be 
identical.  These expectations have been exploded as experiments are able to
probe individual quark and anti-quark distributions more precisely.
Increasingly good data and more sophisticated analysis techniques can
hopefully resolve some of the issues presented here.  This field is obviously
very data-driven.  More surprises undoubtedly await those who dive deeper
into the ``cloudy'' depths of the nucleon sea.

{\bf Acknowledgments} I would like to thank W.M. Alberico, D. Armstrong,
A.O. Bazarko, S.J. Brodsky, W. Seligman, and W. Melnitchouk 
for discussions.  I would also like to thank all the authors who provided
figures for this review, including W.M. Alberico, V. Barone,
A. Bodek, C. Boros, H. Christiansen, X. Ji, M. Leitch, J. Magnin, 
A.D. Martin, P.L. McGaughey, F. Navarra, and F.M. Steffens.


\begin{thebibliography}{999}
\bibitem{xf} The definition of $x_F$ is $x_F = (x_1 - x_2) = 2 p_l/\sqrt{s}$,
             where $p_l$ is the longitudinal momenta of the muon pair
             in the center of mass and $s$ is the total center of mass
             energy squared.
\bibitem{distribs}
The parton distributions used in the analysis of Ref.~\cite{CCFRNLO}
are parameterized at a reference scale $\mu_0^2$ by 
$x q_{NS}(x,\mu_0^2)  =   A_{NS} x^{\eta_1}(1-x)^{\eta_2} +
B_{NS}x^{\eta_3}$,
$x q_{SI}(x,\mu_0^2)  =   x q_{NS}(x,\mu_0^2) + A_{\rm sea}
(1-x)^{\eta_{\rm sea}}$ and
$x g(x,\mu^2_0)  =  A_g(1-x)^{\eta_g}$,  
and evolved to all other values of $\mu^2$ using
the DGLAP \cite{DGLAP}
evolution equations. 
$A_{NS}$ is constrained by the fermion conservation sum rule
$\int {dx\over x} \: xq_{NS}(x) = 3$,
and $A_g$ is constrained by the momentum sum rule
$\int dx \:\left[ xq_{SI}(x) + xg(x)\right] = 1$.
The final fit parameter is the strong coupling constant, given by
$\Lambda_{\overline{\rm MS}}$, for a total of eight fit parameters.
At $\mu_0^2=1\;{\rm GeV^2/c^2}$, the parameters are 
$\Lambda_{\overline{\rm MS}} = 0.210 \pm 0.021$ GeV/c,
$\eta_1=0.947 \pm 0.027$,
$\eta_2=3.81 \pm 0.059$,
$\eta_3= 0.405 \pm 0.047$,
$B_{NS}= 0.0210 \pm 0.0039$,
$A_{\rm sea}=1.404 \pm 0.047$,
$\eta_{\rm sea}=6.49 \pm 0.22$,
$\eta_g=3.18\pm 0.70$, where the errors are statistical only.
\end{thebibliography}
\end{document}